\documentclass[twocolumn]{aastex63}

\usepackage{textcomp}
\usepackage{gensymb}
\usepackage{graphicx}
\usepackage{amsmath}
\usepackage{amssymb}
\usepackage{csquotes}
\usepackage{multirow}

\newcommand{\payne}{ZETA-PAYNE}
\newcommand{\teff}{$T_{\rm eff}$}
\newcommand{\logg}{$\log\,g$}
\newcommand{\vsini}{$v\,\sin\,i$}
\newcommand{\kms}{km\,s$^{-1}$}

\received{December 6, 2021}
\revised{December 6, 2021}
\accepted{\today}
\submitjournal{AJ}

\shorttitle{\payne}
\shortauthors{Straumit et al.}
\graphicspath{{./}{figures/}}

\begin{document}

\title{\payne: a fully automated spectrum analysis algorithm for the Milky Way Mapper program of the SDSS-V survey}

\correspondingauthor{Ilya Straumit}
\email{ilya.straumit@kuleuven.be}

\author{Ilya Straumit}
\affiliation{The Department of Astronomy and Center of Cosmology and AstroParticle Physics, The Ohio State University, Columbus, OH 43210, USA}
\affiliation{Institute of Astronomy, KU Leuven, Celestijnenlaan 200D, 3001 Leuven, Belgium}

\author{Andrew Tkachenko}
\affiliation{Institute of Astronomy, KU Leuven, Celestijnenlaan 200D, 3001 Leuven, Belgium}

\author{Sarah Gebruers}
\affiliation{Institute of Astronomy, KU Leuven, Celestijnenlaan 200D, 3001 Leuven, Belgium}
\affiliation{Max Planck Institute for Astronomy, K\"onigstuhl 17, 69117 Heidelberg, Germany}

\author{Jeroen Audenaert}
\affiliation{Institute of Astronomy, KU Leuven, Celestijnenlaan 200D, 3001 Leuven, Belgium}

\author{Maosheng Xiang}
\affiliation{Max Planck Institute for Astronomy, K\"onigstuhl 17, 69117 Heidelberg, Germany}

\author{Eleonora Zari}
\affiliation{Max Planck Institute for Astronomy, K\"onigstuhl 17, 69117 Heidelberg, Germany}

\author{Conny Aerts}
\affiliation{Institute of Astronomy, KU Leuven, Celestijnenlaan 200D, 3001 Leuven, Belgium}
\affiliation{Max Planck Institute for Astronomy, K\"onigstuhl 17, 69117 Heidelberg, Germany}
\affiliation{Department of Astrophysics, IMAPP, Radboud University Nijmegen, PO Box 9010, 6500 GL Nijmegen, The Netherlands}

\author{Jennifer A. Johnson}
\affiliation{The Department of Astronomy and Center of Cosmology and AstroParticle Physics, The Ohio State University, Columbus, OH 43210, USA}

\author{Juna A. Kollmeier}
\affiliation{The Observatories of the Carnegie Institution for Science, 813 Santa Barbara St., Pasadena, CA 91101}

\author{Hans-Walter Rix}
\affiliation{Max Planck Institute for Astronomy, K\"onigstuhl 17, 69117 Heidelberg, Germany}

\author{Rachael L. Beaton}
\affiliation{Department of Astrophysical Sciences, Princeton University, 4 Ilvy Lane, Princeton, NJ 08544}
\affiliation{The Observatories of the Carnegie Institution for Science, 813 Santa Barbara St., Pasadena, CA 91101}

\author{Jennifer L. Van Saders}
\affiliation{Institute for Astronomy, University of Hawai'i, 2680 Woodlawn Drive, Honolulu, HI 96822, USA}

\author{Johanna Teske}
\affiliation{Earth and Planets Laboratory, Carnegie Institution for Science, 5241 Broad Branch Road, NW, Washington, DC 20015, USA}

\author{Alexandre Roman-Lopes}
\affiliation{Departamento de Astronomia, Facultad de Ciencias, Universidad de La Serena. Av. Juan Cisternas 1200, La Serena, Chile}

\author{Yuan-Sen Ting}
\affiliation{Research School of Astronomy \& Astrophysics, Australian National University, Cotter Rd., Weston, ACT 2611, Australia}
\affiliation{School of Computing, Australian National University, Acton ACT 2601, Australia}

\author{Carlos G. Román-Zúñiga}
\affiliation{Universidad Nacional Autónoma de México, Instituto de Astronomía, AP 106, Ensenada 22800, BC, México}

\begin{abstract}

The Sloan Digital Sky Survey has recently initiated its 5th survey generation (SDSS-V), with a central focus on stellar spectroscopy. In particular, SDSS-V’s Milky Way Mapper program will deliver multi-epoch optical and near-infrared spectra for more than $5\times10^6$ stars across the entire sky, covering a large range in stellar mass, surface temperature, evolutionary stage, and age. About 10\% of those spectra will be of hot stars of OBAF spectral types, for whose analysis no established survey pipelines exist. Here we present the spectral analysis algorithm, \payne, developed specifically to obtain stellar labels from SDSS-V spectra of stars with these spectral types and drawing on machine learning tools. We provide details of the algorithm training, its test on artificial spectra, and its validation on two control samples of real stars. Analysis with \payne\ leads to only modest internal uncertainties in the near-IR with APOGEE (optical with BOSS): 3-10\% (1-2\%) for \teff, 5-30\% (5-25\%) for \vsini, 1.7-6.3 \kms(0.7-2.2 \kms) for RV, $<0.1$~dex ($<0.05$~dex) for \logg, and 0.4-0.5 dex (0.1 dex) for [M/H] of the star, respectively. We find a good agreement between atmospheric parameters of OBAF-type stars when inferred from their high- and low-resolution optical spectra. For most stellar labels the APOGEE spectra are (far) less informative than the BOSS spectra of these stars, while \logg, \vsini, and [M/H] are in most cases too uncertain for meaningful astrophysical interpretation. This makes BOSS low-resolution optical spectra better for stellar labels of OBAF-type stars, unless the latter are subject to high levels of extinction.

\end{abstract}



\section{Introduction}\label{sec:intro}

Recent successes of space-based astrometric \citep[e.g., Gaia,][]{2016A&A...595A...1G} and photometric \citep[CoRoT, {\it Kepler\/}, K2, and TESS,][respectively]{Auvergne2009, 2010Sci...327..977B,2014PASP..126..398H,2019ESS.....433312V} missions implied a boost for stellar astrophysics. Nevertheless, (ground-based) stellar spectroscopy still occupies an important niche, thanks to the complementary nature of information it adds to the photometric and astrometric measurements. Indeed, even low-resolution (R$\sim$2000-5000) spectroscopy offers unprecedented level of detail in the analysis of stellar atmospheres as compared to (broad-band) space-based photometric measurements, allowing us to resolve important diagnostic spectral lines and their blends.

Though space-based (photometric) missions deliver high-quality, high-duty cycle, and nearly uninterrupted time-series of data full of information, a considerable amount of stellar astrophysics applications, both at the level of individual objects and their ensembles, require precise atmospheric parameters and chemical compositions of stars, as well as estimates of their surface rotation and radial velocities. For example, (transiting) exoplanet studies often rely on ground-based spectroscopic measurements for the inference of planetary masses and properties of their host stars \citep[e.g.,][]{Danielski2021}. The study of planetary atmospheres through the method of transmission spectroscopy is another application \citep[e.g.,][]{2018haex.bookE.100K, 10.1117/12.2562371}. Eclipsing binary studies also require spectroscopic observations to deduce masses and radii of both components with precision and accuracy better than 3\% level \citep[e.g.,][]{Torres2010,2021MNRAS.504.5221T,2021A&ARv..29....4S}, enabling stringent tests of stellar structure and evolution theory \citep[e.g.,][and references therein]{Claret2019,Tkachenko2020}. Furthermore, detailed asteroseismic studies of pulsating stars require knowledge of their atmospheric chemical compositions and precise locations in the (\teff-\logg) Kiel or ($T_{\rm eff}$-$\log\,L$) Hertzsprung-Russel (HR) diagram, to enable observational probing of the physical conditions in the deep interiors of stars \citep[e.g.,][]{Aerts2010,Aerts2021}. 


Although detailed studies of individual objects are extremely important to assess the precision and accuracy of numerous theories and to provide recipes for their improvement, large-scale studies of (single, binary, and high-order multiple system) stars and their ensembles (open and globular clusters, star-forming regions, Galactic bulge, etc.) are vital to understand the structure, dynamics, and evolution of galaxies \citep[e.g.,][]{Zari2021}. To that end, large-scale, all-sky, multi-epoch ground-based spectroscopic surveys are irreplaceable both as a stand-alone mechanism for astrophysical studies as well as a complement to all-sky astrometric and photometric space missions like Gaia and TESS, respectively. To mention a few, the Sloan Digital Sky Survey (SDSS) has a two-decades long tradition in large-scale ground-based observations, starting with the SDSS-I survey devoted to imaging and spectroscopy of galaxies and quasars \citep{2002AJ....123.2945R, 2002AJ....124.1810S}, and now proceeding into its fifth phase/generation (SDSS-V, see Section~\ref{sect:SDSS-V}) that will survey over five million stars in the Milky Way, study interstellar gas in the Galaxy and Local Group, and will track evolution of massive black holes growing at the centers of galaxies \citep{Kollmeier2017}. The Large Sky Area Multi-Object Fibre Spectroscopic Telescope (LAMOST) provides a large collection of low-resolution (R$\sim$1800) spectra of stars of all spectral types through its Experiment for Galactic Understanding and Exploration (LEGUE) survey of the Milky Way structure \citep[e.g.,][]{2012RAA....12..735D}. Furthermore, the Gaia-ESO spectroscopic survey \citep{2012Msngr.147...25G} targets over 10$^5$ stars in the Milky Way, complementing Gaia astrometric observations and providing the first homogeneous overview of the distributions of kinematics and elemental abundances in the Galaxy. The GALactic Archaeology with HERMES  \citep[GALAH,][]{2015MNRAS.449.2604D} survey aims to survey over a million of stars of different ages and at different locations in the Milky Way to uncover its formation and evolution history. Last but not least, the WHT Enhanced Area Velocity Explorer \citep[WEAVE,][]{2016sf2a.conf..267B} and the 4-metre Multi-Object Spectroscopic Telescope \citep[4MOST,][]{2019Msngr.175....3D} surveys allow for multi-object low- to medium-resolution spectroscopic observations, enabling chemical and kinematic studies of all components of the Milky Way, extragalactic science through observations of quasars, etc.

Analysis of this large volume of data is hardly a manageable task when relying on human power only. However, various machine learning (ML) based applications have proven the task to be rather easily accomplished by computers, provided respective algorithms can be properly trained. For example, the Cannon APOGEE spectrum analysis pipeline \citep{2015ApJ...808...16N, 2016arXiv160303040C} creates from the spectra of reference stars with known stellar labels (e.g., \teff, \logg, and [M/H]) a flexible generative model that describes a probability density function for continuum normalized stellar flux as a function of the above-mentioned labels. The algorithm assumes that the continuum normalized flux varies smoothly with the stellar labels, enabling fast and precise determination of stellar parameters from the previously unseen normalized spectra of the surveyed stars. The Cannon algorithm is often referred to as a ``data-driven approach’’ owing to the limited model dependency that occurs only at the stage of the algorithm training. Another example of, this time, a fully model-dependent ML-based approach is the {\sc (Hot)Payne} \citep{2019ApJ...879...69T,Xiang2021} algorithm that employs a neural network as an efficient predictor of (continuum normalized) synthetic spectra. When teamed up with an optimization algorithm, the approach offers a fast way to analyze large volumes of spectroscopic data in the parameters space the neural network has been trained in. Even a higher level of data analysis is achieved with ML-based algorithms that employ the method of domain adaptation and offer unique opportunity to improve theoretical models by learning from actual observations, as for example realized in the {\sc cycle-starnet} algorithm \citep{2021ApJ...906..130O}.

Any spectrum analysis starts with the most suitable processing of raw stellar spectra that includes optimal extraction (bias subtraction, flat fielding, wavelength calibration, order merging in case of echelle spectra, etc.) and (optionally) normalization to the local continuum of the star. In medium- to large-scale surveys, where thousands to millions of stellar spectra have to be processed, optimal data processing requires dedicated data reduction pipelines, and the SDSS-V Milky Way Mapper (MWM) Survey is no exception. As discussed in \citet{Kollmeier2017} and Section 2, MWM relies on the optical BOSS \citep{2013AJ....146...32S} and near-infrared APOGEE \citep{2019PASP..131e5001W} spectrographs to execute its science. Each of those instruments has a dedicated data reduction pipeline that deliver one-dimensional, merged, wavelength and flux calibrated stellar spectra ready for a subsequent detailed astrophysical analysis and interpretation \citep{Stoughton2002,Nidever2015}.

Compared to most previous stellar spectroscopic surveys, MWM will take spectra of many different stars, from red giants to OBA-type stars, to cool YSOs and X-ray binaries. These cannot be modeled astrophysically with a single pipeline. Therefore, the MWM Survey employs an overarching software framework called {\sc astra}\footnote{https://github.com/sdss/astra}. In a nutshell, {\sc astra} takes the 1D optimally reduced spectra as input, passes those through fully automated classifiers, makes a machine learning based probabilistic decision on which data analysis pipeline(s) to call for that particular input spectrum, and collects and formats the pipeline's output, i.e. the stellar ``labels'' (e.g., \teff, \logg, abundances, multiplicity, etc.). To minimize the impact of possibly imprecise and/or erroneous classification, the three highest probability classes are considered for each stellar spectrum that passes through the classification module of {\sc astra}. 


In this paper, we build upon the legacy of the {\sc (Hot)Payne} algorithm \citep{2019ApJ...879...69T,Xiang2021} to develop a
machine-learning based (ML-based) spectrum analysis method for the Milky Way Mapper program of the SDSS-V survey, with a particular focus on intermediate- and high-mass stars of spectral types O, B, A, and F, irrespective of whether those are observed with the BOSS low-resolution optical or the APOGEE medium-resolution near-IR instrument. Our algorithm is implemented in the {\sc astra} software framework and is employed for the analysis of spectra of OBAF-type stars targeted by MWM.

Section~\ref{sect:SDSS-V} provides a brief introduction to the SDSS-V survey and its Milky Way Mapper program, as well as a summary of the primary science questions that will be addressed with the sample of OBA(F)-type stars. Our spectrum analysis method is presented in Section~\ref{sec:methods}, where we describe and justify the changes made to the original {\sc (Hot)Payne} algorithm, discuss the training process in detail, and introduce the optimization and statistical uncertainty determination framework. The algorithm is tested on simulated APOGEE and BOSS spectra in Section~\ref{sec:sim_tests} and is further validated on low- to high-resolution spectra of real stars in Section~\ref{sec:control_samples_tests}. We conclude the paper with a discussion and future prospects presented in Section~\ref{sec:conclusions}.

\section{The SDSS-V and Milky Way Mapper surveys}\label{sect:SDSS-V}

The SDSS-V survey has kicked off its observations in the fall of 2020 and will remain on sky for a total survey duration of up to five years. SDSS-V is an all-sky, multi-epoch survey whose spectroscopic observations will be acquired in the optical and near-IR wavelength domains with matched infrastructures in both hemispheres \citep{Kollmeier2017}. The survey serves as an umbrella for its three overarching scientific projects, called ``Mappers’’: Milky Way Mapper (MWM), Local Volume Mapper (LVM), and Black Hole Mapper (BHM). In brief, the LVM survey\footnote{\url{https://www.sdss.org/dr15/future/lvm}} is an optical, integral-field spectroscopic survey that will target the mid plane of the Milky Way, Orion, and the Magellanic Clouds, using newly built telescopes operating at a resolving power of R$\sim$4000. The ultimate goal of LVM is to address questions of star formation and physics of the interstellar medium through mapping the interstellar gas emission with unprecedented spatial resolution: sub-parsec in the Galaxy, 10 parsec in the Magellanic Clouds, and $<$100 parsec out to distances of several Mpc. The BHM survey\footnote{\url{https://www.sdss.org/dr15/future/bhm}} will employ 2.5-meter telescopes  in both hemispheres, at Apache Point Observatory \citep[][]{2006AJ....131.2332G} and at Las Campanas Observatory \citep[][]{Bowen:73}, to acquire multi-epoch, optical low-resolution (R$\sim$2000) BOSS spectroscopy of some 300 000 quasars. The ultimate goal of BHM is to understand the masses, accretion physics, and growth and evolution of supermassive black holes in the centers of galaxies.

The Milky Way Mapper survey\footnote{\url{https://www.sdss.org/dr15/future/mwm}} of SDSS-V will employ both the BOSS low-resolution (R$\sim$2000) optical and the APOGEE medium-resolution (R$\sim$22500) near-IR spectroscopy. The survey will target over 4 million objects to provide a dense and contiguous stellar map across the sky but largely focused on low Galactic latitudes. High signal-to-noise ratio (S/N), medium-resolution near-IR spectra will be used to deduce stellar parameters and surface chemical composition of each star included in the program, providing the means to understand the dominant formation mechanisms of the Milky Way and its place in a cosmological context. The MWM survey will also make use of both optical and near-IR spectrographs to acquire multi-epoch observations of tens of thousands multi-star and planetary systems to understand formation, shaping, and evolution of (sub-)stellar multi-companion systems. Last but not least, MWM will target stellar objects in a high-dimensional  parameter space, including stellar mass, age, evolutionary status, chemical composition, rotation, and internal structure. Among the groups of stars that will be observed are young stellar objects, main-sequence stars, red giants, white dwarfs, low- ($M\lesssim$1.2~M$_{\odot}$), intermediate- (1.2~M$_{\odot}\lesssim M \lesssim$ 8~M$_{\odot}$), and high-mass ($M\gtrsim 8$~M$_{\odot}$) stars, etc. The MWM survey will address a wide range of scientific questions, such as: true relationships between masses, radii, rotation, ages, and internal mixing of intermediate- to high-mass stars; precise age and chemical composition measurements of giant stars with asteroseismic detections; improved understanding of evolution of white dwarfs and their return to the interstellar matter; observations of deeply embedded stellar clusters and a volume-limited (within $\sim$100~pc) census of stars in the solar neighborhood. We refer the reader to \citet{Kollmeier2017} for more information about the SDSS-V survey and its Mappers.

\subsection{Intermediate- to high-mass OBA(F)-type stars}

The MWM survey of SDSS-V has a large science component devoted to intermediate- and high-mass stars of spectral types O, B, and A(F). Sample selection is done in the spirit of the SDSS-V requirement of a known and well-defined selection function, and is therefore based on Gaia EDR3 photometry and astrometry combined with the 2MASS photometric information. First of all, the sample is restricted to sources with Gaia $G_{\rm mag}<16$ due to the spectral-level requirement of the MWM survey that S/N of $\sim75$ has to be reached with 15 min exposures with the BOSS instrument. Secondly, all objects whose absolute magnitude in the $K$ photometric band is smaller than zero are selected, which roughly corresponds to a late B-type main-sequence star. Ultimately, several color cuts are applied to clean the sample from intrinsically bright red giant- and asymptotic giant-branch stars, as well as from objects with unnaturally blue colors. The final catalog comprises some 0.9M objects, where the fraction of O- and B-type stars is estimated to be close to 50\%. The other half of the sample is largely comprised of A-type stars, though a small contamination from F-type stars cannot be excluded. More details about the target selection and estimation of the purity and completeness of the catalog are provided in \citet{Zari2021}.

Following its science requirements, the MWM survey will deliver atmospheric properties of OBA(F)-type stars with the precision better than 5-10\% for the effective temperature \teff, better than some 25-30\% for the projected rotational velocity \vsini, $\sim$0.1~dex for the surface gravity \logg, and $\sim$0.15~dex for the bulk metallicity [M/H] and surface abundances of critical chemical elements such as He, C, N, O, Si, and Mg (perhaps more limited for the hottest O-type stars). With such a large sample of multi-epoch spectra we can probe: 
\begin{itemize}
    \item The intrinsic variability (stellar pulsations, rotational modulation, quasi-periodic variability, etc.), internal properties (interior rotation and mixing, convective core masses, etc.), and ages of intermediate- to high-mass stars, thanks to synergy with the {\it Kepler\/} and TESS space-based photometric surveys. For example, we will be able to better assess fractions of Ae/Be stars and magnetic intermediate- to high-mass stars, where the latter can be unraveled by indirect means through photometric detections of rotational modulation and spectroscopic inferences of surface chemical abundance anomalies. Stellar pulsations will be used in combination with spectroscopically inferred atmospheric parameters to learn about internal physical properties of stars, using well-established asteroseismic methods 
    \citep{2018ApJS..237...15A,2019ARA&A..57...35A,Aerts2021}. In particular, it was  recently demonstrated that internal rotation and mixing properties of B-type main-sequence stars can be readily inferred from a combined asteroseismic and spectroscopic analysis \citep[e.g.,][]{Papics2017,Pedersen2018,Pedersen2021}. Moreover, \citet{Bowman2019,Bowman2020} demonstrate that high-mass O-type stars and evolved B-type supergiants also hold strong asteroseismic potential even at metallicities as low as that of the LMC, thanks to the observational detection of low-frequency stochastic variability such as the one caused by internal gravity waves excited at the interface of convective and radiative regions near the stellar core \citep{Edelmann2019,Horst2020};
    \item binary and multiplicity fractions among intermediate- to high-mass stars across the sky and as a function of metallicity (Galaxy vs. Magellanic Clouds). \citet{Sana2012,Sana2013} find a large fraction of binaries among massive stars and hypothesize that almost all high-mass stars have gone through some sort of binary interactions in the course of their evolution. Notably, \citet{Almeida2017} report some 60\% observed binary fraction among O-type stars in the 30~Dor region of the LMC, while \citet{Bodensteiner2021} and \citet{Banyard2021} find similar binary fractions for B-type stars in the SMC and Galaxy, respectively. \citet{Luo2021} report a comparable binary fraction of $\sim$40\% from their study of some 330 OB-type stars observed by LAMOST having at least three spectroscopic epochs. The Milky Way Mapper Survey of SDSS-V will deliver optical and near-IR measurements for a three orders of magnitude larger sample of OBA(F)-type stars, thus allowing for a homogeneous search for binary and higher-order multiple systems in the Galaxy and Magellanic Clouds;
    \item the structure of the Galaxy as revealed by its young stellar components, in particular the kinematics, dynamics, and nature of the Galactic spiral arms. \citet{Zari2021} demonstrate that the structure of the Milky Way as traced with young OB-type stars is not necessarily the same as deduced from observations of red giants, from analysis of the distribution of dust in the Galaxy, or from the distributions of Cepheids and/or masers (see also \citet[][]{2021A&A...651A.104P}). The authors find that the distribution of OBA-type stars in the plane is highly structured, with pronounced over- and under-densities, and conclude that young stars in the Galaxy are not neatly organized into distinct spiral arms. In agreement with previous studies of OB-type stars in literature, \citet{Zari2021}'s findings might point either to a more flocculent structure of the Milky Way at optical wavelengths or to the fact that star formation occurs in a clumpy and patchy fashion. The authors emphasize that better quality Gaia DR3 data and spectroscopic information from the SDSS-V survey will allow them to assess different models of spiral arms and shed light on their nature.
\end{itemize}

\begin{figure*}
    \centering
    \includegraphics[width=0.43\textwidth]{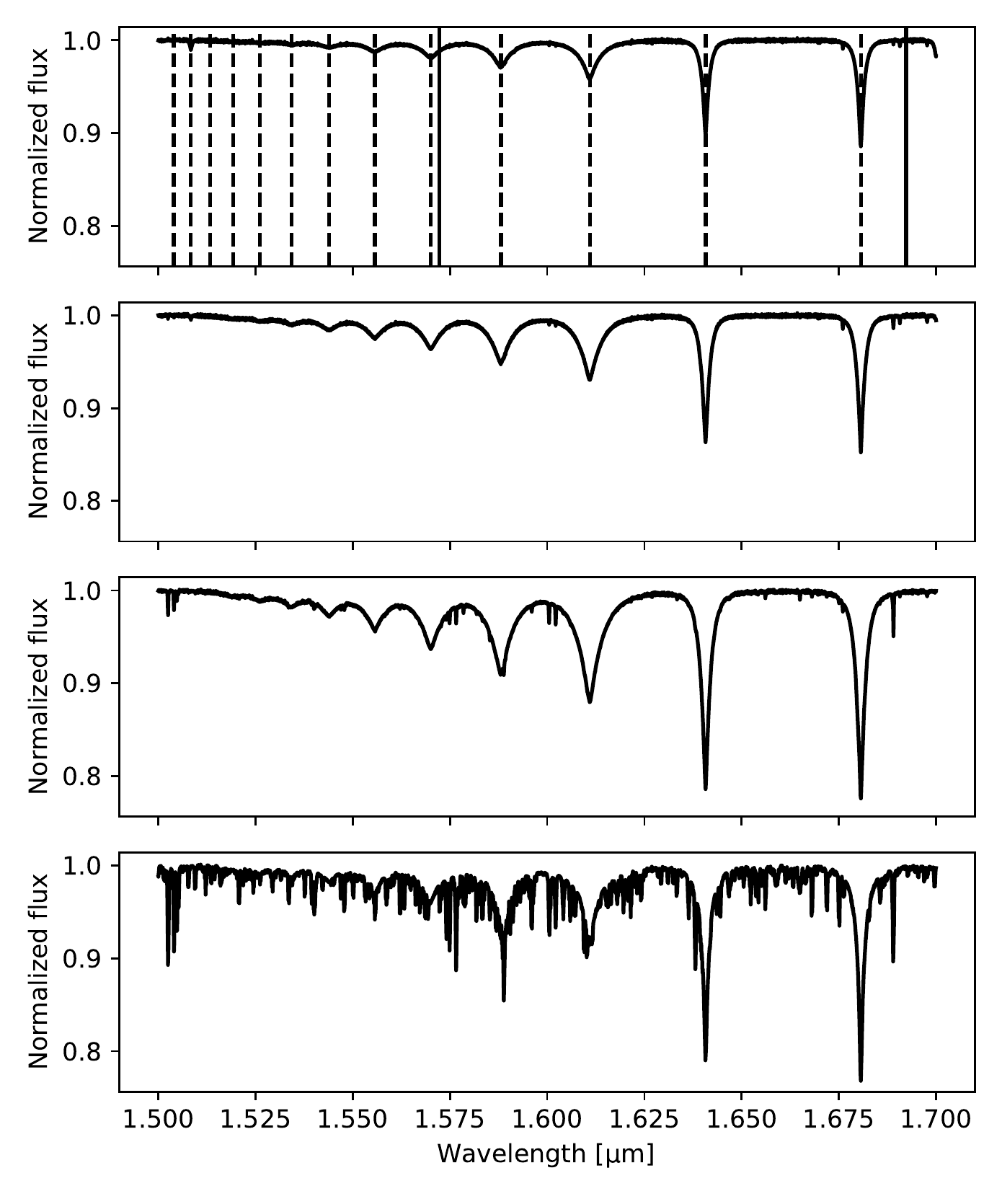}
    \includegraphics[width=0.45\textwidth]{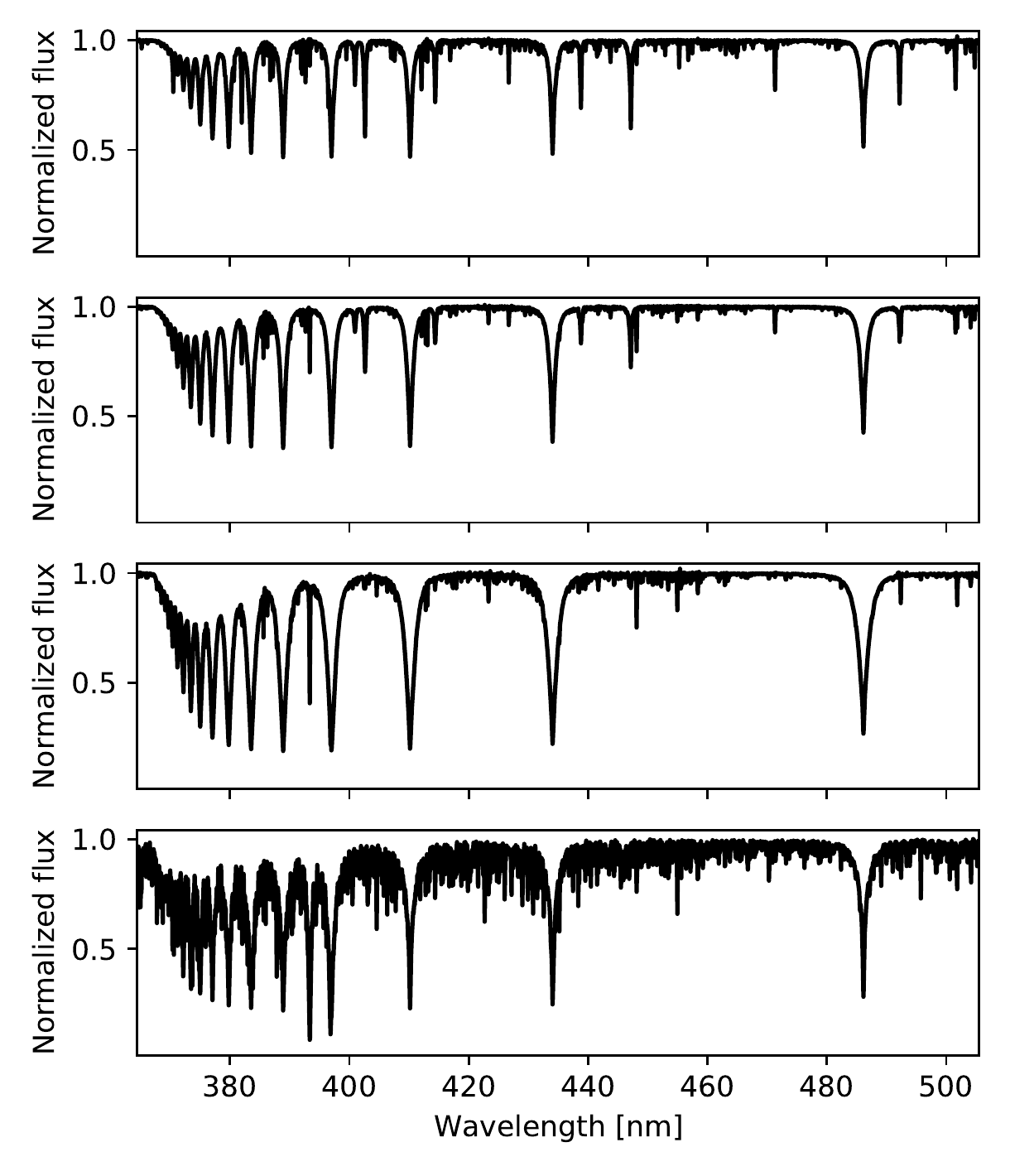}
    \caption{Left: normalized synthetic spectra of (from top to bottom) OBAF-type stars computed with the {\sc gssp} software package in the APOGEE wavelength range. Vertical dashed lines in the top plot indicate positions of hydrogen lines of the Brackett series, while vertical solid lines mark positions of the \ion{He}{2} lines at 1.5722 \micro m and 1.6923 \micro m, whose wavelengths are taken from \citet{2004A&A...422..275L}. Right: normalized synthetic spectra for the same stellar parameters in the BOSS wavelength range (only part of the full range is shown for clarity).}
    \label{fig:SynthAPOGEESpectra}
\end{figure*}

\section{Spectrum analysis algorithm}\label{sec:methods}

As it was briefly mentioned in Section~\ref{sec:intro}, we aim to develop an efficient spectrum analysis algorithm for the Milky Way Mapper survey of SDSS-V, with the primary focus on intermediate- to high-mass stars of spectral types O, B, A, and F. One of the main requirements at this stage is a wide applicability range of the algorithm in terms of the wavelength coverage and resolving power of the obtained spectra, and without the need for human intervention between data reduction pipelines and spectrum analysis itself. Here, we use the heritage of one of the publicly available ML-based methods, namely the {\sc (Hot)Payne} algorithm \citep{2019ApJ...879...69T,Xiang2021}, to which we introduce a number of modifications to comply with the aforementioned primary requirement of MWM.

\begin{table*}
    \centering
    \caption{Parameters of the training sets used for training neural networks in this study. See text for details.}
    \begin{tabular}{llll}
    \hline\hline
    \multirow{2}{*}{Parameter} & near-IR & \multicolumn{2}{c}{optical} \\
    & APOGEE & BOSS & HERMES \\
    \hline
    Wavelength range, \AA & 15\,000...17\,000 & 3\,600...10\,400 & 4\,200...5\,800 \\
    Wavelength step, \AA & 0.05 & 0.06 & 0.015 \\
    Number of models & \multicolumn{3}{c}{5000 Sobol + 5000 Gaussian} \\
    $\mathrm{T_{eff}}$, K & \multicolumn{3}{c}{Sobol(6000...25000) + Gaussian(6000, 3000)} \\
    $\log\,g$, dex & \multicolumn{3}{c}{Sobol(3.0...5.0)} \\
    $v\,\sin\,i$, km\,s$^{-1}$ & \multicolumn{3}{c}{Sobol(0...400) + Gaussian(0, 25)} \\
    $\mathrm{[M/H]}$, dex & \multicolumn{3}{c}{Sobol(-0.8...0.8)} \\
    \hline\hline
    \end{tabular}
    \label{tab:grids_params}
\end{table*}

It is quite common to use optical low- to high-resolution spectra to deduce atmospheric parameters of intermediate- to high-mass OBAF-type stars, unless there is an interest in, e.g., specific UV lines to study winds of massive stars, etc. On one hand, optical spectra of late-type stars will often contain important diagnostic lines of hydrogen (the Balmer series) and/or helium, a large number of metal lines (e.g. C, N, O, Mg, Si, Al, etc.), and (in many cases) quite well-defined regions of pseudo-continuum, making it possible to achieve high-quality spectrum normalization. To give a few examples, methods such as Spectroscopy Made Easy\footnote{\url{https://www.stsci.edu/~valenti/sme.html}} \citep[SME,][]{Valenti1996,Piskunov2017}, Grid Search in Stellar Parameters\footnote{\url{https://fys.kuleuven.be/ster/meetings/binary-2015/gssp-software-package}} \citep[GSSP,][]{Tkachenko2015}, the {\sc detail} \citep{Giddings1981} and {\sc surface} \citep{Butler1984} suite of codes, are all designed to work with continuum-normalized observed spectra and are most often applied in the optical wavelength range. On the other hand, use of near-IR spectra for the analysis of intermediate- to high-mass stars is rather scarce owing to few diagnostic (metal) lines occurring at those wavelengths and to problems associated with often uncertain normalization of spectra to the pseudo-continuum. \citet{Roman-Lopes2018} demonstrate that the APOGEE near-IR spectra (wavelength coverage from 1.5 to 1.7~\textmu m) of OB-type stars are rather featureless and display exclusively hydrogen lines of the Brackett series and at most two \ion{He}{2} lines at higher temperatures corresponding to O-type stars. The authors, alongside \citet{Ramirez-Preciado2020}, develop a (semi-)empirical spectral classification method for hot OB-type stars based on equivalent width (EW) measurements of key spectral lines in the APOGEE spectra, and making use of the SPM instrument medium-resolution and LAMOST survey low-resolution optical spectra, respectively, to benchmark their relations. Their method has recently been extended and applied to the APOGEE spectra of cooler A-type stars (Ramirez-Preciado et al., in revision).

Here, we aim to develop a spectrum analysis approach that can be applied to APOGEE spectra of OBAF-type stars (where high density of hydrogen lines prevents precise normalization of spectra to the local continuum) and to BOSS optical spectra without the need for a substantial modification of the algorithm. That said, specifics of the APOGEE near-IR spectra of OBAF-type stars is what drives our definition of the model spectrum as described in Section~\ref{sect:model_spectrum}. However, it does not mean that we value the medium-resolution near-IR spectra more than their low-resolution optical counterpart. On the contrary, as we demonstrate and conclude in Sections~\ref{sec:control_samples_tests} and \ref{sec:conclusions}, respectively, low-resolution optical spectra of OBAF-type stars will often contain more information than medium-resolution near-IR spectra of these objects, hence both types of data deserve our attention in equal proportions.

\begin{figure*}
    \centering
    \includegraphics[width=0.44\textwidth]{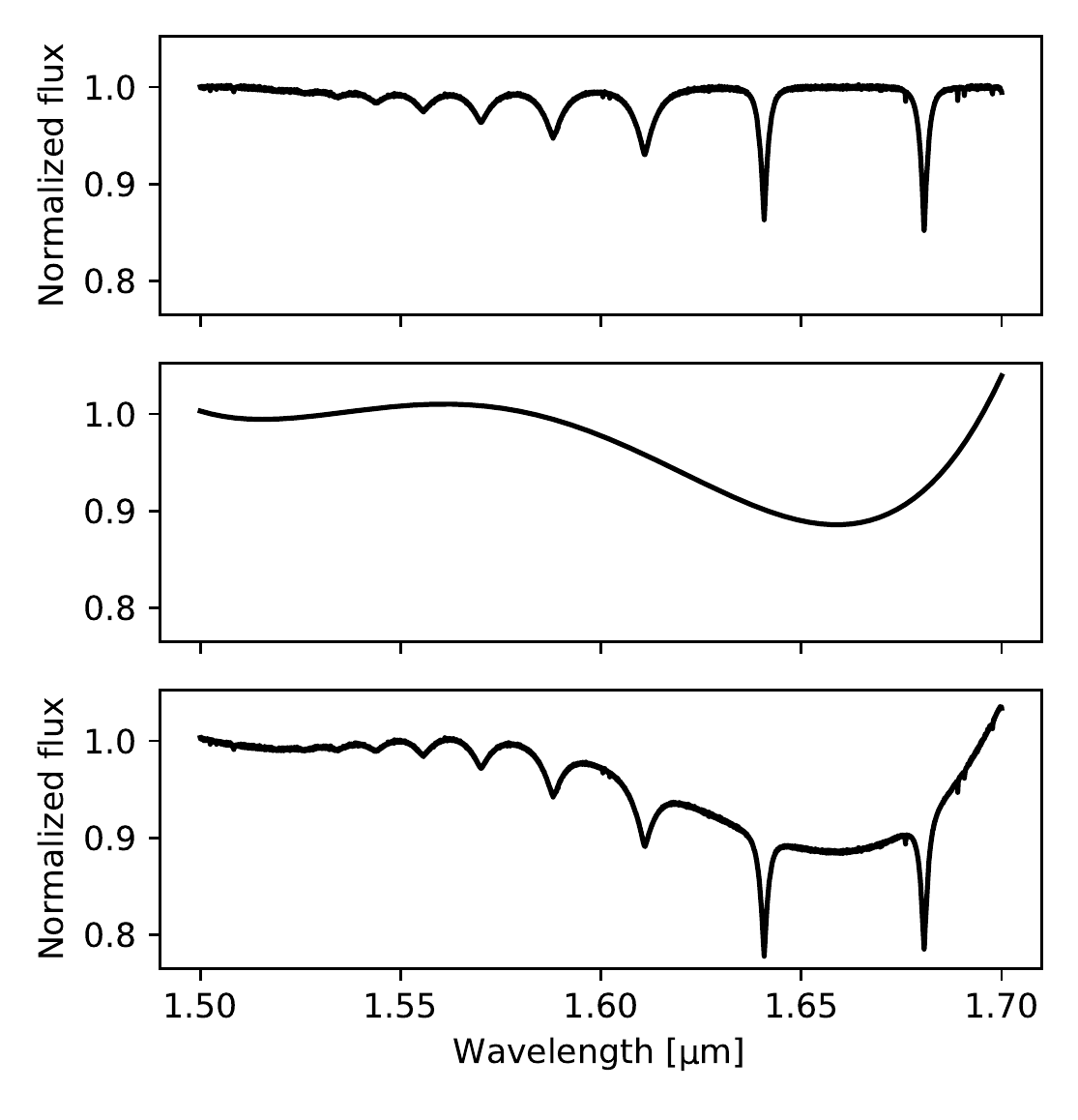}
    \includegraphics[width=0.46\textwidth]{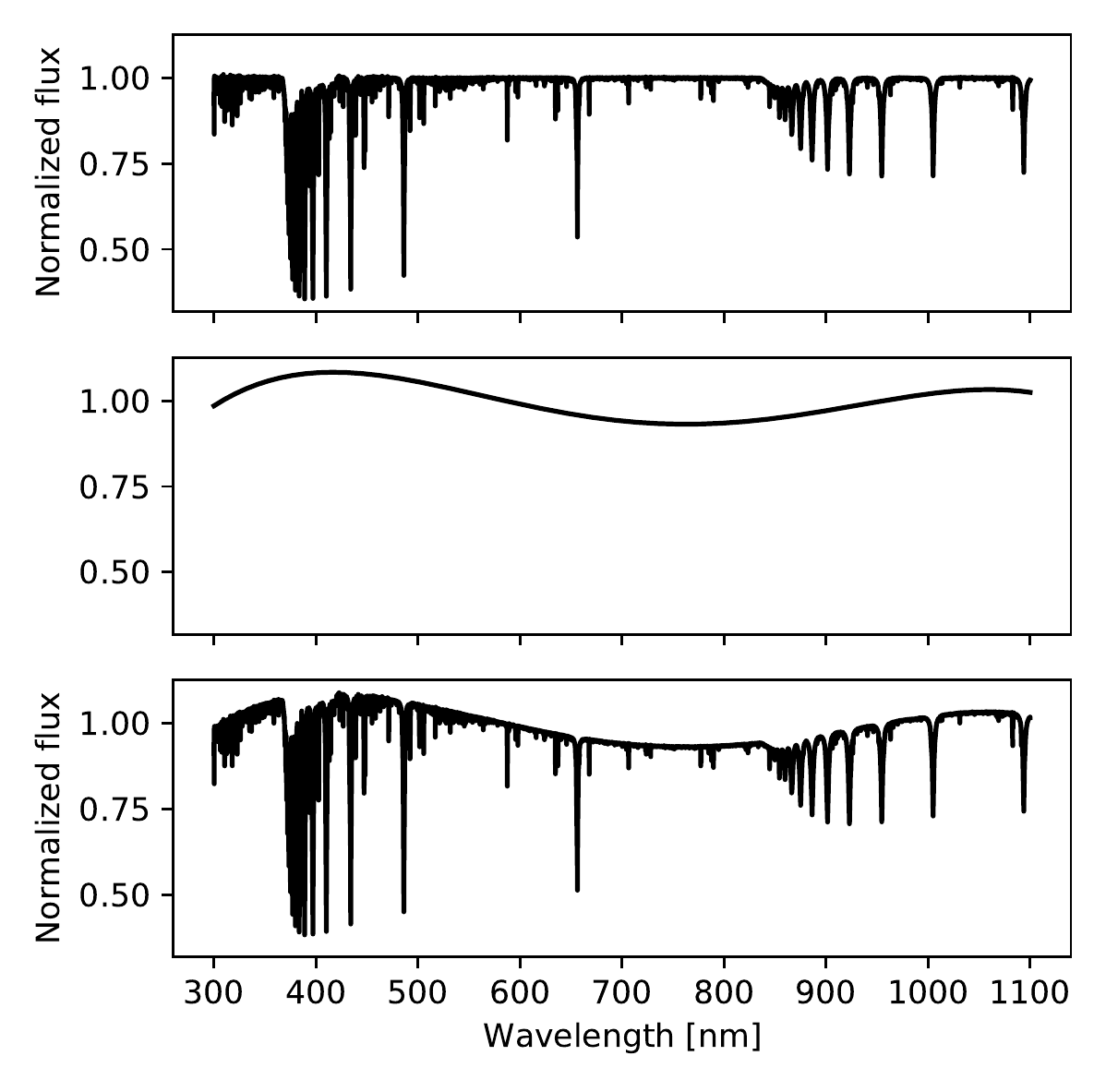}
    \caption{From top to bottom: synthetic APOGEE (left) or BOSS (right) spectrum of a B-type star, random fifth-order Chebyshev polynomial (model of the unknown response function), and a product of the synthetic spectrum with the model of the response function. }
    \label{fig:ModelSpectrumRepresentation}
\end{figure*}

\subsection{Representation of a model spectrum}\label{sect:model_spectrum}

A fully automated analysis of the APOGEE near-IR and BOSS optical spectra of OBAF-type stars in the entire wavelength range requires a careful treatment of pseudo-continuum in the model. As demonstrated in Figure~\ref{fig:SynthAPOGEESpectra} (left panel), the high density of hydrogen lines in the near-IR part of the stellar spectrum, combined with the (typically) moderate- to high rotation of OBAF-type stars, makes determination and placement of the pseudo-continuum extremely difficult, and in the case of hotter OB-type stars often impossible. The same is true for the BOSS optical spectra when the most blue and red parts of the spectrum are included in the analysis, where hydrogen lines of, respectively, the Balmer and Paschen series start merging together. In addition, even when the above-mentioned wavelength intervals are omitted in the analysis, the remaining wavelength range of the BOSS instrument is sufficiently wide to observe significant variations that are instrumental in origin. Therefore, we incorporate a theoretical function into the model spectrum that aims to account for instrumental effects, possible interstellar reddening, and local stellar continuum, thus eliminating the need to process the input observed spectrum in a way beyond its optimal extraction. The instrumental effects that are taken into account in the model are: (i) the wavelength-dependent residual response function of the instrument that produces a large-scale distortion of the observed spectrum, and (ii) the line-spread function (LSF) that causes blurring of spectral lines in the observed spectrum. Observed spectra in the APOGEE entire wavelength range and the BOSS red part of the spectrum are rich in telluric contributions that are removed by the data reduction pipeline using observations of telluric standard stars \citep[typically, rapidly rotating stars of spectral types B and A;][]{Nidever2015}. Thus, a general model of a stellar spectrum, as it comes out of the SDSS data reduction pipeline(s), can be written down as follows:
\begin{equation}
    \mathrm{Flux} = {\rm LSF} \ast [ \prod_i \mathrm{Line_i(T_{eff}, log(g), ...)} ] \times \mathrm{Response}(\lambda).
\end{equation}
Here, LSF is the line-spread function, ${\rm Line}_i$ are the spectral lines formed in the stellar photosphere characterized by a set of parameters \teff, \logg, \vsini, [M/H], $v_{micro}$ (microturbulent velocity), and Response($\lambda$) is a wavelength-dependent theoretical function that accounts for instrumental effects, interstellar reddening, and local stellar continuum contributions. For simplicity and from now onwards, we dub the Response($\lambda$) the ``residual response function''. We include two options to account for the line-spread function of the instrument, a simplified approach where the LSF is represented by a wavelength-independent Gaussian kernel with the full width at half maximum corresponding to a given resolving power {\it R} of the instrument, and a detailed wavelength-dependent model of the LSF that will also be variable from fiber to fiber. The latter is usually estimated from a spectrum of the wavelength calibration unit and is available for both instruments (i.e., BOSS and APOGEE) of the MWM survey. The residual response function is modelled as a series of Chebyshev polynomials, i.e.:
\begin{equation}
    \label{eq:resp}
    \mathrm{Response(\lambda)} = \sum_i c_i T_i(\lambda),
\end{equation}
where $c_i$ are the coefficients of the series, $T_i$ are the Chebyshev polynomials of the first kind, and $\lambda$ is the wavelength. Figure~\ref{fig:ModelSpectrumRepresentation} shows a representative example of the model spectrum of a B-type star in the APOGEE (left) and BOSS (right) wavelength ranges. The normalized synthetic spectrum, a model of the residual response function, and their product are shown in the top, middle, and bottom panel, respectively. 

\subsection{Neural network configuration, training, and validation}\label{sec:NN_train}
Following \citet{2019ApJ...879...69T}, we use a neural network to approximate a grid of model spectra, where the neural network essentially acts as an efficient interpolator in the parameter space defined by the pre-computed grid of models. The main purpose of using a neural network is to make the model spectra differentiable by the stellar parameters, which allows us to perform optimization instead of a full grid search. Aside from that, the neural network is trained on a quasi-random grid as defined in \citet{Sobol1967}, which is more time efficient than computing a full grid of model spectra with the same parameter range. At present, we use the {\sc gssp} software package \citep{Tkachenko2015} to compute a grid of models required for the neural network training. {\sc gssp} employs a 
grid of plane-parallel atmosphere models
pre-computed with the {\sc LLmodels} code \citep{Shulyak2004} coupled to the {\sc SynthV} \citep{Tsymbal1996} line formation code to calculate synthetic spectra in an arbitrary wavelength range. Both codes rely on the local thermodynamical equilibrium (LTE) approximation and include the option to compute the atmospheric structure and detailed line formation for a user-specified chemical composition pattern (including vertical stratification of elements in the stellar atmosphere, if necessary).

The neural network consists of two layers of neurons with ``leaky ReLU'' (leaky rectifier linear unit) activation function, which is defined as follows:
\begin{equation}
    g(x) = \begin{cases}
    x & \text{if } x > 0, \\
    0.01x & \text{otherwise}.
\end{cases}
\end{equation}
Once the neural network has been trained, it takes stellar parameters normalized to the range $[-0.5, 0.5]$ as input and returns the corresponding synthetic spectrum as the output. This means that the number of the neural network outputs is equal to the number of spectral channels in the training set of synthetic spectra. Theoretical spectra for the training set are computed at infinite resolving power and cover wavelength ranges of the SDSS-V BOSS and APOGEE instruments, i.e. 3\,600 - 10\,400~\AA\ and 1.5 - 1.7~\textmu m, respectively. Table~\ref{tab:grids_params} provides a summary of the properties of the training set in terms of the total number of models used, wavelength range coverage, and definition of the parameter space. The network is implemented and trained with the {\sc torch} framework using the {\sc RAdam} optimization algorithm discussed in detail in \citet{2019arXiv190803265L}.

We note that the optical wavelength range specified for the HERMES instrument in Table~\ref{tab:grids_params} that we use for the algorithm validation in Section~\ref{sec:control_samples_tests} is shorter than its full wavelength coverage from 3\,770~\AA\ to 9\,000~\AA. This is due to the memory limitation of the computing device currently used for the training, as the large number of models and the fine wavelength step of 0.015~\AA\ dictated by the high-resolution of the HERMES instrument do not allow us to train a neural network for the full wavelength range. We choose to work with the wavelength interval from 4\,200~\AA\ to 5\,800~\AA\ for the following reasons: 1) it includes spectral lines of hydrogen from the Balmer series ($H_{\rm \gamma}$, and $H_{\rm \beta}$), and in case of hotter OB-type stars several spectral lines of helium, that are usually employed as the main diagnostic lines for determination of the effective temperature and surface gravity of OBAF-type stars \citep[][]{2009ApJ...692..618M}, 2) it also includes a plethora of metal lines that serve as the main diagnostic for the inference of the metallicity and projected rotational velocity (and extra line broadening parameters, if applicable) of the star, and 3) parts of the HERMES optical spectrum blue- and red-wards the chosen wavelength interval suffer from high levels of noise and large contributions from telluric lines, respectively. Finally, because we (in particular) perform a comparative analysis between high- and low-resolution optical spectroscopy in this work (see Section~\ref{sec:control_samples_tests} for details), the exact overlap between the two types of data in terms of the considered wavelength interval is much more critical than the exact length of the interval used, unless it is unrealistically short (which is not the case here).


\begin{figure*}
    \centering
    \includegraphics[width=0.41\textwidth]{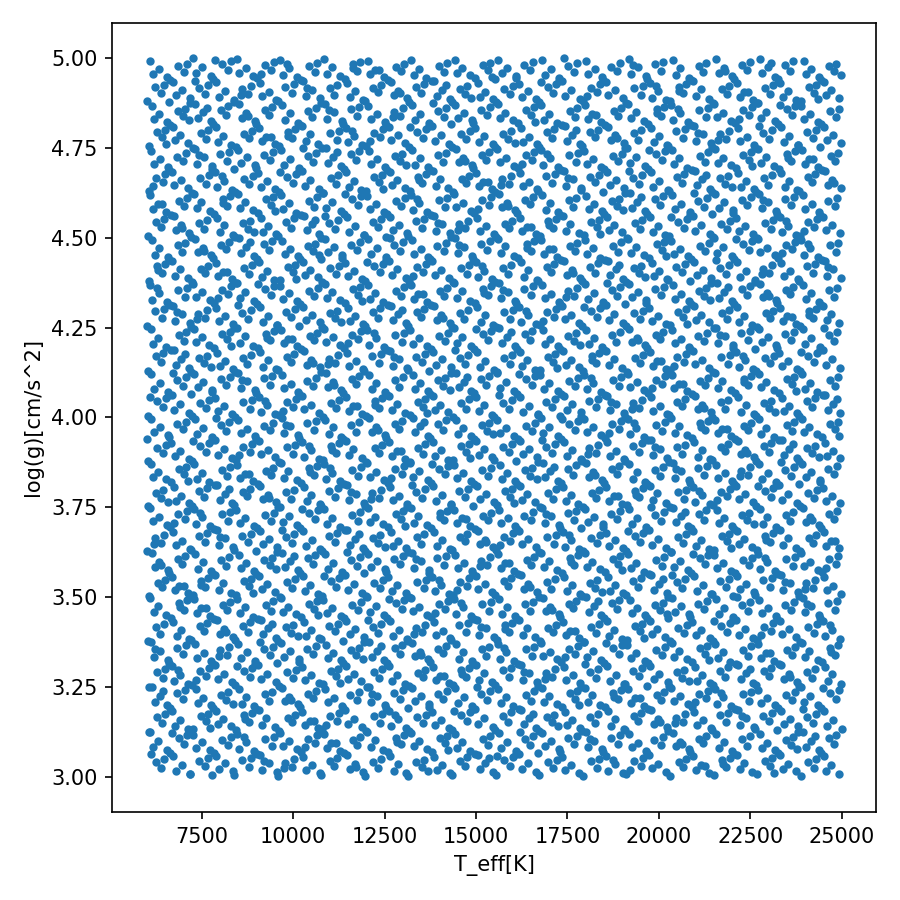}\hspace{5mm}
    \includegraphics[width=0.47\textwidth]{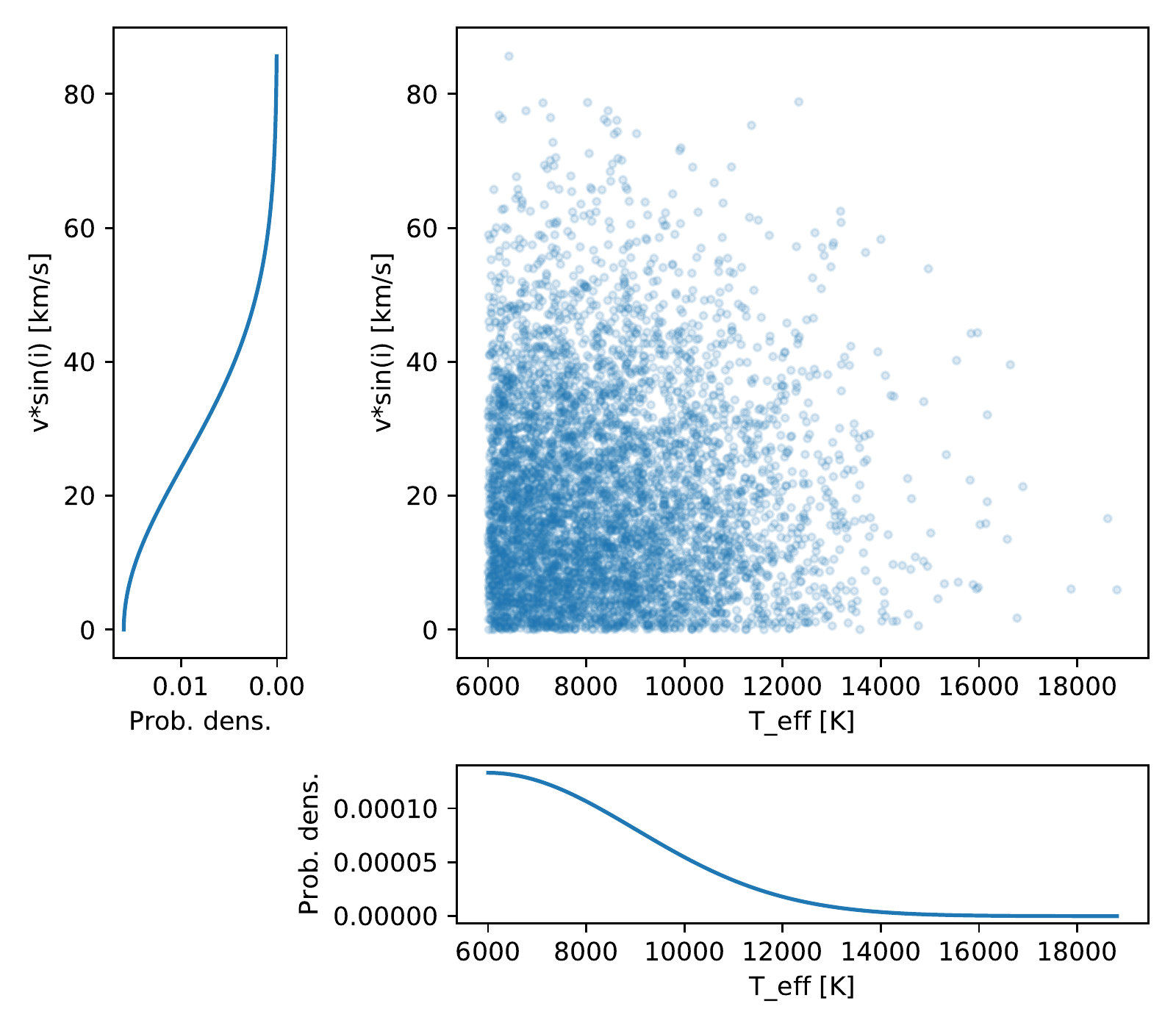}
    \caption{Left: a quasi-random grid of synthetic APOGEE spectra, projected onto the \teff-\logg\ plane. Right: an additional random sample of models to increase the density of the grid in the low \teff-\vsini\ region. The probability density functions for \teff\ and \vsini\ from which the points are sampled are shown in the left and bottom panels, respectively.}
    \label{fig:NNTrainingGrid}
\end{figure*}

\begin{figure}
    \centering
    \includegraphics[width=0.4\textwidth]{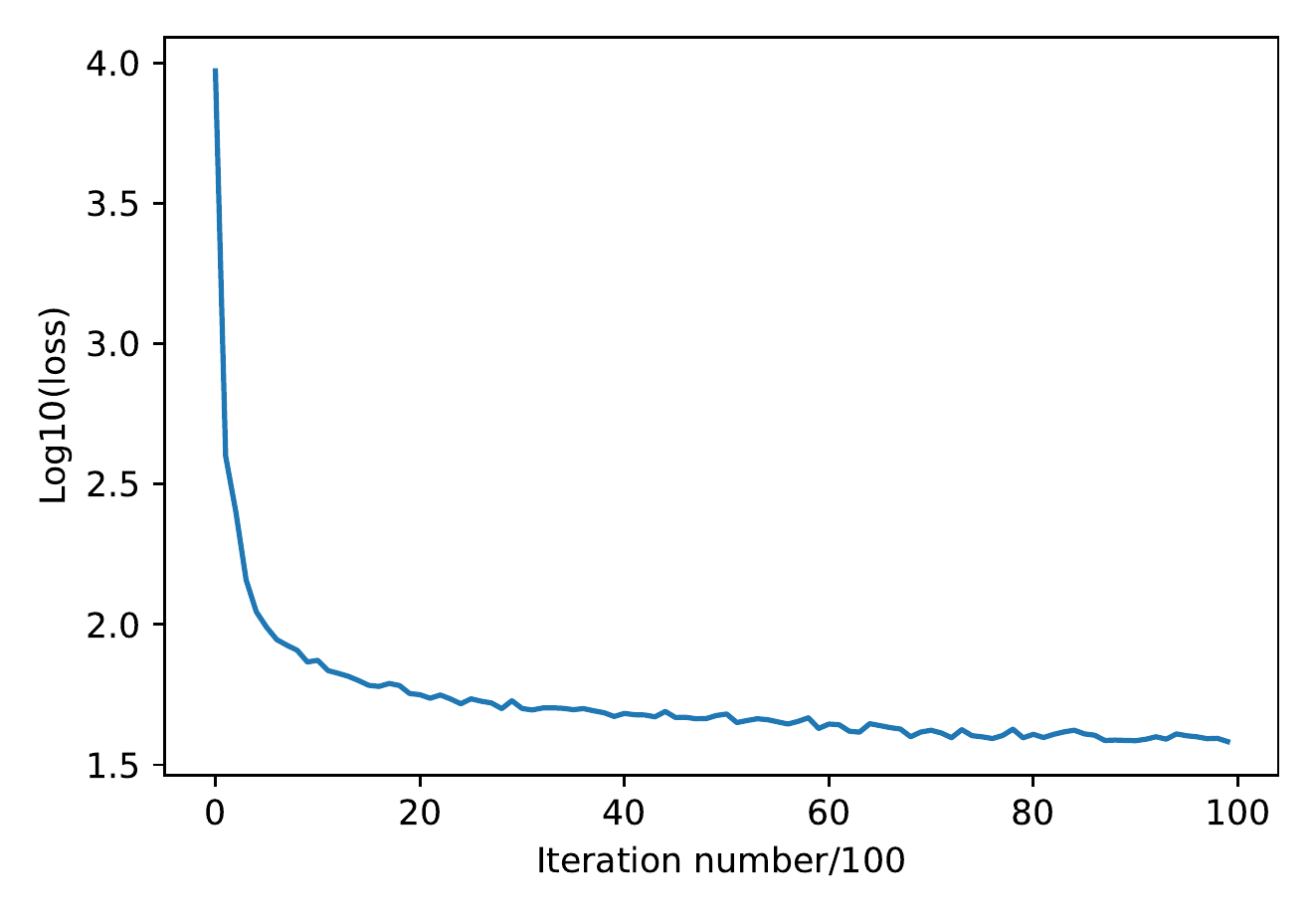}
    \caption{Score as a function of iteration number during the training of a neural network.}
    \label{fig:nn_training}
\end{figure}

For the training set, we ultimately use a hybrid grid that consists of a combination of a quasi-random Sobol grid that covers the entire parameter space of interest and an additional random grid sampled from a Gaussian distribution in a restricted region of the parameter space, as demonstrated in Figure~\ref{fig:NNTrainingGrid}. A quasi-random Sobol grid has the useful property that it covers the parameter space uniformly regardless of the number of points in the grid, unlike a true random uniform distribution, which may produce `clumps' of points and regions of the parameter space that are poorly sampled. The need for a hybrid grid arises from inferior performance of the neural network in the APOGEE wavelength range in the regime of slowly rotating late A- to F-type stars ($T_{\rm eff} \lesssim$ 10000~K and $v\,\sin\,i \lesssim$ 30~\kms) when it is trained on a quasi-random Sobol grid only. The low performance of the neural network in this regime is associated with a high (morphological) complexity of stellar spectra that are found to exhibit a large number of narrow spectral lines of metals as compared to spectra of hotter and more rapidly rotating stars (see, e.g., a comparison between APOGEE spectra of B- and F-type stars in Figure~\ref{fig:SynthAPOGEESpectra}). When trained exclusively on a quasi-random Sobol grid (see left panel in Figure~\ref{fig:NNTrainingGrid}), the neural network does not have enough examples of highly complex spectra and experiences performance difficulties in the corresponding region of the parameter space. Adding a random sample of models from a Gaussian distribution $T_{\rm eff} \sim N(6000, 3000)$, $v\,\sin\,i \sim N(0, 25)$ (see right panel in Figure~\ref{fig:NNTrainingGrid}) resolves the issue and equalizes the neural network performance across the entire parameter space. The grid used for training the neural network eventually consists of 10\,000 models, equally split between the Sobol quasi-random and extra random grids. 


Training of the neural network is performed in an iterative fashion. The combined grid of 10\,000 models is split into training and validation sets, where the former comprises 90\% of models while the latter contains the remaining 10\%. Note that the validation set is different from the test set, which is used to evaluate performance of the network after it is trained (Section \ref{sec:sim_tests}); the test set is generated separately from the validation set. At every iteration, the neural network coefficients are optimized based on the training set using the {\sc RAdam} optimization algorithm. The training is done in batches of 1000 models, selected randomly from the training set of models; the number of batches used at every iteration equals the total number of models in the training set divided by the batch size. Every 100-th iteration, a score is calculated based on the validation set, and if the new score is better than any of the previous scores, the network weights are saved into a file. This process is run for a fixed number of $10^4$ iterations that we found experimentally to be sufficient for the convergence. Figure~\ref{fig:nn_training} shows the evolution of the score during the neural network training process.

\subsection{Model fitting and parameter statistical uncertainty}
The best stellar parameters for an input observed spectrum are found by performing minimization of the $\chi^2$ merit function defined as:
\begin{equation}
    \chi^2 = \sum_i \left( \frac{f_i - F(\theta, \lambda_i)}{\sigma_i} \right)^2.
\end{equation}
Here, $f_i$ represents the observed fluxes with errors $\sigma_i$, $\theta$ is the stellar parameter vector and $\lambda_i$ is the wavelength grid. $F(\theta)$ is the model spectrum defined as:
\begin{equation}
    F(\theta) = \{ LSF \ast D[\mathrm{NN}(\theta, \lambda_i), v_r] \} \times R(\lambda_i),
\end{equation}
where $\mathrm{NN}(\theta, \lambda_i)$ is the neural network output for stellar parameters $\theta$, the operator $D[\cdot, v_r]$ performs a Doppler shift of the input observed spectrum according to the radial velocity $v_r$, and $R(\lambda_i)={\rm Response}(\lambda)$ represents the residual response function as defined in Eq.\,(\ref{eq:resp}).

Optimization of the objective function is done using the ``Trust Region Reflective'' method discussed in \citet{Branch1999}. In the first instance, we use the central point of the parameter space of the neural network training set to initialize the optimization algorithm, and investigate the algorithm convergence properties based on artificial spectra of OBAF-stars. These tests reveal a non-negligible number of cases where the optimization algorithm gets stuck in a local minimum located in a wrong region of the parameter space. The most common failures of the optimization algorithm are associated with it getting stuck above or below \teff\ of some 10\,000~K, and this divergence of the algorithm is most pronounced for the APOGEE near-IR spectra. We solve the issue by providing the algorithm with a better (lower $\chi^2$) initial guess for the (stellar) parameters vector instead of consistently using the central point of the neural network training grid. In order to find a better starting point for the optimization, we first perform a global (pre-)search by visiting a number of points in the stellar parameters space, passing the value of the parameters at a given point to the neural network and checking the value of $\chi^2$ at every point, with the aim to find the point that achieves the lowest $\chi^2$. In order to make the search scalable for different numbers of points (currently we set it to 4000), we use the Sobol algorithm to generate the set of points to visit, so that 
it covers the parameter space uniformly.

The preliminary search is performed in the space of stellar parameters and radial velocity. Because the main purpose of the pre-search step is to locate a sensible initial guess in the fastest possible way, we decouple modeling of the residual response function from the stellar parameter and radial velocity estimation at this stage. This way, we first perform a run of the optimization algorithm starting from a point in the center of the parameter space, which provides a first estimate for a vector of the Chebyshev coefficients that describe the residual response function. This set of coefficients is then used to perform a pre-search in the stellar parameter space (including radial velocity). After that, a point corresponding to the lowest $\chi^2$ merit function is identified as the most suitable initial guess for the input (observed) spectrum. The identified initial guess is then applied in a final round of optimization in the combined space of stellar parameters and Chebyshev coefficients, to derive their optimal values and functional form for the observed spectrum in question.

The statistical uncertainty of the stellar parameters is calculated by fitting a second-degree polynomial to the $\chi^2$ distribution for each of the parameters, which is equivalent to assuming a Gaussian form of the posterior probability distribution. Due to the finite resolving power of the instrument, the spectrum has less degrees of freedom than the number of spectral bins. To account for this, we calculate the effective number of degrees of freedom as follows:
\begin{equation}
    N_{deg} = 4 R \frac{\lambda_{end} - \lambda_{start}}{\lambda_{start} + \lambda_{end}},
\end{equation}
where {\it R} is the resolving power of the instrument, while $\lambda_{start}$ and $\lambda_{end}$ determine the wavelength range covered by the spectrum. The effective number of degrees of freedom $N_{deg}$ is used to compute the 1$\sigma$ statistical uncertainty level in terms of $\chi^2$:
\begin{equation}
    \chi^2_{1\sigma} = 1 + \sqrt{2/N_{deg}}.
\end{equation}
Ultimately, the statistical uncertainty interval for the parameter in question is found from the intersection points of the second order polynomial fitted to the $\chi^2$ distribution with the horizontal line drawn at $\chi^2_{1\sigma}\cdot \chi^2_{min}$, where $\chi^2_{min}$ is the minimum value of the second-order polynomial.

\begin{figure}
    \centering
    \includegraphics[width=0.43\textwidth]{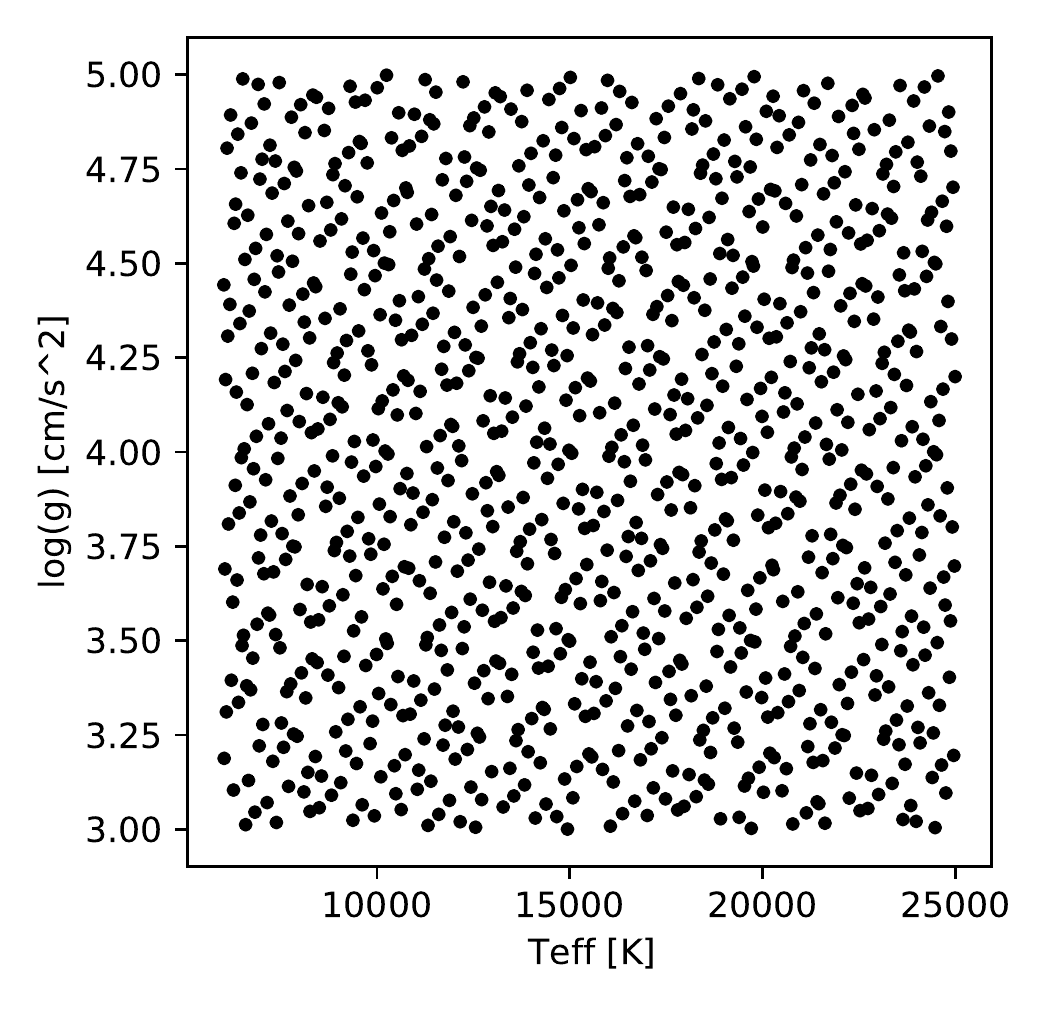}
    \includegraphics[width=0.4\textwidth]{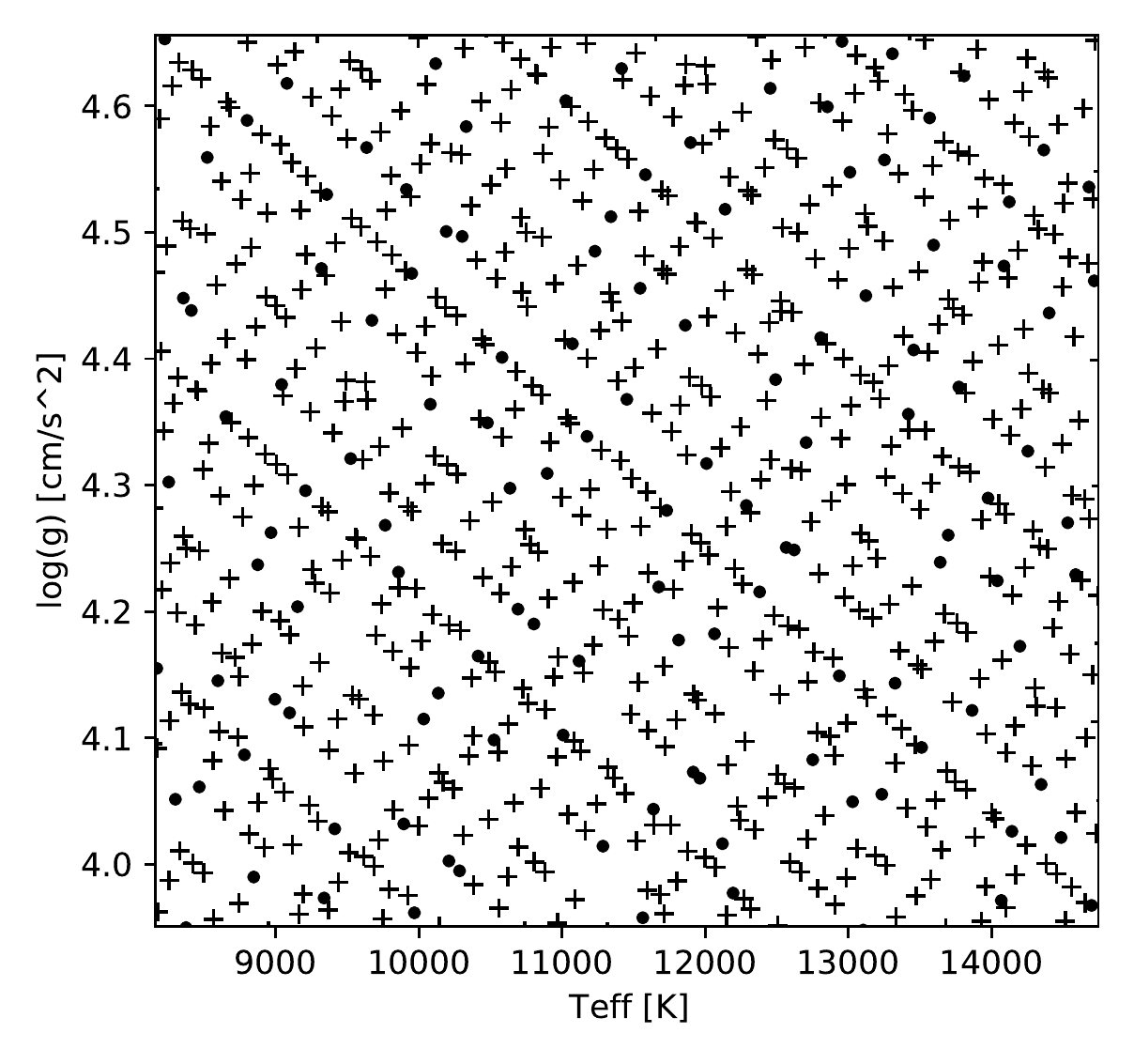}
    \caption{Top: the \teff-\vsini\ distribution of the 1\,000 simulated spectra that are used to estimate the internal uncertainty. Bottom: comparison of the training (crosses) and test (dots) grids. See text for details.}
    \label{fig:APOGEE_sim_test_Teff_vsini}
\end{figure}

\section{Performance of the ``\payne'' algorithm on simulated data}\label{sec:sim_tests}

In order to evaluate the internal uncertainty for stellar parameters intrinsic to the developed modelling framework
in Section~\ref{sec:methods}, we perform a simulation study where a 1\,000 artificial APOGEE and BOSS spectra are generated and subsequently processed as if they were observations of real stars. The simulated spectra are created on the actual APOGEE and BOSS wavelength grids, covering a wavelength range from 1.5 \micro m to 1.7 \micro m and from 3\,600~\AA\ to 10\,400~\AA, respectively. For APOGEE spectra, we also introduce two wavelength gaps between CCD sensors to resemble real observations as close as possible. The spectra for these simulated datasets are generated with the {\sc gssp} software package and for a random set of stellar parameters. The spectra are shifted in wavelength according to a random radial velocity (uniformly distributed from -50 to 50 km$^{-1}$), convolved with a Gaussian function to simulate limited spectral resolution of the instrument (R$\approx$22\,500 and 2\,000 for APOGEE and BOSS, respectively), and multiplied by a random fifth-order Chebyshev series imitating the instrumental response function. Finally, Poisson noise is added to the spectrum to simulate varying quality levels of observations in terms of signal-to-noise-ratio (S/N), where we consider the cases of S/N = $\infty$ (noiseless artificial data), 100, and 50. 

\begin{figure*}
    \centering
    \includegraphics[width=0.9\textwidth]{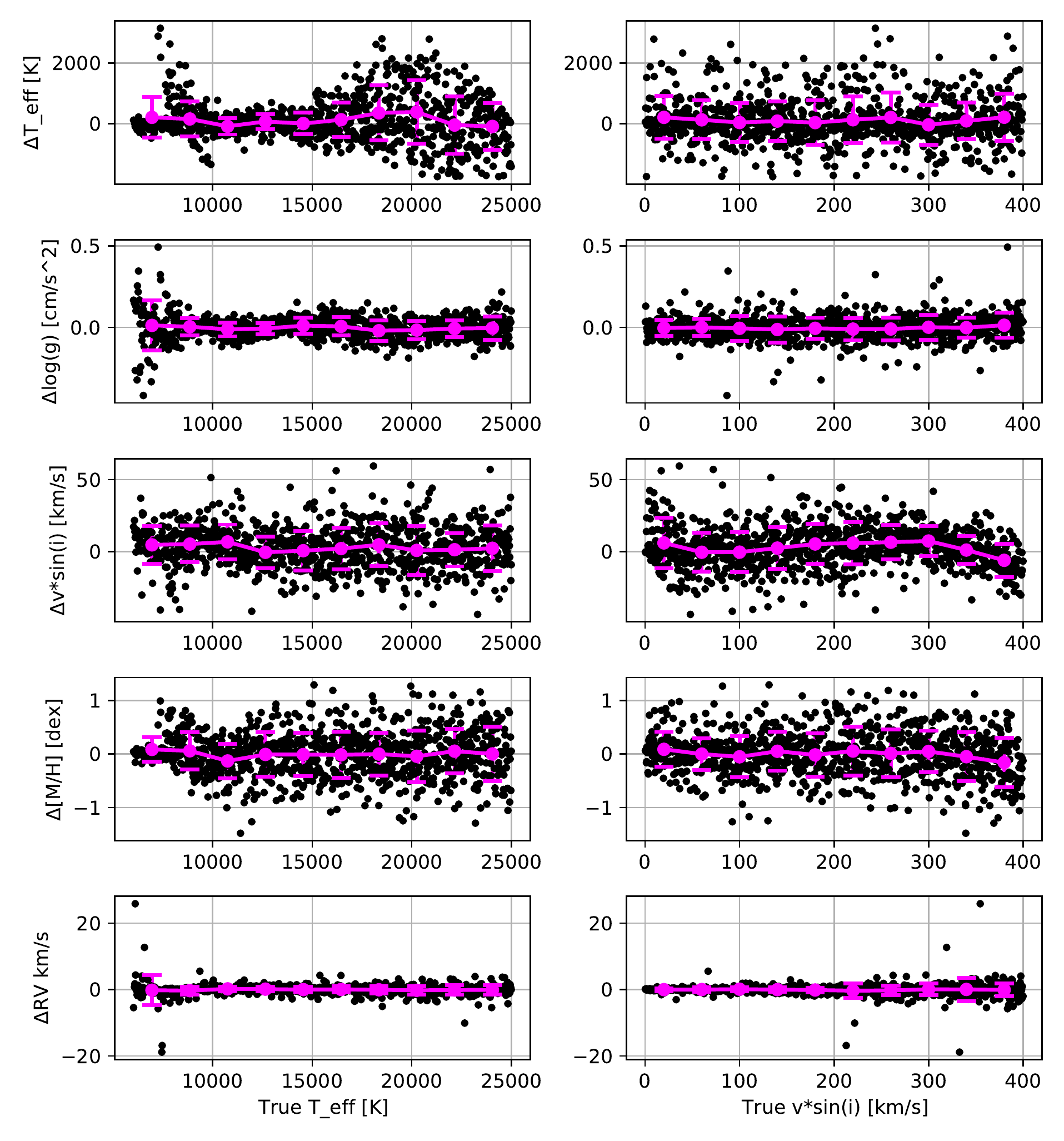}
    \caption{Differences between the parameters inferred with the \payne\ algorithm from the simulated noiseless APOGEE spectra and their true values (black dots), as a function of the true \teff\ (left) and \vsini\ (right). From top to bottom: the differences are shown for \teff, \logg, \vsini, [M/H] and RV. The internal uncertainties per parameter bin are shown with the magenta full circles with error bars; definition of the individual parameter bins is given in Table~\ref{tab:sys_uncert_bin}.}
    \label{fig:NN_test_APOGEE}
\end{figure*}

The sets of simulated APOGEE and BOSS spectra are generated to have a uniform density in the parameter space. The sampling is done using a quasi-random (Sobol) algorithm in the stellar parameter space and is assumed to be the same for both instruments (see top panel in Figure~\ref{fig:APOGEE_sim_test_Teff_vsini} for the projection of the dataset on the \teff-\vsini\ plane). This way, we are not only able to quantify the parameter internal uncertainties in both wavelength ranges, but also to perform a cross-validation between the medium-resolution near-IR and low-resolution optical regimes. We also make sure that the test dataset has zero overlap with the neural network training set (see bottom panel in Figure~\ref{fig:APOGEE_sim_test_Teff_vsini}).

The mock spectra are analyzed with the \payne\ algorithm and the internal uncertainty is calculated as follows: (i) a difference in the true versus predicted parameter value $\Delta \theta$ is calculated and the $16^{th}$, $50^{th}$ and $84^{th}$ percentiles (further labelled as $P_{16}$, $P_{50}$ and $P_{84}$) of the obtained distributions are computed; (ii) models with the parameter value differences satisfying the criterion $4(P_{50} + P_{16}) < \Delta \theta < 4 (P_{50} + P_{84})$ are selected as achieving a satisfactory fit of the model to the mock spectrum, while the rest are classified as failing to converge to the correct model. The internal uncertainty is then calculated based on the converged models. Thus the performance of the method is characterized by the internal uncertainty as well as by the reliability metric, i.e. the probability that the optimization algorithm converges to a correct value. The internal uncertainty reflects (i) how well the neural network is able to predict synthetic fluxes for a given set of stellar labels, and (ii) how good the performance of the chosen minimization algorithm is in finding and converging to the global minimum in the parameter space.

\begin{table*}
    \centering
    \caption{Root mean square errors (internal uncertainty) per bin for the stellar parameters recovered by the developed {\sc \payne} algorithm from simulated noiseless APOGEE and BOSS spectra (mimicking infinite S/N value). The uncertainties are reported for both instruments and as a function of \teff\ and \vsini\ of the spectrum.}    
    \begin{tabular}{lllllllllll}
    \hline\hline
    \multirow{2}{*}{Parameter} & \multicolumn{10}{c}{Internal uncertainty (RMS)}  \\
    & Bin 1 & Bin 2 & Bin 3 & Bin 4 & Bin 5 & Bin 6 & Bin 7 & Bin 8 & Bin 9 & Bin 10 \\
    \hline
    & \multicolumn{10}{c}{APOGEE instrument} \\
    & \multicolumn{10}{c}{as function of \teff} \\
    Bin, kK & 6-7.9 & 7.9-9.8 & 9.8-11.7 & 11.7-13.6 & 13.6-15.5 & 15.5-17.4 & 17.4-19.3 & 19.3-21.2 & 21.2-23.1 & 23.1-25 \\
    $T_{\rm eff}$, K & 670 & 579 & 269 & 249 & 354 & 563 & 908 & 1048 & 949 & 764 \\
    $\log\,g$, dex & 0.15 & 0.05 & 0.04 & 0.03 & 0.05 & 0.06 & 0.06 & 0.06 & 0.05 & 0.07 \\
    $v\,\sin\,i$, km\,s$^{-1}$  & 13 & 13 & 12 & 11 & 14 & 15 & 15 & 17 & 12 & 16 \\
    ${\rm [M/H]}$, dex & 0.23 & 0.35 & 0.32 & 0.42 & 0.40 & 0.43 & 0.40 & 0.49 & 0.41 & 0.51 \\
    RV, km\,s$^{-1}$  & 4.5 & 1.1 & 0.7 & 0.7 & 0.9 & 0.9 & 1.1 & 1.2 & 1.4 & 1.4 \\
    & \multicolumn{10}{c}{as function of $v\,\sin\,i$} \\
    Bin, km\,s$^{-1}$ & 0-40 & 40-80 & 80-120 & 120-160 & 160-200 & 200-240 & 240-280 & 280-320 & 320-360 & 360-400 \\
    $T_{\rm eff}$, K & 709 & 643 & 644 & 653 & 728 & 764 & 821 & 657 & 609 & 776 \\
    $\log\,g$, dex & 0.05 & 0.05 & 0.08 & 0.08 & 0.06 & 0.07 & 0.07 & 0.08 & 0.06 & 0.08 \\
    $v\,\sin\,i$, km\,s$^{-1}$  & 18 & 14 & 14 & 15 & 14 & 15 & 12 & 11 & 10 & 12 \\
    ${\rm [M/H]}$, dex & 0.32 & 0.30 & 0.38 & 0.37 & 0.41 & 0.46 & 0.44 & 0.38 & 0.46 & 0.46 \\
    RV, km\,s$^{-1}$  & 0.4 & 0.7 & 0.6 & 0.7 & 0.7 & 2.1 & 1.4 & 1.8 & 3.5 & 2.0 \\
    & \multicolumn{10}{c}{BOSS instrument} \\
    & \multicolumn{10}{c}{as function of $T_{\rm eff}$} \\
    Bin, kK & 6-7.9 & 7.9-9.8 & 9.8-11.7 & 11.7-13.6 & 13.6-15.5 & 15.5-17.4 & 17.4-19.3 & 19.3-21.2 & 21.2-23.1 & 23.1-25 \\
    $T_{\rm eff}$, K & 50 & 80 & 96 & 83 & 98 & 146 & 197 & 209 & 222 & 268 \\
    $\log\,g$, dex & 0.09 & 0.06 & 0.04 & 0.02 & 0.02 & 0.02 & 0.02 & 0.02 & 0.02 & 0.04 \\
    $v\,\sin\,i$, km\,s$^{-1}$ & 19 & 16 & 15 & 11 & 11 & 13 & 12 & 10 & 11 & 15 \\
    ${\rm [M/H]}$, dex & 0.06 & 0.11 & 0.11 & 0.07 & 0.11 & 0.08 & 0.09 & 0.09 & 0.08 & 0.12 \\
    RV, km\,s$^{-1}$ & 1.1 & 1.2 & 0.5 & 0.4 & 0.2 & 0.2 & 0.4 & 0.4 & 0.3 & 0.7\vspace{1mm} \\
    & \multicolumn{10}{c}{as function of $v\,\sin\,i$} \\
    Bin, km\,s$^{-1}$ & 0-40 & 40-80 & 80-120 & 120-160 & 160-200 & 200-240 & 240-280 & 280-320 & 320-360 & 360-400 \\
    $T_{\rm eff}$, K & 174 & 144 & 158 & 166 & 157 & 149 & 152 & 175 & 159 & 180 \\
    $\log\,g$, dex & 0.03 & 0.04 & 0.04 & 0.05 & 0.04 & 0.04 & 0.04 & 0.05 & 0.04 & 0.06 \\
    $v\,\sin\,i$, km\,s$^{-1}$ & 19 & 16 & 15 & 16 & 8 & 9 & 9 & 10 & 9 & 10 \\
    ${\rm [M/H]}$, dex & 0.08 & 0.08 & 0.08 & 0.11 & 0.07 & 0.08 & 0.09 & 0.13 & 0.09 & 0.11 \\
    RV, km\,s$^{-1}$ & 0.3 & 0.4 & 0.3 & 0.4 & 0.6 & 0.5 & 0.7 & 1.0 & 0.7 & 1.0\vspace{3mm} \\
    \hline
    \end{tabular}
    \label{tab:sys_uncert_bin}
\end{table*}

\begin{figure*}
    \centering
    \includegraphics[width=0.9\textwidth]{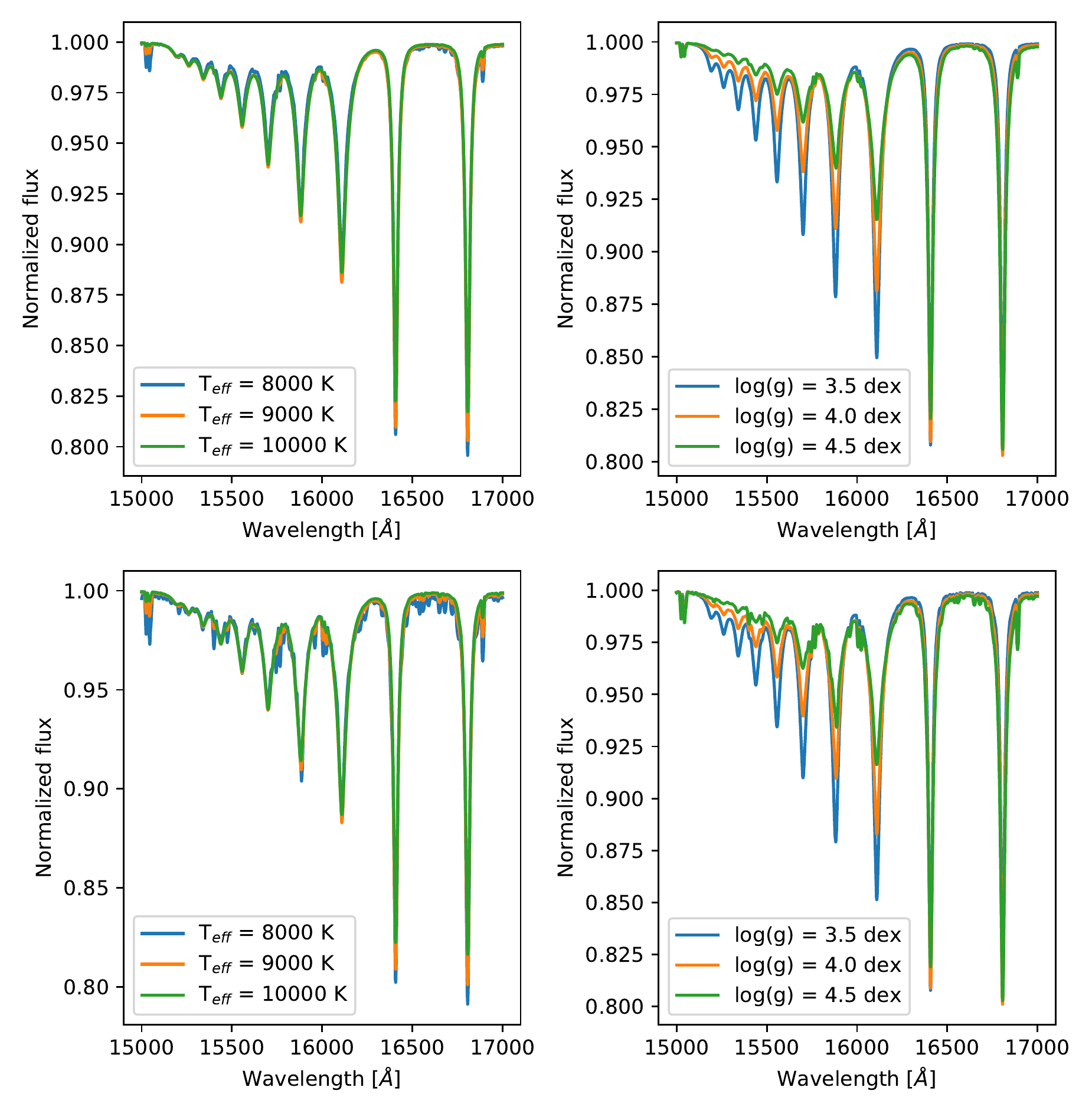}
    \caption{Effect of the \teff\ and \logg\ parameter changes on the appearance of stellar spectra of late A- to early F-type stars in the wavelength range of the APOGEE instrument. Top and bottom rows show the effect at significantly sub-solar ([M/H] = -0.8~dex) and solar ([M/H] = 0.0~dex) metallicity, respectively. Left column: blue, orange, and green lines represent models with \teff\ = 8\,000, 9\,000, and 10\,000~K, respectively; \logg\ and \vsini\ are kept fixed at 4.0~dex and 150~\kms, respectively. Right column: blue, orange, and green lines represent models with \logg\ = 3.5, 4.0, and 4.5~dex, respectively; \teff\ and \vsini\ are kept fixed at 9\,000~K and 150~\kms, respectively.}
    \label{fig:APOGEE_TeffLogg_effect}
\end{figure*}

\begin{table*}
    \centering
    \caption{Root mean square errors (internal uncertainty) averaged over the entire parameter space for the stellar parameters as inferred from the simulated APOGEE and BOSS spectra with realistic noise levels.}    
    \begin{tabular}{lllllll}
    \hline\hline
    & \multicolumn{6}{c}{Internal uncertainty (RMS)}  \\
    Parameter & \multicolumn{3}{c}{APOGEE} & \multicolumn{3}{c}{BOSS} \\
    & noiseless & S/N = 100 & S/N = 50 & noiseless & S/N = 100 & S/N = 50 \\
    \hline
    $T_{\rm eff}$, K            & 707   & 736   & 791   & 163   & 177   & 206 \\
    $\log\,g$, dex              & 0.07  & 0.08  & 0.09  & 0.043 & 0.044 & 0.045 \\
    $v\,\sin\,i$, km\,s$^{-1}$  & 14    & 17    & 23    & 14    & 15    & 16 \\
    ${\rm [M/H]}$, dex          & 0.41  & 0.46  & 0.50  & 0.10  & 0.10  & 0.11 \\
    RV, km\,s$^{-1}$            & 1.67  & 3.65  & 6.30  & 0.65  & 1.29  & 2.19 \\
    \hline
    Reliability & 96.2 \% & 96.5 \% & 95.6 \% & 95.6 \% & 97.4\% & 97.3\% \\
    \hline
    \end{tabular}
    \label{tab:sys_uncert_final}
\end{table*}

Results of the application of the {\sc \payne} algorithm to the set of 1000 artificial noiseless APOGEE spectra are summarized in Figure~\ref{fig:NN_test_APOGEE}, where we compare the parameters inferred from the simulated data with their true values. The differences between the inferred and true parameter values are presented for (from top to bottom) $T_{\rm eff}$, $\log\,g$, $v\,\sin\,i$, and [M/H], and as a function of true $T_{\rm eff}$ (left column) and $v\,\sin\,i$ (right column). In each panel in Figure~\ref{fig:NN_test_APOGEE}, we divide the corresponding dataset into ten equal width bins and compute the {internal} uncertainty in each of those bins as described above. The resulting internal uncertainties per parameter bin are shown as magenta full circles with error bars in Figure~\ref{fig:NN_test_APOGEE}, with the corresponding numerical values listed in the top part of Table~\ref{tab:sys_uncert_bin} (designated as the ``APOGEE instrument'').
The most notable feature seen in Figure~\ref{fig:NN_test_APOGEE} is a ``tail'' of models in the \teff\ range between some 8\,000~K and 10\,000~K in the top left panel, where we record about 25 of the \payne\ best fit models that show \teff\ discrepancies of above some 600~K and up to some 2\,500~K with the true values. The discrepancy is confined to that specific $\sim$2\,000~K wide \teff\ interval and occurs in the low [M/H] - high \vsini\ range of the parameter space. As demonstrated in the top left panel in Figure~\ref{fig:APOGEE_TeffLogg_effect}, the APOGEE spectra are dominated by the Brackett series of hydrogen lines having very low sensitivity to the \teff\ variations in that particular \teff\ interval. Therefore, metal lines represent an important \teff\ diagnostic. However, the cumulative effect of low metallicity and high projected rotational velocity of the star makes metal lines appear weak and shallow. Altogether, the overly weak metal lines and low sensitivity of the Brackett series to \teff\ variations act as a source of confusion (hence large internal uncertainty) for the spectrum analysis algorithm in the \teff\ bin under consideration. The degeneracy gets progressively smaller with increasing metallicity (and decreasing projected rotational velocity), so that the internal uncertainty in \teff\ becomes comparable to adjacent \teff\ bins as [M/H] approaches the solar value (see bottom left panel in Figure~\ref{fig:APOGEE_TeffLogg_effect} where the growth of strength of metal lines compared to the low [M/H] case is visible). 
We also note a larger internal uncertainty in the effective temperature of the star at $T_{\rm eff}\gtrsim 15\,000$~K (top left panel in Figure~\ref{fig:NN_test_APOGEE}) that we attribute to the fact that the APOGEE spectra of late O- and B-type stars are largely featureless and dominated by the Brackett series of broad hydrogen lines. The spectra additionally suffer from two wavelength gaps present in the APOGEE data that reduce the amount of available information and thus contribute to the internal uncertainty. A lower internal uncertainty in metallicity of the star is also seen in the low $T_{\rm eff}$ ($\lesssim 8\,000$~K) and low $v\,\sin\,i$ ($\lesssim 40$~\kms) region of the parameter space (see penultimate panel in the left column in Figure~\ref{fig:NN_test_APOGEE}). This result can be explained by larger number of (narrow) metal lines available in those spectra for the inference of stellar metallicity, while the lines either disappear or get significantly broadened at higher effective temperatures and projected rotational velocities, respectively.
    
Table~\ref{tab:sys_uncert_final} (left column, designated as ``APOGEE'') lists the internal uncertainties for all five stellar parameters ($T_{\rm eff}$, $\log\,g$, $v\,\sin\,i$, [M/H], and RV) as inferred from the entire simulated APOGEE dataset, which makes them representative of the parameter space in consideration. We also note that a similar exercise was performed for the simulated APOGEE datasets characterized by S/N = 50 and 100 as an indication of the effect of Poisson noise on the resulting internal uncertainties. The results are summarized in the APOGEE instrument ``S/N = 50'' and ``S/N = 100'' columns in Table~\ref{tab:sys_uncert_final}. One can see that overall uncertainty in \teff\ and \vsini\ increases by some 15\% and 50\%, respectively, between the noiseless and S/N=50 spectra, while there is hardly any change in the uncertainty for \logg\ of the star. The largest increase by a factor of $\sim$5 is recorded for the uncertainty in the RV of the star (1.25~\kms\ in the noiseless case as compared to 6.27~\kms\ for S/N=50), while the increase of some 0.1~dex is observed for the uncertainty in stellar metallicity. 
    
\begin{figure*}
    \centering
    \includegraphics[width=1.0\textwidth]{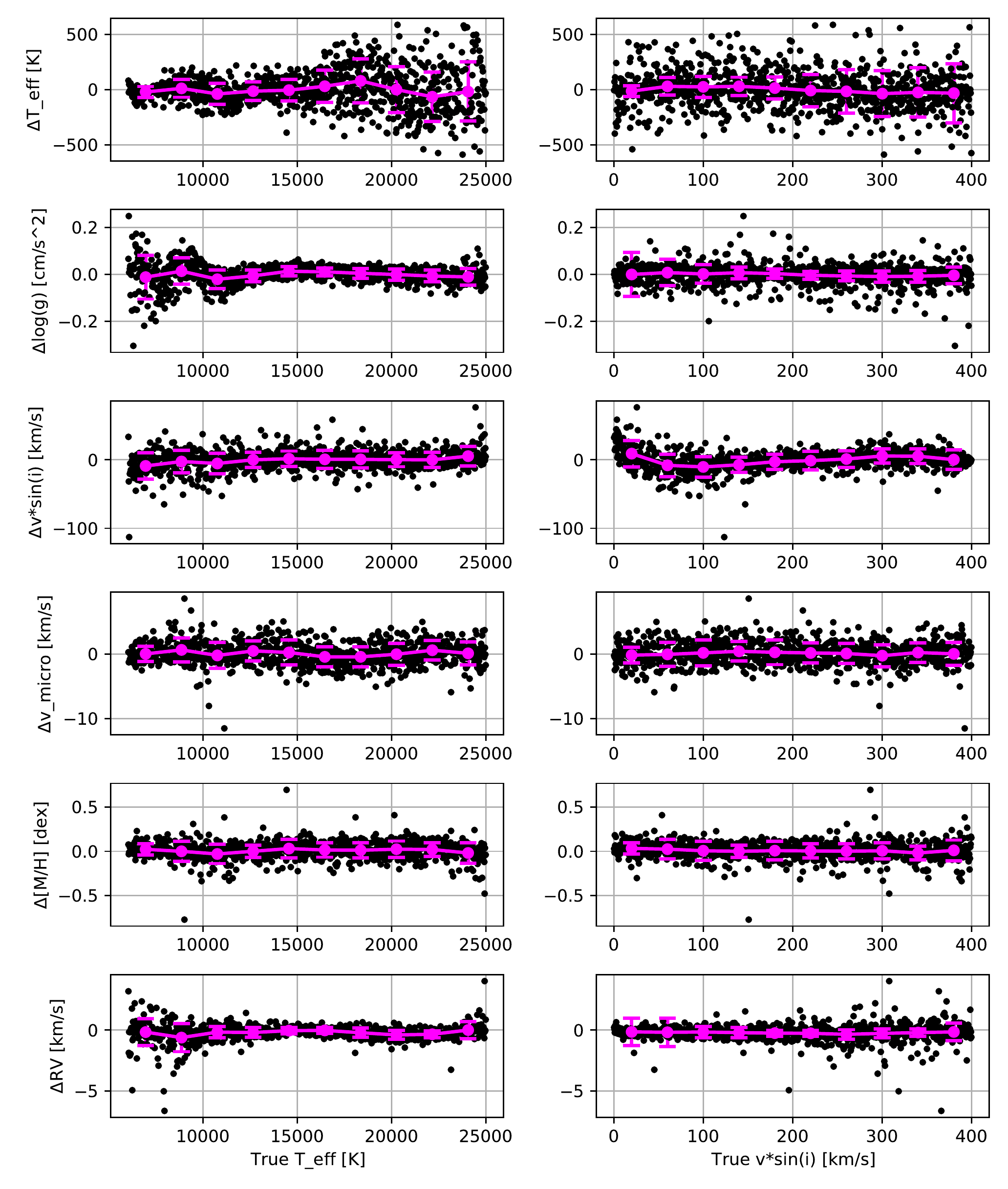}
    \caption{Same as Figure~\ref{fig:NN_test_APOGEE} but for the simulated noiseless BOSS data.}
    \label{fig:NN_test_BOSS}
\end{figure*}

The above-described approach is also applied to the BOSS simulated spectra; the results obtained for the noiseless dataset are presented in Figure~\ref{fig:NN_test_BOSS} and in Tables~\ref{tab:sys_uncert_bin} and \ref{tab:sys_uncert_final}. In addition to the expected increase in the internal uncertainty for \teff\ of the star towards higher effective temperatures (top panel in the left column in Figure~\ref{fig:NN_test_BOSS}), a larger scatter is also recorded for \logg\ of the star at \teff\ values below some 10\,000~K (second top panel in the left column in Figure~\ref{fig:NN_test_BOSS}). This effect is explained by low sensitivity of the merit function employed in the optimization algorithm to \logg\ variations in the \teff\ range of late A- to F-type stars. Indeed, as discussed in detail, e.g., in \citet[][see their Section 4.1]{Gebruers2021}, metal lines along with the central cores of the Balmer lines represent the main diagnostic for the inference of the surface gravity of these stars at optical wavelengths. Owing to the low spectral resolution of the BOSS instrument, the number of spectral bins that appear to be sensitive to \logg\ variations is small relative to the total number of bins that contribute to the merit function. Coupled with partial degeneracy between the \logg\ and \teff\ parameters in the spectroscopic analysis of intermediate spectral type stars, this results in larger internal uncertainty for \logg\ in the \teff\ regime of late A- to F-type stars. Finally, we observe an increase of the uncertainty for \teff\ and \vsini\ of the star by some 25\% and 15\%, respectively, when degrading the quality of input data from the noiseless case to S/N of 50. An increase by a factor of $\sim$2 is also observed for the RV uncertainty, while the uncertainties for \logg\ and [M/H] remain largely unchanged (see Table~\ref{tab:sys_uncert_final}).
 
\begin{figure*}
    \centering
    \includegraphics[width=\textwidth]{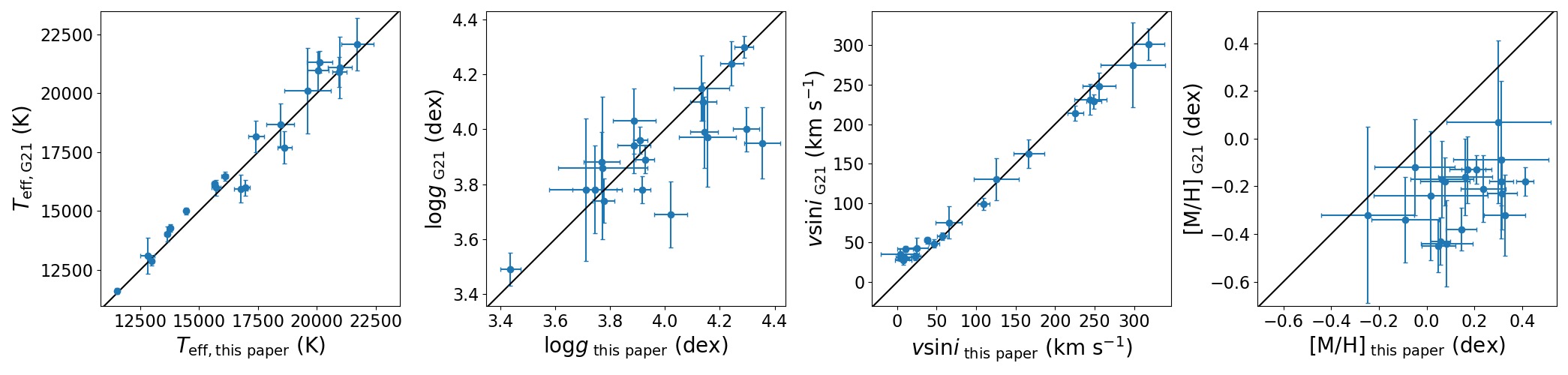}
    \caption{Comparison between stellar parameters of 20 $Kepler$ SPB-type stars as inferred from their high-resolution optical spectra in \citet{Gebruers2021} and in this study. From left to right: \teff, \logg, \vsini, and [M/H]. The solid black line indicates the one-to-one correspondence between the parameters and is shown to help guide the eye. The shown errors bars reflect 1$\sigma$ statistical uncertainties and do not include the internal uncertainty reported in Sect.~\ref{sec:sim_tests}.}
    \label{fig:Gebruers_vs_ThisWork}
\end{figure*}

\begin{table*}
    \centering
    \caption{Stellar parameters of the 20 SPB stars from \citet{Gebruers2021} as inferred in this study from their HERMES high-resolution and LAMOST low-resolution optical spectra (2 of these stars only have HERMES spectra). The quoted parameter uncertainties reflect 1$\sigma$ statistical uncertainties and do not account for the internal uncertainties reported in Sect.~\ref{sec:sim_tests}.}
    \begin{tabular}{lllllllll}
    \hline
    \multirow{2}{*}{KIC number} & \multicolumn{2}{c}{\teff\ (K)} & \multicolumn{2}{c}{\logg\ (dex)} & \multicolumn{2}{c}{\vsini\ (\kms)} & \multicolumn{2}{c}{[M/H] (dex)} \\
    & HERMES & LAMOST & HERMES & LAMOST & HERMES & LAMOST & HERMES & LAMOST \\
    \hline
10285114 & 15994 $\pm$ 262 & 16429 $\pm$ 649 & 4.04 $\pm$ 0.05 & 4.09 $\pm$ 0.13 & 247 $\pm$ 24 & 236 $\pm$ 72 & -0.04 $\pm$ 0.11 & -0.21 $\pm$ 0.34 \\
10536147 & 21225 $\pm$ 711 & 21013 $\pm$ 1126 & 3.81 $\pm$ 0.09 & 3.71 $\pm$ 0.14 & 174 $\pm$ 17 & 139 $\pm$ 47 & -0.20 $\pm$ 0.13 & -0.08 $\pm$ 0.20 \\
11360704 & 17856 $\pm$ 398 & 17608 $\pm$ 957 & 3.91 $\pm$ 0.06 & 3.90 $\pm$ 0.13 & 307 $\pm$ 26 & 301 $\pm$ 64 & -0.12 $\pm$ 0.13 & -0.16 $\pm$ 0.35 \\
12258330 & 16320 $\pm$ 149 & 16432 $\pm$ 601 & 4.27 $\pm$ 0.04 & 4.12 $\pm$ 0.13 & 118 $\pm$ 7 & 78 $\pm$ 66 & -0.10 $\pm$ 0.06 & -0.37 $\pm$ 0.29 \\
3240411 & 21985 $\pm$ 300 & 20708 $\pm$ 1014 & 4.22 $\pm$ 0.05 & 4.09 $\pm$ 0.14 & 36 $\pm$ 4 & 31 $\pm$ 65 & -0.12 $\pm$ 0.04 & 0.01 $\pm$ 0.17 \\
3756031 & 17302 $\pm$ 297 & 16925 $\pm$ 659 & 3.95 $\pm$ 0.06 & 3.90 $\pm$ 0.12 & 14 $\pm$ 6 & 21 $\pm$ 39 & -0.35 $\pm$ 0.12 & -0.33 $\pm$ 0.43 \\
3839930 & 17436 $\pm$ 239 & 17237 $\pm$ 660 & 4.32 $\pm$ 0.05 & 4.34 $\pm$ 0.13 & 28 $\pm$ 6 & 30 $\pm$ 61 & 0.02 $\pm$ 0.07 & -0.27 $\pm$ 0.35 \\
3865742 & 19776 $\pm$ 1242 & 19941 $\pm$ 1210 & 3.89 $\pm$ 0.16 & 3.82 $\pm$ 0.16 & 132 $\pm$ 40 & 110 $\pm$ 65 & 0.27 $\pm$ 0.19 & 0.03 $\pm$ 0.22 \\
5941844 & 14105 $\pm$ 141 & 13806 $\pm$ 445 & 4.33 $\pm$ 0.04 & 4.26 $\pm$ 0.13 & 28 $\pm$ 3 & 24 $\pm$ 98 & 0.05 $\pm$ 0.06 & 0.10 $\pm$ 0.32 \\
6462033 & 18171 $\pm$ 537 & 18645 $\pm$ 772 & 4.22 $\pm$ 0.11 & 4.17 $\pm$ 0.13 & 73 $\pm$ 18 & 28 $\pm$ 103 & -0.08 $\pm$ 0.18 & -0.60 $\pm$ 0.45 \\
6780397 & 13150 $\pm$ 141 & 13068 $\pm$ 373 & 3.77 $\pm$ 0.04 & 3.63 $\pm$ 0.10 & 57 $\pm$ 6 & 61 $\pm$ 53 & -0.07 $\pm$ 0.06 & -0.19 $\pm$ 0.26 \\
7760680 & 11570 $\pm$ 97 & 11858 $\pm$ 209 & 3.91 $\pm$ 0.03 & 4.22 $\pm$ 0.08 & 72 $\pm$ 6 & 0 $\pm$ 30 & 0.18 $\pm$ 0.08 & -0.02 $\pm$ 0.27 \\
8057661 & 23795 $\pm$ 661 & 21693 $\pm$ 1219 & 4.40 $\pm$ 0.12 & 4.52 $\pm$ 0.17 & 33 $\pm$ 9 & 33 $\pm$ 77 & -0.22 $\pm$ 0.09 & -0.14 $\pm$ 0.24 \\
8087269 & 13198 $\pm$ 471 & 13043 $\pm$ 421 & 3.73 $\pm$ 0.12 & 3.59 $\pm$ 0.10 & 287 $\pm$ 83 & 257 $\pm$ 78 & 0.11 $\pm$ 0.23 & -0.21 $\pm$ 0.41 \\
8381949 & 21288 $\pm$ 517 & 22300 $\pm$ 1148 & 3.93 $\pm$ 0.07 & 4.23 $\pm$ 0.13 & 220 $\pm$ 19 & 206 $\pm$ 60 & -0.10 $\pm$ 0.10 & -0.34 $\pm$ 0.42 \\
8714886 & 19148 $\pm$ 623 & 18430 $\pm$ 675 & 4.24 $\pm$ 0.07 & 4.44 $\pm$ 0.13 & 19 $\pm$ 7 & 35 $\pm$ 38 & -0.05 $\pm$ 0.09 & -0.17 $\pm$ 0.30 \\
8766405 & 14498 $\pm$ 156 & 14427 $\pm$ 680 & 3.49 $\pm$ 0.03 & 3.22 $\pm$ 0.14 & 209 $\pm$ 14 & 227 $\pm$ 70 & -0.27 $\pm$ 0.08 & -0.63 $\pm$ 0.44 \\
9964614 & 21387 $\pm$ 532 & 21049 $\pm$ 1033 & 4.00 $\pm$ 0.08 & 4.22 $\pm$ 0.14 & 46 $\pm$ 10 & 1 $\pm$ 57 & -0.20 $\pm$ 0.10 & -0.34 $\pm$ 0.27 \\
11971405& 15084 $\pm$ 136 & --- & 3.89 $\pm$ 0.03 & --- & 223 $\pm$ 13 & --- & -0.10 $\pm$ 0.06 & --- \\
8459899 & 16311 $\pm$ 135 & --- & 3.89 $\pm$ 0.03 & --- & 47 $\pm$ 4 & --- & 0.00 $\pm$ 0.05 & --- \\
    \hline
    \end{tabular}
    \label{tab:SPB_params_HERMES_vs_LAMOST}
\end{table*}

\section{Application of \payne~ to control stellar samples}\label{sec:control_samples_tests}

\begin{table*}
    \centering
    \caption{Parameters inferred with the \payne\ algorithm from the input artificial spectra that assume \teff, \logg, \vsini, and [M/H] fixed to 13\,000~K, 4.0~dex, 100~\kms, and 0.0~dex, respectively, and variable microturbulent velocity $\xi$ parameter as indicated in the first column. The quoted parameter errors are 1$\sigma$ statistical uncertainties.}
    \begin{tabular}{lllll}
    \hline
    Input spectrum & \multicolumn{4}{c}{Inferred parameters} \\
    $\xi$ (\kms) & $T_{\rm{eff}}$ (K) & $\log{g}$ (dex) & $v\sin{i}$ (km s$^{-1}$) & $[M/H]$ (dex) \\
    \hline
    2 & 13\,110 $\pm$ 22 & 4.01 $\pm$ 0.01 & 94 $\pm$ 3 & +0.02 $\pm$ 0.02 \\
    4  & 13\,109 $\pm$ 31 & 4.02 $\pm$ 0.01 & 95 $\pm$ 4 & +0.15 $\pm$ 0.03 \\
    6 & 13\,057 $\pm$ 45 & 4.03 $\pm$ 0.02 & 99 $\pm$ 8 & +0.29 $\pm$ 0.03 \\
    8 & 13\,039 $\pm$ 52 & 4.05 $\pm$ 0.03 & 95 $\pm$ 4 & +0.42 $\pm$ 0.03 \\
    10 & 13\,014 $\pm$ 62 & 4.04 $\pm$ 0.03 & 95 $\pm$ 5 & +0.49 $\pm$ 0.04 \\
    \hline
    \end{tabular}
    \label{tab:vmicro_effect}
\end{table*}

\begin{figure}
    \centering
    \includegraphics[width=0.45\textwidth]{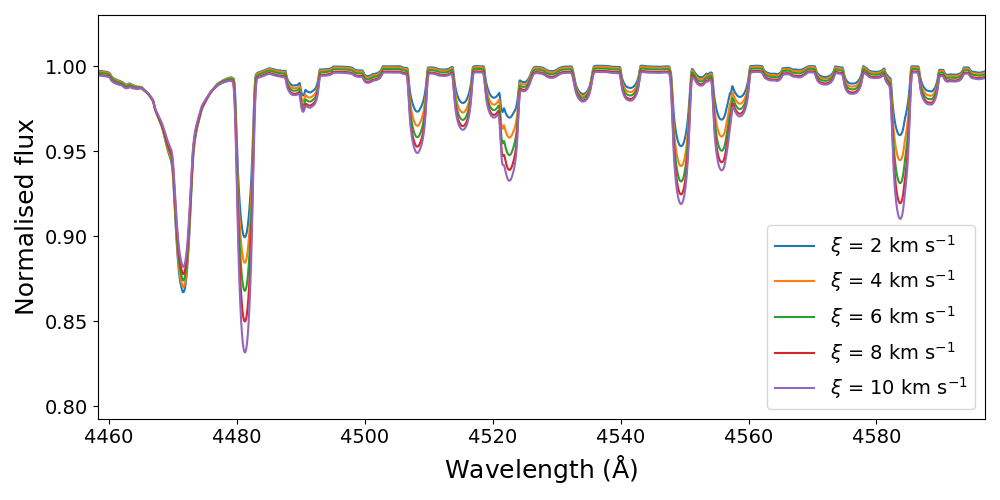}
    \caption{Synthetic spectra computed for $T_{\rm{eff}}$ = 13\,000\,K, $\log{g}$ = 4.0\,dex, $v\sin{i}$ = 100\,km\,s$^{-1}$, $[M/H]$ = 0.0\,dex and varying microturbulent velocity $\xi$ from 2 to 10~\kms.}
    \label{fig:vmicro_effect}
\end{figure}

Aside from the successful performance test on simulated 
spectra discussed in the previous Section, we apply the {\sc \payne} algorithm to spectra of two control samples of real stars. First, we employ the sample of Slowly Pulsating B-type (SPB) stars analyzed in \citet{Gebruers2021} to test if we can reproduce their results with respect to the atmospheric parameters of the stars inferred from both high-resolution (R$\approx$85\,000) HERMES \citep{Raskin2011} and low-resolution (R$\approx$1\,800) LAMOST optical spectra. At the time of writing, we do not have any OBAF-type stars observed with the BOSS instrument in SDSS-V that would be bright enough for observations with the HERMES instrument at the 1.2-m Mercator telescope and with the APOGEE instrument (see below). Therefore, we decide to use LAMOST spectra instead, given a number of similarities with the BOSS instrument, among which are the wavelength coverage, resolving power, spectrum reduction pipelines, etc. This exercise allows us to quantify potential differences in the inferred stellar parameters with respect to those derived in \citet{Gebruers2021}, as well as to unravel potential systematic effects in the inferred atmospheric parameters associated with a factor of $\sim$40 reduction in the spectral resolution (HERMES vs. LAMOST). Secondly, we cross-match a sample of OBAF-type star candidates observed with the APOGEE instrument as part of the SDSS-V Pathfinder program \citep[][their Section~5]{Kollmeier2017} with archival HERMES high-resolution optical observations, and analyze both datasets with the {\sc \payne} algorithm. These tests offer a cross-validation between the parameter inference from medium-resolution near-IR and high-resolution optical spectra. When combined with the results of the previous exercise based on the sample of \citet{Gebruers2021}, this also allows us to close the loop of cross-validation between the SDSS-V APOGEE and BOSS instruments and high-resolution optical spectroscopy of OBAF-type stars that cannot be assembled for the large SDSS-V sample.

\citet{Gebruers2021} used an earlier version of the \payne\ algorithm to analyze a sample of 111 pulsating B- and F-type stars in the $Kepler$ field based on high-resolution (R=85\,000) optical HERMES spectra. Two important differences between the version of the \payne\ algorithm presented in this study and the version used in \citet{Gebruers2021} are that the latter study: 1) used a training scheme only based on a quasi-random sampling of the training examples after \citet{Sobol1967}, and 2) included the microturbulent velocity as a free parameter in view of the high resolution of the spectra used in their analysis. Difference 1) has important consequences for slowly rotating stars where extra care has to be taken during the training process to fully capture the rapidly increasing morphological complexity of stellar spectra compared to the cases of moderate to high projected rotational velocities (Section~\ref{sec:NN_train}). The inclusion of the microturbulent velocity as a free parameter is expected to result in a notable difference in the inferred stellar metallicity because of  non-negligible correlations between those two parameters.

\begin{figure*}
    \centering
    \includegraphics[width=\textwidth]{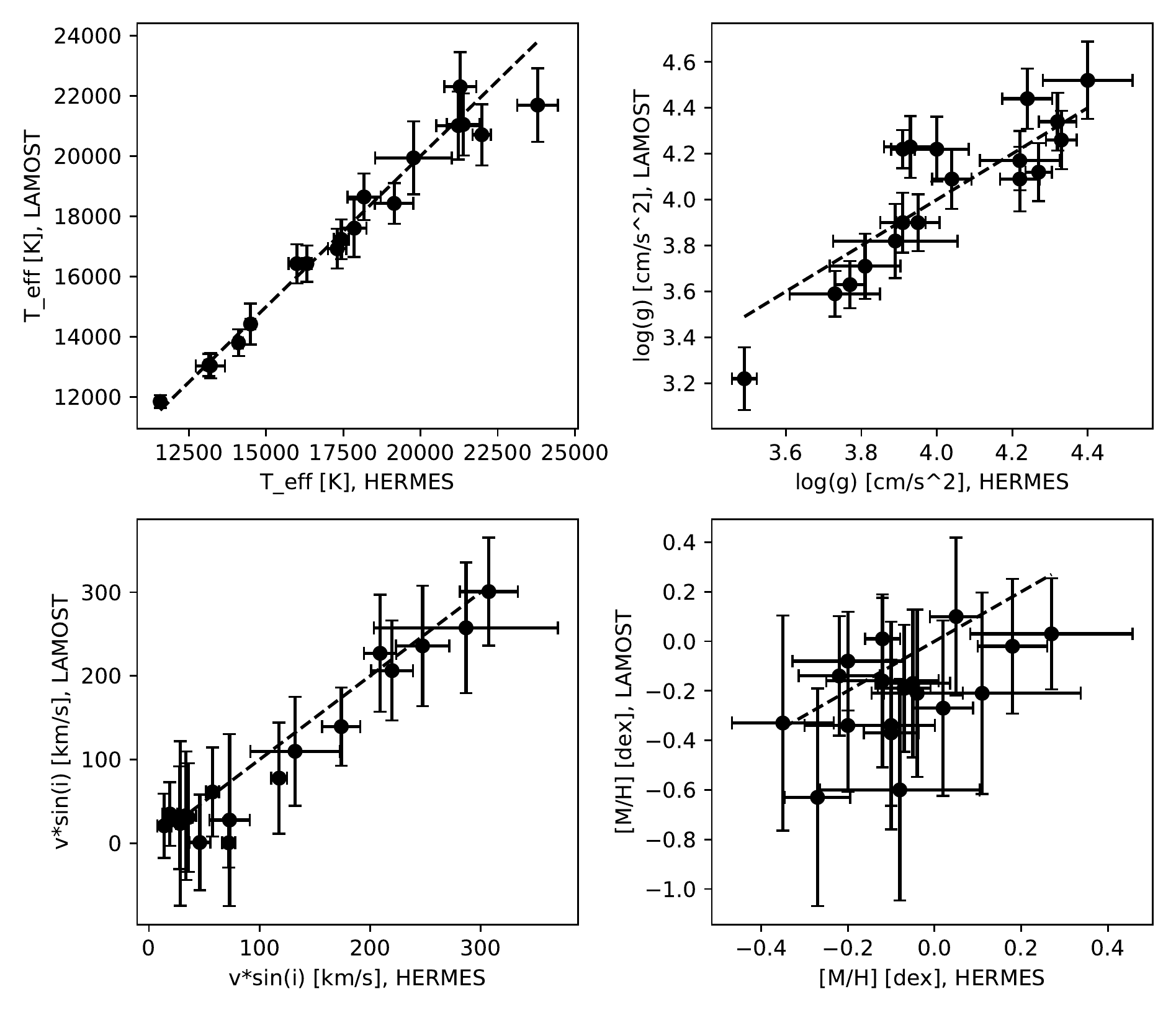}
    \caption{Comparison between stellar parameters inferred with the \payne\ algorithm from the HERMES high-resolution and LAMOST low-resolution optical spectra of a sample of 18 SPB stars from \citet{Gebruers2021}. The dashed black line indicates the one-to-one correspondence between the parameters and is shown to help guide the eye. The error bars reflect 1$\sigma$ statistical uncertainties of the fit and do not account for the internal parameter uncertainties.}
    \label{fig:LAMOST_vs_HERMES}
\end{figure*}

Figure~\ref{fig:Gebruers_vs_ThisWork} shows a comparison between the \teff, \logg, \vsini, and [M/H] parameters as inferred from the HERMES high-resolution spectra in \citet{Gebruers2021} and in this study (see also Table~\ref{tab:SPB_params_HERMES_vs_LAMOST} for numerical values; columns designated as ``HERMES''). We observe a good agreement within the quoted 1$\sigma$ statistical uncertainties (i.e. not taking the internal uncertainty into account) for \teff\ of the star, for \logg\ and \vsini\ for most part of the sample. However [M/H] derived here is systematically higher than the corresponding values obtained in \citet{Gebruers2021}. The small but statistically significant discrepancy observed for \vsini\ for the slowest rotators in the sample is explained by a more sophisticated and precise training of the neural network performed in this study in the corresponding region of the parameters space, as discussed in detail in Section~\ref{sec:NN_train}. There are also five objects for which we find systematically larger \logg\ values than reported in \citet{Gebruers2021} (see second panel in Figure~\ref{fig:Gebruers_vs_ThisWork}). Those are the five slowest rotators in the sample for which we also find discrepant values of the projected rotational velocity compared with \citet{Gebruers2021} due to their sub-optimal training of the neural network.
The systematic offset in the derived metallicity of the stars is associated with our exclusion of the microturbulent velocity $\xi$ from the free parameter vector as we kept it fixed to 2.0~\kms. Indeed, as demonstrated in Figure~\ref{fig:vmicro_effect}, the depths of most of the spectral lines of metals (and to a lesser extent of helium) steadily increase with increasing microturbulent velocity. This way, when synthetic spectra are computed for 2.0~\kms\ fixed value of the microturbulent velocity while the star in reality shows a larger value of microturbulence, the spectrum analysis algorithm tends to compensate for the observed difference in the line depths of metals, which is most easily achieved by increasing the [M/H] parameter in the models. The effect is quantitatively demonstrated in Table~\ref{tab:vmicro_effect}, where we summarize the results of the \payne\ analysis of five artificial spectra, each one computed with a different value of the microturbulent velocity parameter (from 2~\kms\ to 10~\kms\ in steps of 2~\kms) but with fixed values of \teff = 13\,000~K, \logg = 4.0~dex, \vsini = 100~\kms, and [M/H] = 0.0~dex. One can see that while we successfully recover \teff, \logg, and \vsini\ from the input spectrum within the quoted 1$\sigma$ statistical uncertainties in all five test cases, the discrepancy between the inferred and assumed metallicity steadily increases with the microturbulent velocity and reaches some 0.5~dex for the most extreme considered case of $\xi=10.0$~\kms.

We further proceed with the analysis of the LAMOST spectra of exactly the same sample of $Kepler$ SPB-type stars. We note that two out of twenty stars do not have LAMOST spectra, hence this particular analysis is restricted to 18 stars for which low-resolution spectra could be found in the LAMOST data archive. Our goal here is to quantify the effects of a factor $\sim$40 reduction in the resolving power of the instrument and substantially different properties of its response function on the inferred atmospheric parameters of the star. The results of our analysis are summarized in Figure~\ref{fig:LAMOST_vs_HERMES} and Table~\ref{tab:SPB_params_HERMES_vs_LAMOST} (columns designated as ``LAMOST''). We find an overall good agreement within the quoted 1$\sigma$ statistical uncertainties between the \teff, \vsini, and [M/H] parameters derived from the HERMES high-resolution and LAMOST low-resolution optical spectra. There is a small population of stars whose \logg\ values as inferred from the LAMOST spectra exceed those derived from the HERMES high-resolution spectra, where the agreement occurs only at $2\sigma$ instead of $1\sigma$ statistical uncertainty level (see top right panel in Figure~\ref{fig:LAMOST_vs_HERMES}). 
These discrepant cases are located in the region of high surface gravity, i.e. \logg~$\gtrsim$~3.9~dex, and are in agreement with each other when internal uncertainties are also taken into account.

\begin{figure*}
    \centering
    \includegraphics[width=0.28\textwidth]{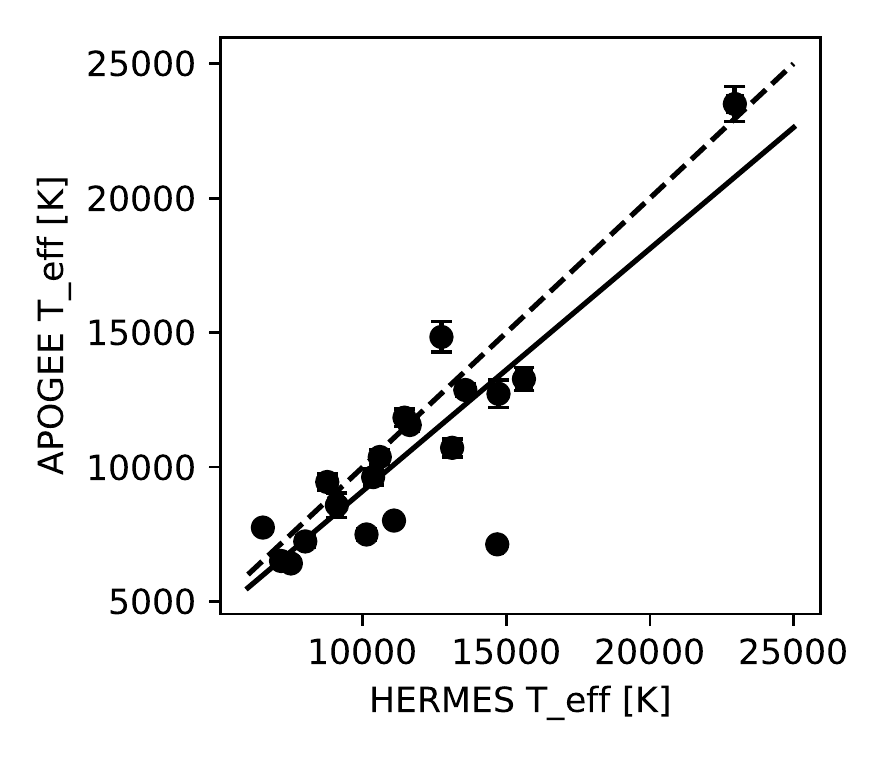}
    \includegraphics[width=0.34\textwidth]{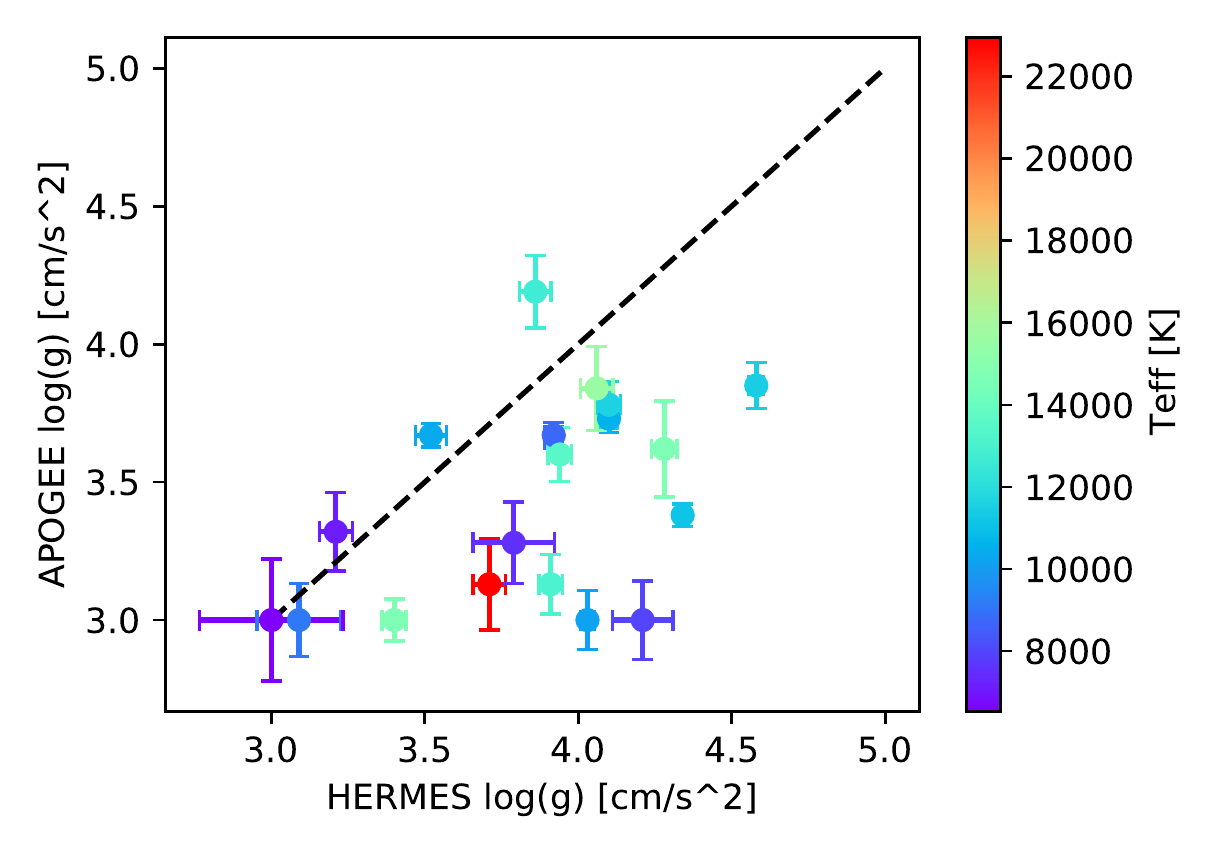}
    \includegraphics[width=0.34\textwidth]{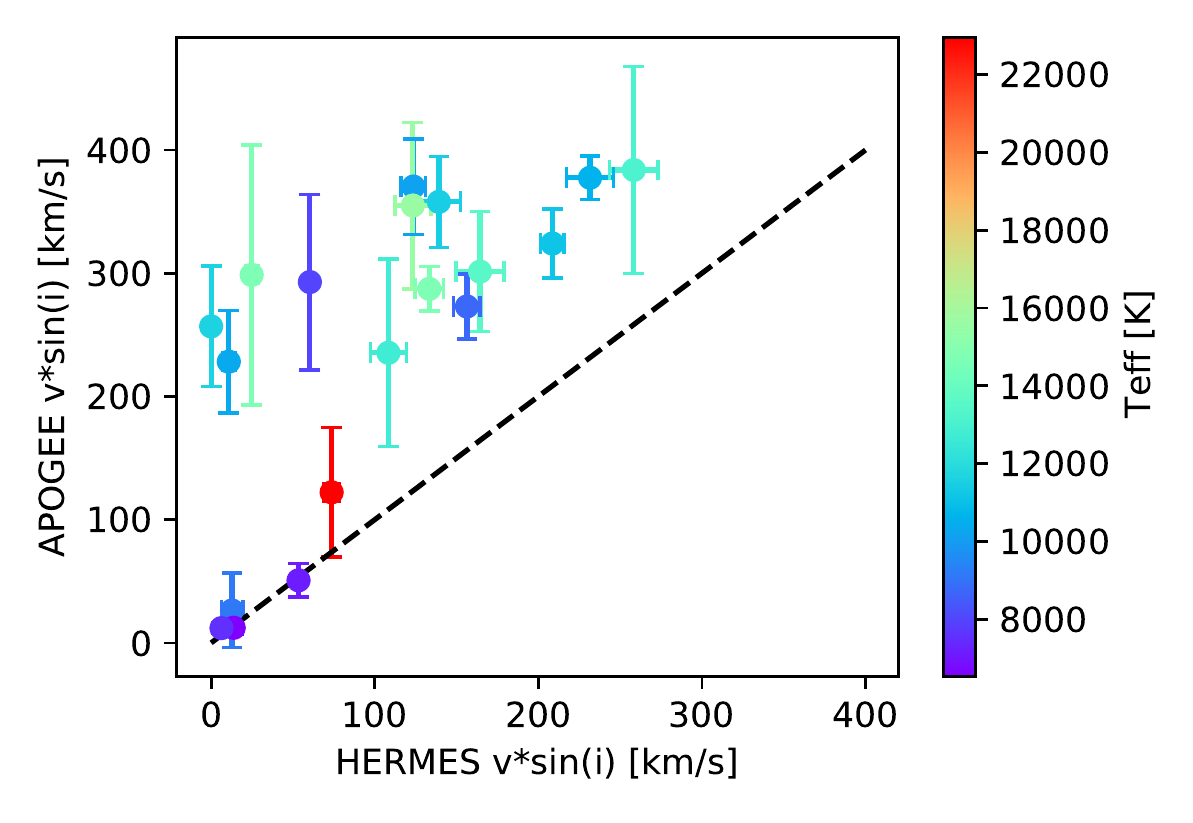}
    \caption{Comparison between atmospheric parameters derived from the medium-resolution NIR APOGEE and high-resolution optical HERMES spectra. From left to right: \teff, \logg, and [M/H] of the star. The dashed and solid lines represent the one-to-one correspondence and the best fit linear regression model, respectively. The error bars shown reflect 1$\sigma$ statistical uncertainties of the fit.} See text for details.
    \label{fig:HERMES_vs_APOGEE}
\end{figure*}

Finally, the \payne\ algorithm is applied to a sample of B- to F-type stars that were observed with the APOGEE instrument as part of the SDSS-V Pathfinder program in the northern hemisphere and for which high-resolution optical spectra exist in the HERMES data archive. The results of our analysis are illustrated in Figure~\ref{fig:HERMES_vs_APOGEE} and summarized in Table~\ref{tab:BAF_params_HERMES_vs_APOGEE}. As expected, the \vsini\ parameter can realistically be inferred from the APOGEE spectra for the coolest and slowly rotating F-type stars only, owing to a large number of strong metal lines found in the spectra of these objects. For the rest of the sample, the APOGEE spectra are dominated by the broad hydrogen lines of the Brackett series, which results in significant and systematic overestimation of the projected rotational velocity of the star compared to the case of high-resolution optical spectra. Similarly, we observe a significant discrepancy between the surface gravity \logg\ values inferred from the APOGEE and from the high-resolution optical spectra, with no clear dependency on \teff\ or \vsini\ of the star. The observed discrepancy is the result of a degeneracy between the surface gravity of the star and parameters of the residual response function model in the analysis of the near-IR APOGEE spectra. Indeed, as illustrated in Figure~\ref{fig:APOGEE_TeffLogg_effect} (right column), the wavelength region in the APOGEE spectra between some 1.5~$\mu$m and 1.6~$\mu$m (where the density of hydrogen lines steadily increases and they ultimately merge) is the most informative one for the inference of the surface gravity of B- to F-type stars. However, that particular part of the spectrum is also the most uncertain one in fitting the residual response function, leading to the above-mentioned degeneracy. Finally, we find a reasonably good agreement between the \teff\ values inferred from the near-IR APOGEE and high-resolution optical spectra, with a small but statistically significant underestimation of the effective temperature of the star from the near-IR spectra. The linear regression analysis gives an intercept and slope of 104~K and 0.9, respectively, with the null hypothesis of equal values being rejected at the level of a $p$-value of $2.3\times10^{-6}$ (see also left panel in Figure~\ref{fig:HERMES_vs_APOGEE}). We interpret the observed difference in the inferred \teff\ parameter as being due to the cumulative effect of (i) much more limited amount of information in the near-IR spectra of B- to F-type stars as compared to the optical wavelengths, (ii) partial methodological degeneracy between \teff, \logg, and parameters of the residual response function model, and (iii) physical effects like different sensitivity of spectral lines of hydrogen and metals to non-LTE effects at the optical and near-IR wavelengths. We also note that the above-described analysis was performed assuming a fixed value of the stellar metallicity [M/H] = 0.0~dex because of lack of information for the inference of the respective parameter from the near-IR spectra and the associated large internal uncertainty (see Figure~\ref{fig:NN_test_APOGEE} and Table~\ref{tab:sys_uncert_final}).

\begin{table*}
    \centering
    \caption{Stellar parameters of 19 B- to F-type stars as derived in this study from their APOGEE medium-resolution near-IR and HERMES high-resolution optical spectra. The analysis assumes fixed metallicity [M/H] = 0.0~dex. The quoted parameter uncertainties reflect 1$\sigma$ statistical uncertainties and do not account for the internal uncertainties reported in Sect.~\ref{sec:sim_tests}. See text for details.}
    \begin{tabular}{lllllll}
    \hline
    \multirow{2}{*}{Star name} & \multicolumn{2}{c}{\teff\ (K)} & \multicolumn{2}{c}{\logg\ (dex)} & \multicolumn{2}{c}{\vsini\ (\kms)} \\
    & HERMES & APOGEE & HERMES & APOGEE & HERMES & APOGEE \\
    \hline
16371249+7609490 & 7493 $\pm$ 50 & 6413 $\pm$ 140 & 3.79 $\pm$ 0.13 & 3.28 $\pm$ 0.15 & 6 $\pm$ 1 & 12 $\pm$ 4 \\
16462193+7701175 & 15611 $\pm$ 250 & 13274 $\pm$ 424 & 4.06 $\pm$ 0.05 & 3.84 $\pm$ 0.15 & 123 $\pm$ 11 & 355 $\pm$ 68 \\
16472272+6905558 & 11085 $\pm$ 35 & 8008 $\pm$ 75 & 4.34 $\pm$ 0.02 & 3.38 $\pm$ 0.04 & 209 $\pm$ 7 & 324 $\pm$ 28 \\
16524392+7651096 & 7152 $\pm$ 21 & 6506 $\pm$ 142 & 3.21 $\pm$ 0.05 & 3.32 $\pm$ 0.14 & 53 $\pm$ 1 & 51 $\pm$ 14 \\
17033403+5729591 & 11449 $\pm$ 57 & 11836 $\pm$ 329 & 4.58 $\pm$ 0.02 & 3.85 $\pm$ 0.08 & 139 $\pm$ 13 & 358 $\pm$ 37 \\
17122286+5713243 & 10358 $\pm$ 93 & 9625 $\pm$ 298 & 3.52 $\pm$ 0.05 & 3.67 $\pm$ 0.04 & 11 $\pm$ 4 & 228 $\pm$ 41 \\
17264168+5944556 & 14724 $\pm$ 146 & 12722 $\pm$ 507 & 4.28 $\pm$ 0.04 & 3.62 $\pm$ 0.17 & 25 $\pm$ 4 & 299 $\pm$ 105 \\
17503133+5726367 & 10590 $\pm$ 51 & 10372 $\pm$ 271 & 4.10 $\pm$ 0.02 & 3.73 $\pm$ 0.05 & 232 $\pm$ 14 & 377 $\pm$ 18 \\
18132623+6445575 & 13569 $\pm$ 141 & 12864 $\pm$ 240 & 3.94 $\pm$ 0.04 & 3.60 $\pm$ 0.10 & 164 $\pm$ 15 & 301 $\pm$ 49 \\
18440460+6046128 & 9095 $\pm$ 186 & 8578 $\pm$ 463 & 3.09 $\pm$ 0.14 & 3.00 $\pm$ 0.13 & 13 $\pm$ 7 & 26 $\pm$ 30 \\
18520222+5940014 & 13110 $\pm$ 127 & 10713 $\pm$ 333 & 3.91 $\pm$ 0.04 & 3.13 $\pm$ 0.11 & 258 $\pm$ 15 & 384 $\pm$ 84 \\
18534470+6001044 & 14680 $\pm$ 159 & 7124 $\pm$ 89 & 3.40 $\pm$ 0.04 & 3.00 $\pm$ 0.08 & 133 $\pm$ 9 & 287 $\pm$ 18 \\
18585261+6931525 & 12736 $\pm$ 132 & 14837 $\pm$ 568 & 3.86 $\pm$ 0.05 & 4.19 $\pm$ 0.13 & 108 $\pm$ 11 & 235 $\pm$ 76 \\
19094260+6451320 & 6518 $\pm$ 38 & 7747 $\pm$ 102 & 3.00 $\pm$ 0.23 & 3.00 $\pm$ 0.22 & 14 $\pm$ 1 & 12 $\pm$ 5 \\
19162421+6708066 & 22956 $\pm$ 256 & 23502 $\pm$ 657 & 3.71 $\pm$ 0.05 & 3.13 $\pm$ 0.17 & 74 $\pm$ 5 & 122 $\pm$ 52 \\
19181164+6057360 & 11633 $\pm$ 74 & 11563 $\pm$ 248 & 4.10 $\pm$ 0.04 & 3.78 $\pm$ 0.08 & 0 $\pm$ 1 & 257 $\pm$ 49 \\
19224444+7438013 & 8760 $\pm$ 40 & 9442 $\pm$ 291 & 3.92 $\pm$ 0.03 & 3.67 $\pm$ 0.05 & 156 $\pm$ 8 & 273 $\pm$ 26 \\
19373625+6401001 & 7994 $\pm$ 36 & 7232 $\pm$ 210 & 4.21 $\pm$ 0.10 & 3.00 $\pm$ 0.14 & 60 $\pm$ 3 & 293 $\pm$ 71 \\
20000399+6826121 & 10127 $\pm$ 47 & 7490 $\pm$ 214 & 4.03 $\pm$ 0.02 & 3.00 $\pm$ 0.11 & 124 $\pm$ 8 & 370 $\pm$ 39 \\
    \hline
    \end{tabular}
    \label{tab:BAF_params_HERMES_vs_APOGEE}
\end{table*}





\if false
\begin{figure}
    \centering
    \includegraphics[width=0.45\textwidth]{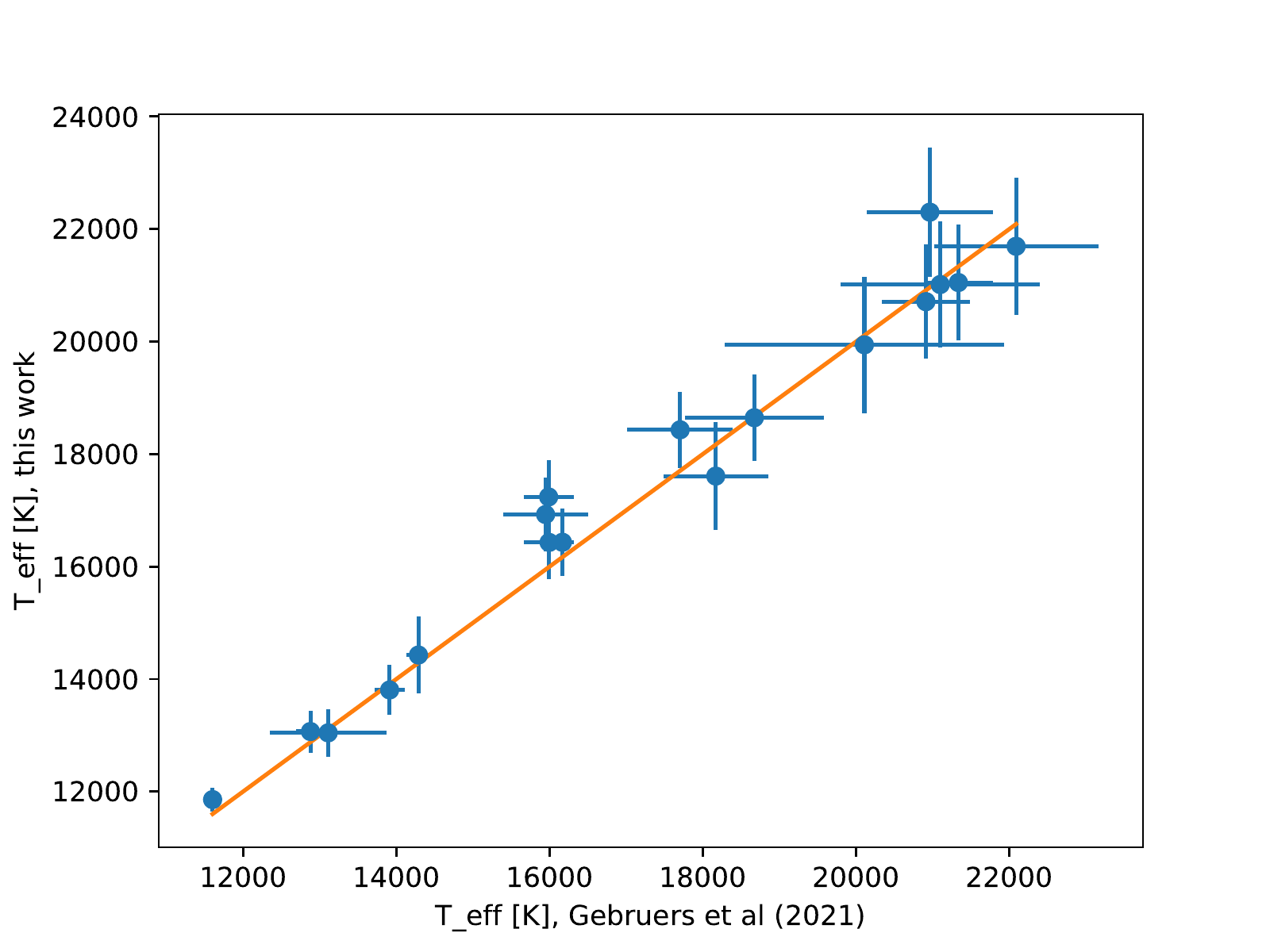}
    \caption{Comparison for $T_{eff}$ from LAMOST spectra of a sample of SPB stars, between Payne output and the values from Gebruers et al (2021).}
    \label{fig:spb_lamost_teff}
\end{figure}

\begin{figure}
    \centering
    \includegraphics[width=0.45\textwidth]{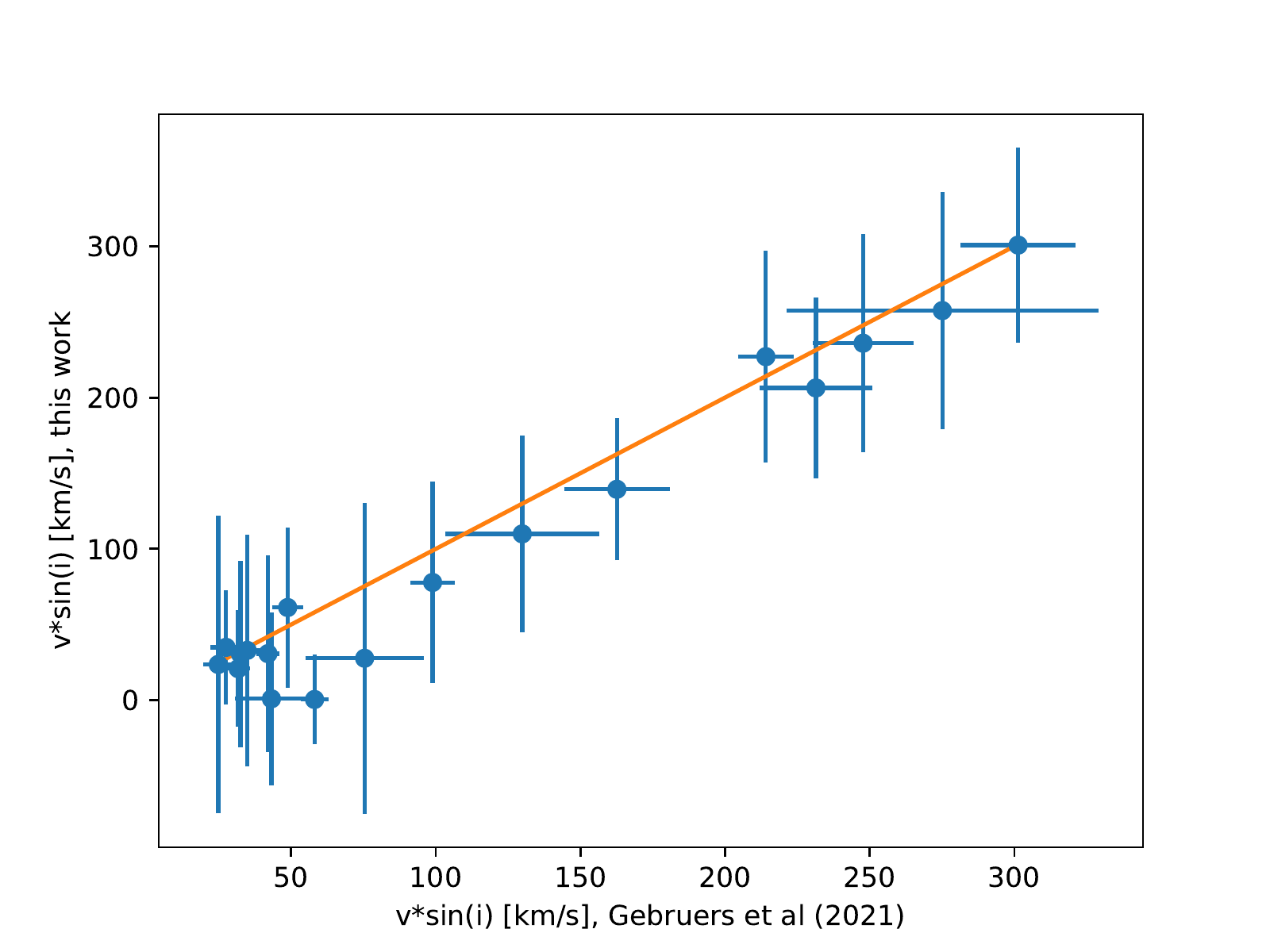}
    \caption{Comparison for $v*sin(i)$ from LAMOST spectra of a sample of SPB stars, between Payne output and the values from Gebruers et al (2021).}
    \label{fig:spb_lamost_vsini}
\end{figure}

\begin{figure}
    \centering
    \includegraphics[width=0.45\textwidth]{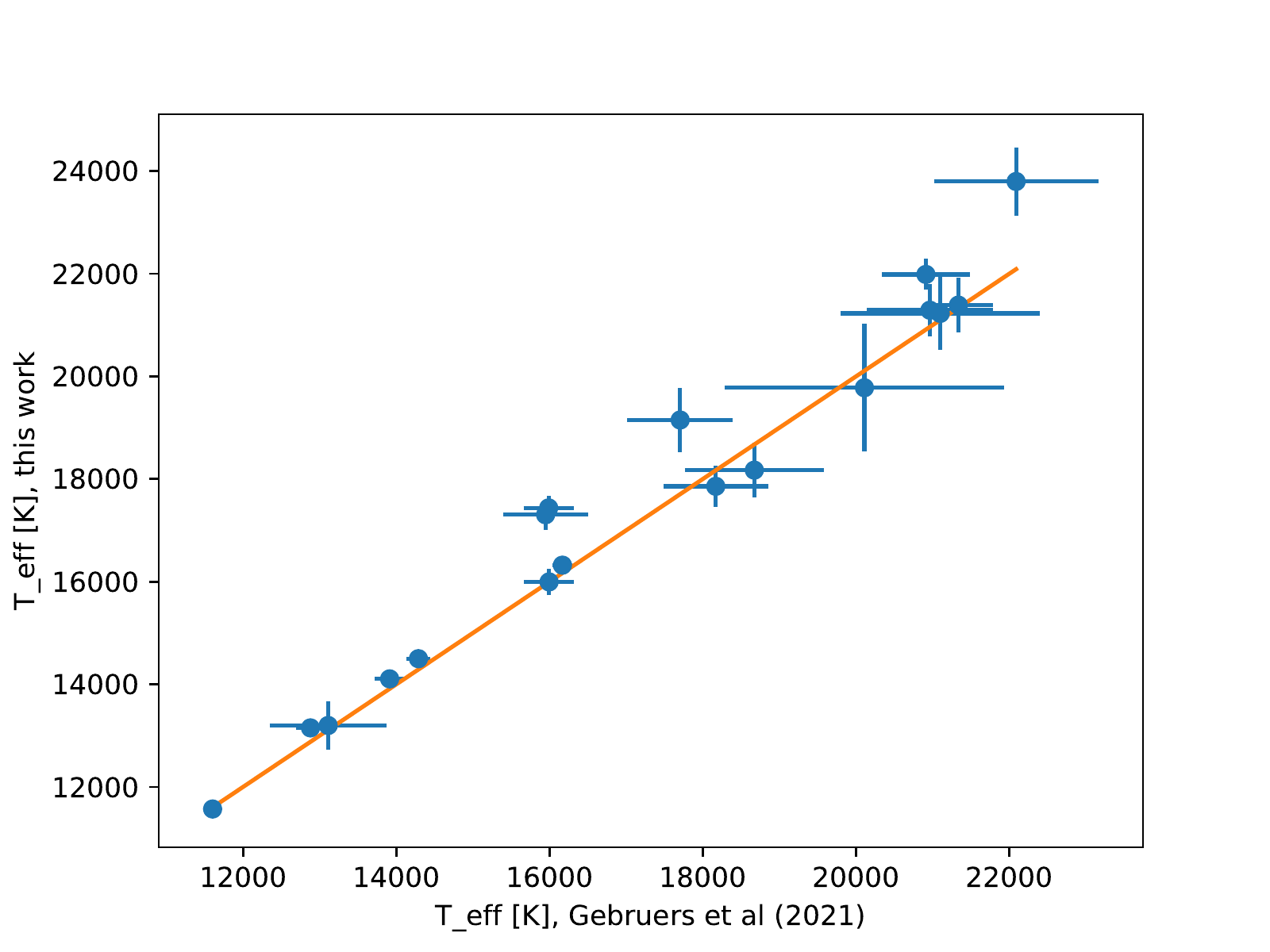}
    \caption{Comparison for $T_{eff}$ from HERMES spectra of a sample of SPB stars, between Payne output and the values from Gebruers et al (2021).}
    \label{fig:spb_hermes_teff}
\end{figure}

\begin{figure}
    \centering
    \includegraphics[width=0.45\textwidth]{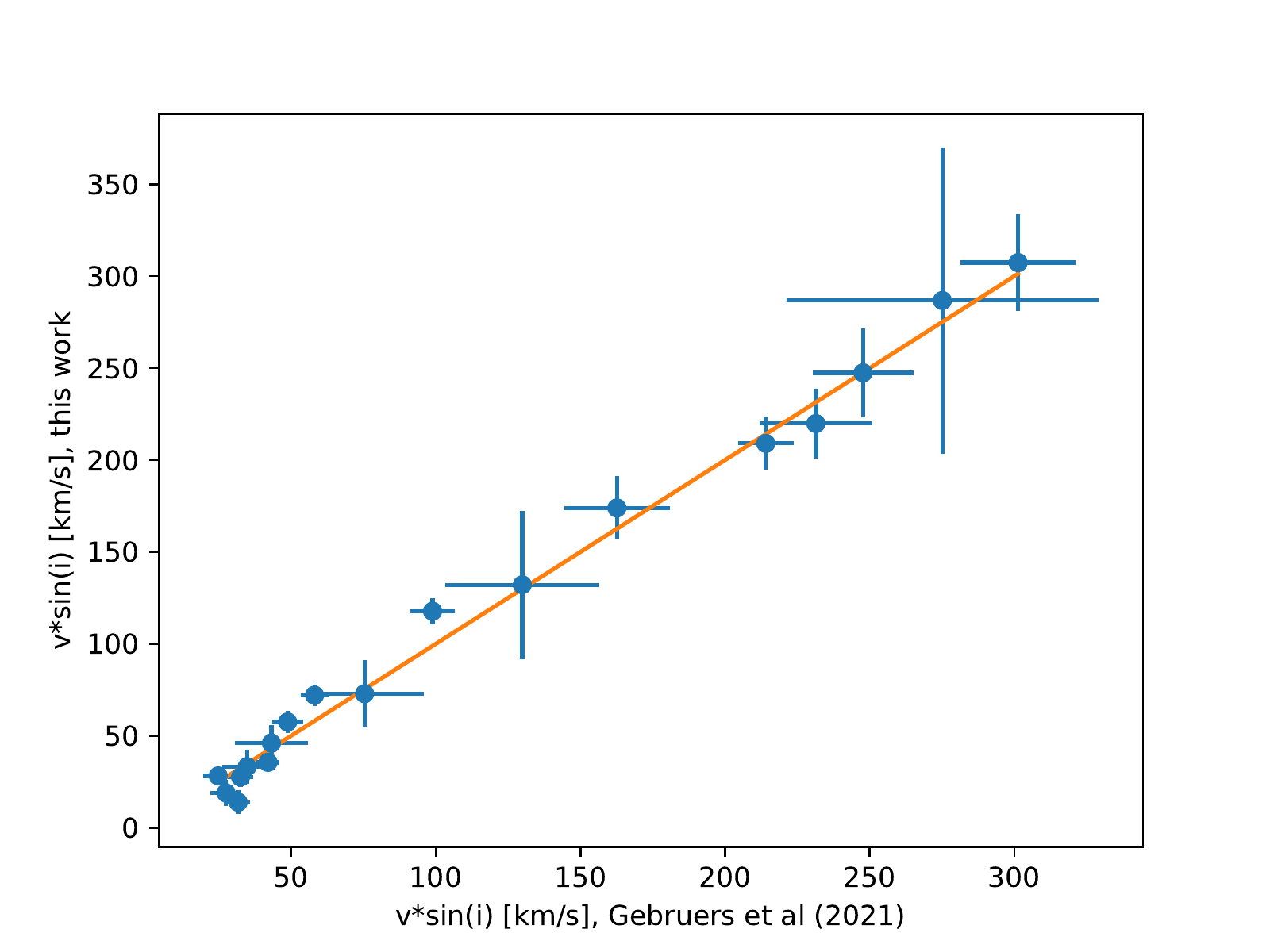}
    \caption{Comparison for $v*sin(i)$ from HERMES spectra of a sample of SPB stars, between Payne output and the values from Gebruers et al (2021).}
    \label{fig:spb_hermes_vsini}
\end{figure}
\fi


\section{Conclusions and future prospects}\label{sec:conclusions}

In this study, we develop a fully automated, machine learning-based spectrum analysis algorithm \payne, whose main purpose is to serve the Milky Way Mapper program of the SDSS-V survey in its daily data analysis routine. Though the algorithm is mainly positioned for the analysis of intermediate- to high-mass stars of spectral types O, B, A, and F, it can easily be extended towards lower stellar effective temperatures and masses, if necessary. This paper presents a detailed description of the spectrum analysis algorithm and the statistical framework it is embedded in. It details the algorithm training, validation, and testing steps. The algorithm tests are performed both on artificial medium-resolution near-IR and low-resolution optical spectra, and on two control samples of real stars for which (HERMES) high- and (LAMOST) low-resolution optical and (APOGEE) medium-resolution near-IR spectra have been acquired. The main results and conclusions of the paper are as follows:
\begin{itemize}
    \item The \payne\ algorithm is purposely generalized to cover data analysis needs of the MWM program that builds on observations with two multi-object instruments operating in different wavelength regimes and at different resolving powers. To make the algorithm readily applicable to both types of data, where APOGEE spectra of OBAF-type stars in particular suffer from the lack of a well-defined pseudo-continuum, we integrate the spectrum normalization step into the analysis framework where the pseudo-continuum of the star is represented by a series of Chebyshev polynomials. Coefficients of the polynomials are optimized along with atmospheric parameters of the star, thus presenting a valuable alternative to the conventional approach where the input spectrum is (pre)normalized to the local continuum prior to its detailed analysis.
    \item For the algorithm training, we employ a mixture of a quasi-random sampling \citep{Sobol1967} in the entire parameter space and an additional random grid sampled from a Gaussian distribution at low effective temperatures (\teff~$\lesssim$~10\,000~K) and projected rotational velocities (\vsini~$\lesssim$~30~\kms). This hybrid training approach allows us to maximally capture the large diversity in morphological complexity of stellar spectra when transitioning from hotter and more rapidly rotating B- and A-type stars to cooler F-type objects whose spectra are typically characterized by more narrow metal lines that are also available in numbers.
    \item The \payne\ algorithm testing on artificial APOGEE medium-resolution near-IR and BOSS low-resolution optical spectra reveals a similar performance at some 96\% level in terms of the reliability metric (defined as the probability that the optimization algorithm converges to a correct set of atmospheric parameters; see Section~\ref{sec:sim_tests}). However, significantly different internal uncertainties occur for the two wavelength regimes. This overall {\b internal} uncertainty is lower in the low-resolution optical spectra than in the APOGEE spectra at S/N values of 100 and 50 by a factor of $\sim$3-4 for \teff\ and RV of the star, and some 0.05~dex and 0.35~dex for its \logg\ and [M/H], respectively. Thus, we conclude that despite a factor ten lower resolving power of the BOSS instrument compared to the APOGEE spectrograph, optical spectra remain a preferred option for the extraction of atmospheric parameters of OBAF-type stars, unless the latter suffer from high extinction so that their flux can only be observed at longer (e.g. near-IR) wavelengths. We note, however, that this conclusion does not concerns inferences of the surface chemical composition of the stars, where resolving individual spectral lines of metals becomes an important factor. The optimal SDSS-V scenario is to observe a star with both the BOSS and APOGEE instruments and rely on the optical and near-IR spectrum for the determination of atmospheric parameters and surface chemical abundances, respectively. An exception would be O- and early B-type stars that do not display spectral lines of metals in the APOGEE spectra (cf. Figure~\ref{fig:SynthAPOGEESpectra}).
    \item Application of the \payne\ algorithm to the HERMES high-resolution and LAMOST low-resolution optical spectra of 18 objects from the sample of SPB-type stars analyzed spectroscopically in \citet{Gebruers2021} reveals a good agreement between the two sets of atmospheric parameters, including \teff, \logg, \vsini, and [M/H] of the star. 
    The present version of the \payne\ algorithm brings an improved performance
    in the regime of slowly rotating late A- to F-type stars compared to an earlier version of the algorithm employed in \citet{Gebruers2021}. We conclude that as long as the LSF
of the instrument is properly accounted for in the analysis of stellar spectra, atmospheric parameters of OBAF-type stars can still be accurately inferred from low-resolution optical spectra, albeit with typically lower precision than from their high-resolution optical spectra. \citet{Xiang2021} come to similar conclusions from the analysis of LAMOST spectra of some 330\,000 OBA-type stars and validation of their {\sc HotPayne} algorithm based on medium- to high-resolution spectra in the literature.
    \item Application of the \payne\ algorithm to the HERMES high-resolution optical and APOGEE medium-resolution near-IR spectra of a sample of 19 BAF-type stars reinforces our conclusions drawn from the tests with simulated data that the APOGEE near-IR spectra are less suitable for the extraction of accurate atmospheric parameters of OBAF-type stars than optical spectra. Owing to the large dominance of the hydrogen lines of the Brackett series in the APOGEE instrument wavelength range, \vsini\ can only be reliably inferred for the coolest F-type stars with slow to moderate rotation. Inference of the stellar surface gravity suffers from strong degeneracy with parameters of the instrument response function model, while the lack of prominent metal lines in the APOGEE spectra of OB(early A)-type stars prevents determination of their atmospheric chemical composition. Finally, the effective temperature of the star is derived with higher accuracy and precision than \vsini\ and \logg. The intercept and slope of the linear regression model fitted to the difference between \teff\ inferred from HERMES and APOGEE spectra are 104~K and 0.9, respectively. From the above results, it may appear that there is limited advantage in using the APOGEE spectra for the analysis OBAF-type stars. However, we emphasize that the results based on the APOGEE spectra obtained with the current version of the \payne\ algorithm are still up for improvement owing to the following two assumptions made: (i) the LSF of the instrument is well known and can be described with a wavelength-independent Gaussian kernel corresponding to R$\sim$22\,500, and (ii) the residual response function is smooth and sufficiently well-behaved such that it can be approximated with a series of Chebyshev polynomials. None of these assumptions holds in reality and we need to gain a better understanding of both the APOGEE LSF and residual response function before any firm conclusions can be drawn about the gain of using the near-IR spectra for the analysis of OBAF-type stars.
\end{itemize}

Although the \payne\ pipeline requires little to no human intervention and serves the basic needs of the Milky Way Mapper spectroscopic survey, there is room for improvement. This concerns the input physics currently used in atmosphere models and generalization of the algorithm beyond the analysis of exclusively stellar spectra of single stars. In particular, the (near-)future prospects that we plan for implementation in the forthcoming second release of the \payne\ pipeline are:
\begin{itemize}
    \item A better understanding and model descriptions of the APOGEE LSF and residual response function.
    \item Generalization of the algorithm towards fully automated detection and subsequent analysis of (composite) spectra of spectroscopic double-lined binary stars.
    \item Algorithm extension to allow for the determination of surfaces abundances of individual chemical elements such as He, C, N, O, Si, Mg, and Fe.
    \item Extension of the currently employed input physics to non-LTE atmosphere models and/or spectral line formation. As discussed in detail in \citet{Nieva2007}, the use of the hybrid approach that employs LTE-based atmosphere models and non-LTE spectral line formation is justified for the spectrum analysis of OB-type dwarf and giant stars. Both, the hybrid and full non-LTE approaches will be implemented by coupling the {\sc tlusty} \citep{Hubeny1995} non-LTE model atmosphere code to the line formation code currently employed in the \payne\ pipeline. For modelling spectra of the hottest and most massive O- and early B-type stars that often have extended atmospheres and winds, we will employ the most recent version of the {\sc fastwind} code \citep{Puls2005,Sundqvist2019} to (re-)train the \payne\ algorithm in the relevant part of the parameter space.
    \item Overall, and for the analysis of APOGEE spectra of OBAF-type stars in particular, it is highly beneficial to include more input options for stellar observables, such as spectral energy distributions (SEDs) and/or photometric colors, Gaia information, etc. The main purpose of this addition is to provide extra observational constraints, in particular for the effective temperature and surface gravity of the star, to break degeneracies between various parameters in the spectroscopic analysis of OBAF-type stars.
\end{itemize}


\section*{Acknowledgments}
Funding for the Sloan Digital Sky Survey V has been provided by the Alfred P. Sloan Foundation, the Heising-Simons Foundation, and the Participating Institutions. SDSS acknowledges support and resources from the Center for High-Performance Computing at the University of Utah. The SDSS web site is \url{www.sdss5.org}.

SDSS is managed by the Astrophysical Research Consortium for the Participating Institutions of the SDSS Collaboration, including the Carnegie Institution for Science, Chilean National Time Allocation Committee (CNTAC) ratified researchers, the Gotham Participation Group, Harvard University, Heidelberg University, The Johns Hopkins University, L'Ecole polytechnique f\'{e}d\'{e}rale de Lausanne (EPFL), Leibniz-Institut f\"{u}r Astrophysik Potsdam (AIP), Max-Planck-Institut f\"{u}r Astronomie (MPIA Heidelberg), Max-Planck-Institut f\"{u}r Extraterrestrische Physik (MPE), Nanjing University, National Astronomical Observatories of China (NAOC), New Mexico State University, The Ohio State University, Pennsylvania State University, Smithsonian Astrophysical Observatory, Space Telescope Science Institute (STScI), the Stellar Astrophysics Participation Group, Universidad Nacional Aut\'{o}noma de M\'{e}xico, University of Arizona, University of Colorado Boulder, University of Illinois at Urbana-Champaign, University of Toronto, University of Utah, University of Virginia, Yale University, and Yunnan University.

The research leading to these results has received funding from the European Research Council (ERC) under the European Union's Horizon 2020 research and innovation programme (grant agreement N$^\circ$670519: MAMSIE), from the KU~Leuven Research Council (grant C16/18/005: PARADISE), from the Research Foundation Flanders (FWO) under grant agreement G0H5416N (ERC Runner Up Project), as well as from the BELgian federal Science Policy Office (BELSPO) through PRODEX grant PLATO. SG gratefully acknowledges support from the Research Foundation Flanders (FWO) by means of a PhD Aspirant mandate under contract No. 11E5620N.
This work is based on observations obtained with the HERMES spectrograph, which is supported by the Research Foundation - Flanders (FWO), Belgium, the Research Council of KU Leuven, Belgium, the Fonds National de la Recherche Scientifique (F.R.S.-FNRS), Belgium, the Royal Observatory of Belgium, the Observatoire de Genève, Switzerland and the Thüringer Landessternwarte Tautenburg, Germany. 
The computational resources and services used in this work were provided by the VSC (Flemish Supercomputer Centre), funded by the Research Foundation - Flanders (FWO) and the Flemish Government. YST acknowledges financial support from the Australian Research Council through DECRA Fellowship DE220101520.
The authors are grateful to Cole Johnston (KU Leuven, Radboud University Nijmegen), Dominic M. Bowman (KU Leuven) and Siemen Burssens (KU Leuven) for performing HERMES observations, some of which were used in this study; to Joel Brownstein (Utah State University) and Maarten Dirickx (KU Leuven) for their help with IT aspects of this work.

\bibliography{sample63}{}

\begin{thebibliography}{}
\expandafter\ifx\csname natexlab\endcsname\relax\def\natexlab#1{#1}\fi
\providecommand{\url}[1]{\href{#1}{#1}}
\providecommand{\dodoi}[1]{doi:~\href{http://doi.org/#1}{\nolinkurl{#1}}}
\providecommand{\doeprint}[1]{\href{http://ascl.net/#1}{\nolinkurl{http://ascl.net/#1}}}
\providecommand{\doarXiv}[1]{\href{https://arxiv.org/abs/#1}{\nolinkurl{https://arxiv.org/abs/#1}}}

\bibitem[{{Aerts}(2021)}]{Aerts2021}
{Aerts}, C. 2021, Reviews of Modern Physics, 93, 015001,
  \dodoi{10.1103/RevModPhys.93.015001}

\bibitem[{{Aerts} {et~al.}(2010){Aerts}, {Christensen-Dalsgaard}, \&
  {Kurtz}}]{Aerts2010}
{Aerts}, C., {Christensen-Dalsgaard}, J., \& {Kurtz}, D.~W. 2010,
  {Asteroseismology} (Springer)

\bibitem[{{Aerts} {et~al.}(2019){Aerts}, {Mathis}, \&
  {Rogers}}]{2019ARA&A..57...35A}
{Aerts}, C., {Mathis}, S., \& {Rogers}, T.~M. 2019, \araa, 57, 35,
  \dodoi{10.1146/annurev-astro-091918-104359}

\bibitem[{{Aerts} {et~al.}(2018){Aerts}, {Molenberghs}, {Michielsen},
  {Pedersen}, {Bj{\"o}rklund}, {Johnston}, {Mombarg}, {Bowman}, {Buysschaert},
  {P{\'a}pics}, {Sekaran}, {Sundqvist}, {Tkachenko}, {Truyaert}, {Van Reeth},
  \& {Vermeyen}}]{2018ApJS..237...15A}
{Aerts}, C., {Molenberghs}, G., {Michielsen}, M., {et~al.} 2018, \apjs, 237,
  15, \dodoi{10.3847/1538-4365/aaccfb}

\bibitem[{{Almeida} {et~al.}(2017){Almeida}, {Sana}, {Taylor}, {Barb{\'a}},
  {Bonanos}, {Crowther}, {Damineli}, {de Koter}, {de Mink}, {Evans}, {Gieles},
  {Grin}, {H{\'e}nault-Brunet}, {Langer}, {Lennon}, {Lockwood}, {Ma{\'\i}z
  Apell{\'a}niz}, {Moffat}, {Neijssel}, {Norman}, {Ram{\'\i}rez-Agudelo},
  {Richardson}, {Schootemeijer}, {Shenar}, {Soszy{\'n}ski}, {Tramper}, \&
  {Vink}}]{Almeida2017}
{Almeida}, L.~A., {Sana}, H., {Taylor}, W., {et~al.} 2017, \aap, 598, A84,
  \dodoi{10.1051/0004-6361/201629844}

\bibitem[{{Auvergne} {et~al.}(2009){Auvergne}, {Bodin}, {Boisnard}, {Buey},
  {Chaintreuil}, {Epstein}, {Jouret}, {Lam-Trong}, {Levacher}, {Magnan},
  {Perez}, {Plasson}, {Plesseria}, {Peter}, {Steller}, {Tiph{\`e}ne}, {Baglin},
  {Agogu{\'e}}, {Appourchaux}, {Barbet}, {Beaufort}, {Bellenger}, {Berlin},
  {Bernardi}, {Blouin}, {Boumier}, {Bonneau}, {Briet}, {Butler}, {Cautain},
  {Chiavassa}, {Costes}, {Cuvilho}, {Cunha-Parro}, {de Oliveira Fialho},
  {Decaudin}, {Defise}, {Djalal}, {Docclo}, {Drummond}, {Dupuis}, {Exil},
  {Faur{\'e}}, {Gaboriaud}, {Gamet}, {Gavalda}, {Grolleau}, {Gueguen},
  {Guivarc'h}, {Guterman}, {Hasiba}, {Huntzinger}, {Hustaix}, {Imbert},
  {Jeanville}, {Johlander}, {Jorda}, {Journoud}, {Karioty}, {Kerjean},
  {Lafond}, {Lapeyrere}, {Landiech}, {Larqu{\'e}}, {Laudet}, {Le Merrer},
  {Leporati}, {Leruyet}, {Levieuge}, {Llebaria}, {Martin}, {Mazy}, {Mesnager},
  {Michel}, {Moalic}, {Monjoin}, {Naudet}, {Neukirchner}, {Nguyen-Kim},
  {Ollivier}, {Orcesi}, {Ottacher}, {Oulali}, {Parisot}, {Perruchot},
  {Piacentino}, {Pinheiro da Silva}, {Platzer}, {Pontet}, {Pradines},
  {Quentin}, {Rohbeck}, {Rolland}, {Rollenhagen}, {Romagnan}, {Russ}, {Samadi},
  {Schmidt}, {Schwartz}, {Sebbag}, {Smit}, {Sunter}, {Tello}, {Toulouse},
  {Ulmer}, {Vandermarcq}, {Vergnault}, {Wallner}, {Waultier}, \&
  {Zanatta}}]{Auvergne2009}
{Auvergne}, M., {Bodin}, P., {Boisnard}, L., {et~al.} 2009, \aap, 506, 411,
  \dodoi{10.1051/0004-6361/200810860}

\bibitem[{{Banyard} {et~al.}(2021){Banyard}, {Sana}, {Mahy}, {Bodensteiner},
  {Villase{\~n}or}, \& {Evans}}]{Banyard2021}
{Banyard}, G., {Sana}, H., {Mahy}, L., {et~al.} 2021, arXiv e-prints,
  arXiv:2108.07814.
\newblock \doarXiv{2108.07814}

\bibitem[{{Bodensteiner} {et~al.}(2021){Bodensteiner}, {Sana}, {Wang},
  {Langer}, {Mahy}, {Banyard}, {de Koter}, {de Mink}, {Evans}, {G{\"o}tberg},
  {Patrick}, {Schneider}, \& {Tramper}}]{Bodensteiner2021}
{Bodensteiner}, J., {Sana}, H., {Wang}, C., {et~al.} 2021, \aap, 652, A70,
  \dodoi{10.1051/0004-6361/202140507}

\bibitem[{{Bonifacio} {et~al.}(2016){Bonifacio}, {Dalton}, {Trager}, {Aguerri},
  {Carrasco}, {Vallenari}, {Abrams}, {Middleton}, \&
  {Say{\`e}de}}]{2016sf2a.conf..267B}
{Bonifacio}, P., {Dalton}, G., {Trager}, S., {et~al.} 2016, in SF2A-2016:
  Proceedings of the Annual meeting of the French Society of Astronomy and
  Astrophysics, ed. C.~{Reyl{\'e}}, J.~{Richard}, L.~{Cambr{\'e}sy},
  M.~{Deleuil}, E.~{P{\'e}contal}, L.~{Tresse}, \& I.~{Vauglin}, 267--270

\bibitem[{{Borucki} {et~al.}(2010){Borucki}, {Koch}, {Basri}, {Batalha},
  {Brown}, {Caldwell}, {Caldwell}, {Christensen-Dalsgaard}, {Cochran},
  {DeVore}, {Dunham}, {Dupree}, {Gautier}, {Geary}, {Gilliland}, {Gould},
  {Howell}, {Jenkins}, {Kondo}, {Latham}, {Marcy}, {Meibom}, {Kjeldsen},
  {Lissauer}, {Monet}, {Morrison}, {Sasselov}, {Tarter}, {Boss}, {Brownlee},
  {Owen}, {Buzasi}, {Charbonneau}, {Doyle}, {Fortney}, {Ford}, {Holman},
  {Seager}, {Steffen}, {Welsh}, {Rowe}, {Anderson}, {Buchhave}, {Ciardi},
  {Walkowicz}, {Sherry}, {Horch}, {Isaacson}, {Everett}, {Fischer}, {Torres},
  {Johnson}, {Endl}, {MacQueen}, {Bryson}, {Dotson}, {Haas}, {Kolodziejczak},
  {Van Cleve}, {Chandrasekaran}, {Twicken}, {Quintana}, {Clarke}, {Allen},
  {Li}, {Wu}, {Tenenbaum}, {Verner}, {Bruhweiler}, {Barnes}, \&
  {Prsa}}]{2010Sci...327..977B}
{Borucki}, W.~J., {Koch}, D., {Basri}, G., {et~al.} 2010, Science, 327, 977,
  \dodoi{10.1126/science.1185402}

\bibitem[{Bowen \& Vaughan(1973)}]{Bowen:73}
Bowen, I.~S., \& Vaughan, A.~H. 1973, Appl. Opt., 12, 1430,
  \dodoi{10.1364/AO.12.001430}

\bibitem[{{Bowman} {et~al.}(2020){Bowman}, {Burssens}, {Sim{\'o}n-D{\'\i}az},
  {Edelmann}, {Rogers}, {Horst}, {R{\"o}pke}, \& {Aerts}}]{Bowman2020}
{Bowman}, D.~M., {Burssens}, S., {Sim{\'o}n-D{\'\i}az}, S., {et~al.} 2020,
  \aap, 640, A36, \dodoi{10.1051/0004-6361/202038224}

\bibitem[{{Bowman} {et~al.}(2019){Bowman}, {Burssens}, {Pedersen}, {Johnston},
  {Aerts}, {Buysschaert}, {Michielsen}, {Tkachenko}, {Rogers}, {Edelmann},
  {Ratnasingam}, {Sim{\'o}n-D{\'\i}az}, {Castro}, {Moravveji}, {Pope}, {White},
  \& {De Cat}}]{Bowman2019}
{Bowman}, D.~M., {Burssens}, S., {Pedersen}, M.~G., {et~al.} 2019, Nature
  Astronomy, 3, 760, \dodoi{10.1038/s41550-019-0768-1}

\bibitem[{Branch {et~al.}(1999)Branch, Coleman, \& Li}]{Branch1999}
Branch, M.~A., Coleman, T.~F., \& Li, Y. 1999, SIAM J. Sci. Comput., 21, 1

\bibitem[{Butler(1984)}]{Butler1984}
Butler, K. 1984, PhD thesis, University of London, UK

\bibitem[{{Casey} {et~al.}(2016){Casey}, {Hogg}, {Ness}, {Rix}, {Ho}, \&
  {Gilmore}}]{2016arXiv160303040C}
{Casey}, A.~R., {Hogg}, D.~W., {Ness}, M., {et~al.} 2016, arXiv e-prints,
  arXiv:1603.03040.
\newblock \doarXiv{1603.03040}

\bibitem[{{Claret} \& {Torres}(2019)}]{Claret2019}
{Claret}, A., \& {Torres}, G. 2019, \apj, 876, 134,
  \dodoi{10.3847/1538-4357/ab1589}

\bibitem[{{Danielski} {et~al.}(2021){Danielski}, {Brucalassi}, {Benatti},
  {Campante}, {Delgado-Mena}, {Rainer}, {Sacco}, {Adibekyan}, {Biazzo},
  {Bossini}, {Bruno}, {Casali}, {Kabath}, {Magrini}, {Micela}, {Morello},
  {Palladino}, {Sanna}, {Sarkar}, {Sousa}, {Tsantaki}, {Turrini}, \& {Van der
  Swaelmen}}]{Danielski2021}
{Danielski}, C., {Brucalassi}, A., {Benatti}, S., {et~al.} 2021, Experimental
  Astronomy, \dodoi{10.1007/s10686-021-09765-1}

\bibitem[{{de Jong} {et~al.}(2019){de Jong}, {Agertz}, {Berbel}, {Aird},
  {Alexander}, {Amarsi}, {Anders}, {Andrae}, {Ansarinejad}, {Ansorge},
  {Antilogus}, {Anwand-Heerwart}, {Arentsen}, {Arnadottir}, {Asplund}, {Auger},
  {Azais}, {Baade}, {Baker}, {Baker}, {Balbinot}, {Baldry}, {Banerji},
  {Barden}, {Barklem}, {Barth{\'e}l{\'e}my-Mazot}, {Battistini}, {Bauer},
  {Bell}, {Bellido-Tirado}, {Bellstedt}, {Belokurov}, {Bensby}, {Bergemann},
  {Bestenlehner}, {Bielby}, {Bilicki}, {Blake}, {Bland-Hawthorn}, {Boeche},
  {Boland}, {Boller}, {Bongard}, {Bongiorno}, {Bonifacio}, {Boudon}, {Brooks},
  {Brown}, {Brown}, {Br{\"u}ggen}, {Brynnel}, {Brzeski}, {Buchert},
  {Buschkamp}, {Caffau}, {Caillier}, {Carrick}, {Casagrande}, {Case}, {Casey},
  {Cesarini}, {Cescutti}, {Chapuis}, {Chiappini}, {Childress}, {Christlieb},
  {Church}, {Cioni}, {Cluver}, {Colless}, {Collett}, {Comparat}, {Cooper},
  {Couch}, {Courbin}, {Croom}, {Croton}, {Daguis{\'e}}, {Dalton}, {Davies},
  {Davis}, {de Laverny}, {Deason}, {Dionies}, {Disseau}, {Doel}, {D{\"o}scher},
  {Driver}, {Dwelly}, {Eckert}, {Edge}, {Edvardsson}, {Youssoufi}, {Elhaddad},
  {Enke}, {Erfanianfar}, {Farrell}, {Fechner}, {Feiz}, {Feltzing}, {Ferreras},
  {Feuerstein}, {Feuillet}, {Finoguenov}, {Ford}, {Fotopoulou}, {Fouesneau},
  {Frenk}, {Frey}, {Gaessler}, {Geier}, {Gentile Fusillo}, {Gerhard},
  {Giannantonio}, {Giannone}, {Gibson}, {Gillingham},
  {Gonz{\'a}lez-Fern{\'a}ndez}, {Gonzalez-Solares}, {Gottloeber}, {Gould},
  {Grebel}, {Gueguen}, {Guiglion}, {Haehnelt}, {Hahn}, {Hansen}, {Hartman},
  {Hauptner}, {Hawkins}, {Haynes}, {Haynes}, {Heiter}, {Helmi}, {Aguayo},
  {Hewett}, {Hinton}, {Hobbs}, {Hoenig}, {Hofman}, {Hook}, {Hopgood},
  {Hopkins}, {Hourihane}, {Howes}, {Howlett}, {Huet}, {Irwin}, {Iwert},
  {Jablonka}, {Jahn}, {Jahnke}, {Jarno}, {Jin}, {Jofre}, {Johl}, {Jones},
  {J{\"o}nsson}, {Jordan}, {Karovicova}, {Khalatyan}, {Kelz}, {Kennicutt},
  {King}, {Kitaura}, {Klar}, {Klauser}, {Kneib}, {Koch}, {Koposov},
  {Kordopatis}, {Korn}, {Kosmalski}, {Kotak}, {Kovalev}, {Kreckel}, {Kripak},
  {Krumpe}, {Kuijken}, {Kunder}, {Kushniruk}, {Lam}, {Lamer}, {Laurent},
  {Lawrence}, {Lehmitz}, {Lemasle}, {Lewis}, {Li}, {Lidman}, {Lind}, {Liske},
  {Lizon}, {Loveday}, {Ludwig}, {McDermid}, {Maguire}, {Mainieri}, {Mali},
  {Mandel}, {Mandel}, {Mannering}, {Martell}, {Martinez Delgado}, {Matijevic},
  {McGregor}, {McMahon}, {McMillan}, {Mena}, {Merloni}, {Meyer}, {Michel},
  {Micheva}, {Migniau}, {Minchev}, {Monari}, {Muller}, {Murphy},
  {Muthukrishna}, {Nandra}, {Navarro}, {Ness}, {Nichani}, {Nichol}, {Nicklas},
  {Niederhofer}, {Norberg}, {Obreschkow}, {Oliver}, {Owers}, {Pai},
  {Pankratow}, {Parkinson}, {Paschke}, {Paterson}, {Pecontal}, {Parry},
  {Phillips}, {Pillepich}, {Pinard}, {Pirard}, {Piskunov}, {Plank},
  {Pl{\"u}schke}, {Pons}, {Popesso}, {Power}, {Pragt}, {Pramskiy}, {Pryer},
  {Quattri}, {Queiroz}, {Quirrenbach}, {Rahurkar}, {Raichoor}, {Ramstedt},
  {Rau}, {Recio-Blanco}, {Reiss}, {Renaud}, {Revaz}, {Rhode}, {Richard},
  {Richter}, {Rix}, {Robotham}, {Roelfsema}, {Romaniello}, {Rosario},
  {Rothmaier}, {Roukema}, {Ruchti}, {Rupprecht}, {Rybizki}, {Ryde}, {Saar},
  {Sadler}, {Sahl{\'e}n}, {Salvato}, {Sassolas}, {Saunders}, {Saviauk},
  {Sbordone}, {Schmidt}, {Schnurr}, {Scholz}, {Schwope}, {Seifert}, {Shanks},
  {Sheinis}, {Sivov}, {Sk{\'u}lad{\'o}ttir}, {Smartt}, {Smedley}, {Smith},
  {Smith}, {Sorce}, {Spitler}, {Starkenburg}, {Steinmetz}, {Stilz}, {Storm},
  {Sullivan}, {Sutherland}, {Swann}, {Tamone}, {Taylor}, {Teillon}, {Tempel},
  {ter Horst}, {Thi}, {Tolstoy}, {Trager}, {Traven}, {Tremblay}, {Tresse},
  {Valentini}, {van de Weygaert}, {van den Ancker}, {Veljanoski}, {Venkatesan},
  {Wagner}, {Wagner}, {Walcher}, {Waller}, {Walton}, {Wang}, {Winkler},
  {Wisotzki}, {Worley}, {Worseck}, {Xiang}, {Xu}, {Yong}, {Zhao}, {Zheng},
  {Zscheyge}, \& {Zucker}}]{2019Msngr.175....3D}
{de Jong}, R.~S., {Agertz}, O., {Berbel}, A.~A., {et~al.} 2019, The Messenger,
  175, 3, \dodoi{10.18727/0722-6691/5117}

\bibitem[{{De Silva} {et~al.}(2015){De Silva}, {Freeman}, {Bland-Hawthorn},
  {Martell}, {de Boer}, {Asplund}, {Keller}, {Sharma}, {Zucker}, {Zwitter},
  {Anguiano}, {Bacigalupo}, {Bayliss}, {Beavis}, {Bergemann}, {Campbell},
  {Cannon}, {Carollo}, {Casagrande}, {Casey}, {Da Costa}, {D'Orazi}, {Dotter},
  {Duong}, {Heger}, {Ireland}, {Kafle}, {Kos}, {Lattanzio}, {Lewis}, {Lin},
  {Lind}, {Munari}, {Nataf}, {O'Toole}, {Parker}, {Reid}, {Schlesinger},
  {Sheinis}, {Simpson}, {Stello}, {Ting}, {Traven}, {Watson}, {Wittenmyer},
  {Yong}, \& {{\v{Z}}erjal}}]{2015MNRAS.449.2604D}
{De Silva}, G.~M., {Freeman}, K.~C., {Bland-Hawthorn}, J., {et~al.} 2015,
  \mnras, 449, 2604, \dodoi{10.1093/mnras/stv327}

\bibitem[{{Deng} {et~al.}(2012){Deng}, {Newberg}, {Liu}, {Carlin}, {Beers},
  {Chen}, {Chen}, {Christlieb}, {Grillmair}, {Guhathakurta}, {Han}, {Hou},
  {Lee}, {L{\'e}pine}, {Li}, {Liu}, {Pan}, {Sellwood}, {Wang}, {Wang}, {Yang},
  {Yanny}, {Zhang}, {Zhang}, {Zheng}, \& {Zhu}}]{2012RAA....12..735D}
{Deng}, L.-C., {Newberg}, H.~J., {Liu}, C., {et~al.} 2012, Research in
  Astronomy and Astrophysics, 12, 735, \dodoi{10.1088/1674-4527/12/7/003}

\bibitem[{{Edelmann} {et~al.}(2019){Edelmann}, {Ratnasingam}, {Pedersen},
  {Bowman}, {Prat}, \& {Rogers}}]{Edelmann2019}
{Edelmann}, P.~V.~F., {Ratnasingam}, R.~P., {Pedersen}, M.~G., {et~al.} 2019,
  \apj, 876, 4, \dodoi{10.3847/1538-4357/ab12df}

\bibitem[{{Gaia Collaboration} {et~al.}(2016){Gaia Collaboration}, {Prusti},
  {de Bruijne}, {Brown}, {Vallenari}, {Babusiaux}, {Bailer-Jones}, {Bastian},
  {Biermann}, {Evans}, {Eyer}, {Jansen}, {Jordi}, {Klioner}, {Lammers},
  {Lindegren}, {Luri}, {Mignard}, {Milligan}, {Panem}, {Poinsignon},
  {Pourbaix}, {Randich}, {Sarri}, {Sartoretti}, {Siddiqui}, {Soubiran},
  {Valette}, {van Leeuwen}, {Walton}, {Aerts}, {Arenou}, {Cropper}, {Drimmel},
  {H{\o}g}, {Katz}, {Lattanzi}, {O'Mullane}, {Grebel}, {Holland}, {Huc},
  {Passot}, {Bramante}, {Cacciari}, {Casta{\~n}eda}, {Chaoul}, {Cheek}, {De
  Angeli}, {Fabricius}, {Guerra}, {Hern{\'a}ndez}, {Jean-Antoine-Piccolo},
  {Masana}, {Messineo}, {Mowlavi}, {Nienartowicz}, {Ord{\'o}{\~n}ez-Blanco},
  {Panuzzo}, {Portell}, {Richards}, {Riello}, {Seabroke}, {Tanga},
  {Th{\'e}venin}, {Torra}, {Els}, {Gracia-Abril}, {Comoretto},
  {Garcia-Reinaldos}, {Lock}, {Mercier}, {Altmann}, {Andrae}, {Astraatmadja},
  {Bellas-Velidis}, {Benson}, {Berthier}, {Blomme}, {Busso}, {Carry},
  {Cellino}, {Clementini}, {Cowell}, {Creevey}, {Cuypers}, {Davidson}, {De
  Ridder}, {de Torres}, {Delchambre}, {Dell'Oro}, {Ducourant}, {Fr{\'e}mat},
  {Garc{\'\i}a-Torres}, {Gosset}, {Halbwachs}, {Hambly}, {Harrison}, {Hauser},
  {Hestroffer}, {Hodgkin}, {Huckle}, {Hutton}, {Jasniewicz}, {Jordan},
  {Kontizas}, {Korn}, {Lanzafame}, {Manteiga}, {Moitinho}, {Muinonen},
  {Osinde}, {Pancino}, {Pauwels}, {Petit}, {Recio-Blanco}, {Robin}, {Sarro},
  {Siopis}, {Smith}, {Smith}, {Sozzetti}, {Thuillot}, {van Reeven}, {Viala},
  {Abbas}, {Abreu Aramburu}, {Accart}, {Aguado}, {Allan}, {Allasia},
  {Altavilla}, {{\'A}lvarez}, {Alves}, {Anderson}, {Andrei}, {Anglada Varela},
  {Antiche}, {Antoja}, {Ant{\'o}n}, {Arcay}, {Atzei}, {Ayache}, {Bach},
  {Baker}, {Balaguer-N{\'u}{\~n}ez}, {Barache}, {Barata}, {Barbier}, {Barblan},
  {Baroni}, {Barrado y Navascu{\'e}s}, {Barros}, {Barstow}, {Becciani},
  {Bellazzini}, {Bellei}, {Bello Garc{\'\i}a}, {Belokurov}, {Bendjoya},
  {Berihuete}, {Bianchi}, {Bienaym{\'e}}, {Billebaud}, {Blagorodnova},
  {Blanco-Cuaresma}, {Boch}, {Bombrun}, {Borrachero}, {Bouquillon}, {Bourda},
  {Bouy}, {Bragaglia}, {Breddels}, {Brouillet}, {Br{\"u}semeister},
  {Bucciarelli}, {Budnik}, {Burgess}, {Burgon}, {Burlacu}, {Busonero}, {Buzzi},
  {Caffau}, {Cambras}, {Campbell}, {Cancelliere}, {Cantat-Gaudin}, {Carlucci},
  {Carrasco}, {Castellani}, {Charlot}, {Charnas}, {Charvet}, {Chassat},
  {Chiavassa}, {Clotet}, {Cocozza}, {Collins}, {Collins}, {Costigan}, {Crifo},
  {Cross}, {Crosta}, {Crowley}, {Dafonte}, {Damerdji}, {Dapergolas}, {David},
  {David}, {De Cat}, {de Felice}, {de Laverny}, {De Luise}, {De March}, {de
  Martino}, {de Souza}, {Debosscher}, {del Pozo}, {Delbo}, {Delgado},
  {Delgado}, {di Marco}, {Di Matteo}, {Diakite}, {Distefano}, {Dolding}, {Dos
  Anjos}, {Drazinos}, {Dur{\'a}n}, {Dzigan}, {Ecale}, {Edvardsson}, {Enke},
  {Erdmann}, {Escolar}, {Espina}, {Evans}, {Eynard Bontemps}, {Fabre},
  {Fabrizio}, {Faigler}, {Falc{\~a}o}, {Farr{\`a}s Casas}, {Faye}, {Federici},
  {Fedorets}, {Fern{\'a}ndez-Hern{\'a}ndez}, {Fernique}, {Fienga}, {Figueras},
  {Filippi}, {Findeisen}, {Fonti}, {Fouesneau}, {Fraile}, {Fraser}, {Fuchs},
  {Furnell}, {Gai}, {Galleti}, {Galluccio}, {Garabato}, {Garc{\'\i}a-Sedano},
  {Gar{\'e}}, {Garofalo}, {Garralda}, {Gavras}, {Gerssen}, {Geyer}, {Gilmore},
  {Girona}, {Giuffrida}, {Gomes}, {Gonz{\'a}lez-Marcos},
  {Gonz{\'a}lez-N{\'u}{\~n}ez}, {Gonz{\'a}lez-Vidal}, {Granvik}, {Guerrier},
  {Guillout}, {Guiraud}, {G{\'u}rpide}, {Guti{\'e}rrez-S{\'a}nchez}, {Guy},
  {Haigron}, {Hatzidimitriou}, {Haywood}, {Heiter}, {Helmi}, {Hobbs},
  {Hofmann}, {Holl}, {Holland}, {Hunt}, {Hypki}, {Icardi}, {Irwin}, {Jevardat
  de Fombelle}, {Jofr{\'e}}, {Jonker}, {Jorissen}, {Julbe}, {Karampelas},
  {Kochoska}, {Kohley}, {Kolenberg}, {Kontizas}, {Koposov}, {Kordopatis},
  {Koubsky}, {Kowalczyk}, {Krone-Martins}, {Kudryashova}, {Kull}, {Bachchan},
  {Lacoste-Seris}, {Lanza}, {Lavigne}, {Le Poncin-Lafitte}, {Lebreton},
  {Lebzelter}, {Leccia}, {Leclerc}, {Lecoeur-Taibi}, {Lemaitre}, {Lenhardt},
  {Leroux}, {Liao}, {Licata}, {Lindstr{\o}m}, {Lister}, {Livanou}, {Lobel},
  {L{\"o}ffler}, {L{\'o}pez}, {Lopez-Lozano}, {Lorenz}, {Loureiro},
  {MacDonald}, {Magalh{\~a}es Fernandes}, {Managau}, {Mann}, {Mantelet},
  {Marchal}, {Marchant}, {Marconi}, {Marie}, {Marinoni}, {Marrese},
  {Marschalk{\'o}}, {Marshall}, {Mart{\'\i}n-Fleitas}, {Martino}, {Mary},
  {Matijevi{\v{c}}}, {Mazeh}, {McMillan}, {Messina}, {Mestre}, {Michalik},
  {Millar}, {Miranda}, {Molina}, {Molinaro}, {Molinaro}, {Moln{\'a}r},
  {Moniez}, {Montegriffo}, {Monteiro}, {Mor}, {Mora}, {Morbidelli}, {Morel},
  {Morgenthaler}, {Morley}, {Morris}, {Mulone}, {Muraveva}, {Musella},
  {Narbonne}, {Nelemans}, {Nicastro}, {Noval}, {Ord{\'e}novic},
  {Ordieres-Mer{\'e}}, {Osborne}, {Pagani}, {Pagano}, {Pailler}, {Palacin},
  {Palaversa}, {Parsons}, {Paulsen}, {Pecoraro}, {Pedrosa}, {Pentik{\"a}inen},
  {Pereira}, {Pichon}, {Piersimoni}, {Pineau}, {Plachy}, {Plum}, {Poujoulet},
  {Pr{\v{s}}a}, {Pulone}, {Ragaini}, {Rago}, {Rambaux}, {Ramos-Lerate},
  {Ranalli}, {Rauw}, {Read}, {Regibo}, {Renk}, {Reyl{\'e}}, {Ribeiro},
  {Rimoldini}, {Ripepi}, {Riva}, {Rixon}, {Roelens}, {Romero-G{\'o}mez},
  {Rowell}, {Royer}, {Rudolph}, {Ruiz-Dern}, {Sadowski}, {Sagrist{\`a}
  Sell{\'e}s}, {Sahlmann}, {Salgado}, {Salguero}, {Sarasso}, {Savietto},
  {Schnorhk}, {Schultheis}, {Sciacca}, {Segol}, {Segovia}, {Segransan},
  {Serpell}, {Shih}, {Smareglia}, {Smart}, {Smith}, {Solano}, {Solitro},
  {Sordo}, {Soria Nieto}, {Souchay}, {Spagna}, {Spoto}, {Stampa}, {Steele},
  {Steidelm{\"u}ller}, {Stephenson}, {Stoev}, {Suess}, {S{\"u}veges}, {Surdej},
  {Szabados}, {Szegedi-Elek}, {Tapiador}, {Taris}, {Tauran}, {Taylor},
  {Teixeira}, {Terrett}, {Tingley}, {Trager}, {Turon}, {Ulla}, {Utrilla},
  {Valentini}, {van Elteren}, {Van Hemelryck}, {van Leeuwen}, {Varadi},
  {Vecchiato}, {Veljanoski}, {Via}, {Vicente}, {Vogt}, {Voss}, {Votruba},
  {Voutsinas}, {Walmsley}, {Weiler}, {Weingrill}, {Werner}, {Wevers},
  {Whitehead}, {Wyrzykowski}, {Yoldas}, {{\v{Z}}erjal}, {Zucker}, {Zurbach},
  {Zwitter}, {Alecu}, {Allen}, {Allende Prieto}, {Amorim},
  {Anglada-Escud{\'e}}, {Arsenijevic}, {Azaz}, {Balm}, {Beck}, {Bernstein},
  {Bigot}, {Bijaoui}, {Blasco}, {Bonfigli}, {Bono}, {Boudreault}, {Bressan},
  {Brown}, {Brunet}, {Bunclark}, {Buonanno}, {Butkevich}, {Carret}, {Carrion},
  {Chemin}, {Ch{\'e}reau}, {Corcione}, {Darmigny}, {de Boer}, {de Teodoro}, {de
  Zeeuw}, {Delle Luche}, {Domingues}, {Dubath}, {Fodor}, {Fr{\'e}zouls},
  {Fries}, {Fustes}, {Fyfe}, {Gallardo}, {Gallegos}, {Gardiol}, {Gebran},
  {Gomboc}, {G{\'o}mez}, {Grux}, {Gueguen}, {Heyrovsky}, {Hoar}, {Iannicola},
  {Isasi Parache}, {Janotto}, {Joliet}, {Jonckheere}, {Keil}, {Kim},
  {Klagyivik}, {Klar}, {Knude}, {Kochukhov}, {Kolka}, {Kos}, {Kutka}, {Lainey},
  {LeBouquin}, {Liu}, {Loreggia}, {Makarov}, {Marseille}, {Martayan},
  {Martinez-Rubi}, {Massart}, {Meynadier}, {Mignot}, {Munari}, {Nguyen},
  {Nordlander}, {Ocvirk}, {O'Flaherty}, {Olias Sanz}, {Ortiz}, {Osorio},
  {Oszkiewicz}, {Ouzounis}, {Palmer}, {Park}, {Pasquato}, {Peltzer}, {Peralta},
  {P{\'e}turaud}, {Pieniluoma}, {Pigozzi}, {Poels}, {Prat}, {Prod'homme},
  {Raison}, {Rebordao}, {Risquez}, {Rocca-Volmerange}, {Rosen}, {Ruiz-Fuertes},
  {Russo}, {Sembay}, {Serraller Vizcaino}, {Short}, {Siebert}, {Silva},
  {Sinachopoulos}, {Slezak}, {Soffel}, {Sosnowska}, {Strai{\v{z}}ys}, {ter
  Linden}, {Terrell}, {Theil}, {Tiede}, {Troisi}, {Tsalmantza}, {Tur},
  {Vaccari}, {Vachier}, {Valles}, {Van Hamme}, {Veltz}, {Virtanen}, {Wallut},
  {Wichmann}, {Wilkinson}, {Ziaeepour}, \& {Zschocke}}]{2016A&A...595A...1G}
{Gaia Collaboration}, {Prusti}, T., {de Bruijne}, J.~H.~J., {et~al.} 2016,
  \aap, 595, A1, \dodoi{10.1051/0004-6361/201629272}

\bibitem[{{Gebruers} {et~al.}(2021){Gebruers}, {Straumit}, {Tkachenko},
  {Mombarg}, {Pedersen}, {Van Reeth}, {Li}, {Lampens}, {Escorza}, {Bowman}, {De
  Cat}, {Vermeylen}, {Bodensteiner}, {Rix}, \& {Aerts}}]{Gebruers2021}
{Gebruers}, S., {Straumit}, I., {Tkachenko}, A., {et~al.} 2021, \aap, 650,
  A151, \dodoi{10.1051/0004-6361/202140466}

\bibitem[{Giddings(1981)}]{Giddings1981}
Giddings, J. 1981, PhD thesis, University of London, UK

\bibitem[{{Gilmore} {et~al.}(2012){Gilmore}, {Randich}, {Asplund}, {Binney},
  {Bonifacio}, {Drew}, {Feltzing}, {Ferguson}, {Jeffries}, {Micela},
  {Negueruela}, {Prusti}, {Rix}, {Vallenari}, {Alfaro}, {Allende-Prieto},
  {Babusiaux}, {Bensby}, {Blomme}, {Bragaglia}, {Flaccomio}, {Fran{\c{c}}ois},
  {Irwin}, {Koposov}, {Korn}, {Lanzafame}, {Pancino}, {Paunzen},
  {Recio-Blanco}, {Sacco}, {Smiljanic}, {Van Eck}, {Walton}, {Aden}, {Aerts},
  {Affer}, {Alcala}, {Altavilla}, {Alves}, {Antoja}, {Arenou}, {Argiroffi},
  {Asensio Ramos}, {Bailer-Jones}, {Balaguer-Nunez}, {Bayo}, {Barbuy},
  {Barisevicius}, {Barrado y Navascues}, {Battistini}, {Bellas Velidis},
  {Bellazzini}, {Belokurov}, {Bergemann}, {Bertelli}, {Biazzo}, {Bienayme},
  {Bland-Hawthorn}, {Boeche}, {Bonito}, {Boudreault}, {Bouvier}, {Brandao},
  {Brown}, {de Bruijne}, {Burleigh}, {Caballero}, {Caffau}, {Calura},
  {Capuzzo-Dolcetta}, {Caramazza}, {Carraro}, {Casagrande}, {Casewell},
  {Chapman}, {Chiappini}, {Chorniy}, {Christlieb}, {Cignoni}, {Cocozza},
  {Colless}, {Collet}, {Collins}, {Correnti}, {Covino}, {Crnojevic}, {Cropper},
  {Cunha}, {Damiani}, {David}, {Delgado}, {Duffau}, {Edvardsson}, {Eldridge},
  {Enke}, {Eriksson}, {Evans}, {Eyer}, {Famaey}, {Fellhauer}, {Ferreras},
  {Figueras}, {Fiorentino}, {Flynn}, {Folha}, {Franciosini}, {Frasca},
  {Freeman}, {Fremat}, {Friel}, {Gaensicke}, {Gameiro}, {Garzon}, {Geier},
  {Geisler}, {Gerhard}, {Gibson}, {Gomboc}, {Gomez}, {Gonzalez-Fernandez},
  {Gonzalez Hernandez}, {Gosset}, {Grebel}, {Greimel}, {Groenewegen},
  {Grundahl}, {Guarcello}, {Gustafsson}, {Hadrava}, {Hatzidimitriou}, {Hambly},
  {Hammersley}, {Hansen}, {Haywood}, {Heber}, {Heiter}, {Held}, {Helmi},
  {Hensler}, {Herrero}, {Hill}, {Hodgkin}, {Huelamo}, {Huxor}, {Ibata},
  {Jackson}, {de Jong}, {Jonker}, {Jordan}, {Jordi}, {Jorissen}, {Katz},
  {Kawata}, {Keller}, {Kharchenko}, {Klement}, {Klutsch}, {Knude}, {Koch},
  {Kochukhov}, {Kontizas}, {Koubsky}, {Lallement}, {de Laverny}, {van Leeuwen},
  {Lemasle}, {Lewis}, {Lind}, {Lindstrom}, {Lobel}, {Lopez Santiago}, {Lucas},
  {Ludwig}, {Lueftinger}, {Magrini}, {Maiz Apellaniz}, {Maldonado}, {Marconi},
  {Marino}, {Martayan}, {Martinez-Valpuesta}, {Matijevic}, {McMahon},
  {Messina}, {Meyer}, {Miglio}, {Mikolaitis}, {Minchev}, {Minniti}, {Moitinho},
  {Momany}, {Monaco}, {Montalto}, {Monteiro}, {Monier}, {Montes}, {Mora},
  {Moraux}, {Morel}, {Mowlavi}, {Mucciarelli}, {Munari}, {Napiwotzki},
  {Nardetto}, {Naylor}, {Naze}, {Nelemans}, {Okamoto}, {Ortolani}, {Pace},
  {Palla}, {Palous}, {Parker}, {Penarrubia}, {Pillitteri}, {Piotto}, {Posbic},
  {Prisinzano}, {Puzeras}, {Quirrenbach}, {Ragaini}, {Read}, {Read}, {Reyle},
  {De Ridder}, {Robichon}, {Robin}, {Roeser}, {Romano}, {Royer}, {Ruchti},
  {Ruzicka}, {Ryan}, {Ryde}, {Santos}, {Sanz Forcada}, {Sarro Baro},
  {Sbordone}, {Schilbach}, {Schmeja}, {Schnurr}, {Schoenrich}, {Scholz},
  {Seabroke}, {Sharma}, {De Silva}, {Smith}, {Solano}, {Sordo}, {Soubiran},
  {Sousa}, {Spagna}, {Steffen}, {Steinmetz}, {Stelzer}, {Stempels},
  {Tabernero}, {Tautvaisiene}, {Thevenin}, {Torra}, {Tosi}, {Tolstoy}, {Turon},
  {Walker}, {Wambsganss}, {Worley}, {Venn}, {Vink}, {Wyse}, {Zaggia},
  {Zeilinger}, {Zoccali}, {Zorec}, {Zucker}, {Zwitter}, \& {Gaia-ESO Survey
  Team}}]{2012Msngr.147...25G}
{Gilmore}, G., {Randich}, S., {Asplund}, M., {et~al.} 2012, The Messenger, 147,
  25

\bibitem[{{Gunn} {et~al.}(2006){Gunn}, {Siegmund}, {Mannery}, {Owen}, {Hull},
  {Leger}, {Carey}, {Knapp}, {York}, {Boroski}, {Kent}, {Lupton}, {Rockosi},
  {Evans}, {Waddell}, {Anderson}, {Annis}, {Barentine}, {Bartoszek}, {Bastian},
  {Bracker}, {Brewington}, {Briegel}, {Brinkmann}, {Brown}, {Carr},
  {Czarapata}, {Drennan}, {Dombeck}, {Federwitz}, {Gillespie}, {Gonzales},
  {Hansen}, {Harvanek}, {Hayes}, {Jordan}, {Kinney}, {Klaene}, {Kleinman},
  {Kron}, {Kresinski}, {Lee}, {Limmongkol}, {Lindenmeyer}, {Long}, {Loomis},
  {McGehee}, {Mantsch}, {Neilsen}, {Neswold}, {Newman}, {Nitta}, {Peoples},
  {Pier}, {Prieto}, {Prosapio}, {Rivetta}, {Schneider}, {Snedden}, \&
  {Wang}}]{2006AJ....131.2332G}
{Gunn}, J.~E., {Siegmund}, W.~A., {Mannery}, E.~J., {et~al.} 2006, \aj, 131,
  2332, \dodoi{10.1086/500975}

\bibitem[{{Horst} {et~al.}(2020){Horst}, {Edelmann}, {Andr{\'a}ssy},
  {R{\"o}pke}, {Bowman}, {Aerts}, \& {Ratnasingam}}]{Horst2020}
{Horst}, L., {Edelmann}, P.~V.~F., {Andr{\'a}ssy}, R., {et~al.} 2020, \aap,
  641, A18, \dodoi{10.1051/0004-6361/202037531}

\bibitem[{{Howell} {et~al.}(2014){Howell}, {Sobeck}, {Haas}, {Still},
  {Barclay}, {Mullally}, {Troeltzsch}, {Aigrain}, {Bryson}, {Caldwell},
  {Chaplin}, {Cochran}, {Huber}, {Marcy}, {Miglio}, {Najita}, {Smith},
  {Twicken}, \& {Fortney}}]{2014PASP..126..398H}
{Howell}, S.~B., {Sobeck}, C., {Haas}, M., {et~al.} 2014, \pasp, 126, 398,
  \dodoi{10.1086/676406}

\bibitem[{{Hubeny} \& {Lanz}(1995)}]{Hubeny1995}
{Hubeny}, I., \& {Lanz}, T. 1995, \apj, 439, 875, \dodoi{10.1086/175226}

\bibitem[{{Kollmeier} {et~al.}(2017){Kollmeier}, {Zasowski}, {Rix}, {Johns},
  {Anderson}, {Drory}, {Johnson}, {Pogge}, {Bird}, {Blanc}, {Brownstein},
  {Crane}, {De Lee}, {Klaene}, {Kreckel}, {MacDonald}, {Merloni}, {Ness},
  {O'Brien}, {Sanchez-Gallego}, {Sayres}, {Shen}, {Thakar}, {Tkachenko},
  {Aerts}, {Blanton}, {Eisenstein}, {Holtzman}, {Maoz}, {Nandra}, {Rockosi},
  {Weinberg}, {Bovy}, {Casey}, {Chaname}, {Clerc}, {Conroy}, {Eracleous},
  {G{\"a}nsicke}, {Hekker}, {Horne}, {Kauffmann}, {McQuinn}, {Pellegrini},
  {Schinnerer}, {Schlafly}, {Schwope}, {Seibert}, {Teske}, \& {van
  Saders}}]{Kollmeier2017}
{Kollmeier}, J.~A., {Zasowski}, G., {Rix}, H.-W., {et~al.} 2017, arXiv
  e-prints, arXiv:1711.03234.
\newblock \doarXiv{1711.03234}

\bibitem[{Kreidberg(2017)}]{2018haex.bookE.100K}
Kreidberg, L. 2017, Exoplanet Atmosphere Measurements from Transmission
  Spectroscopy and Other Planet Star Combined Light Observations (Cham:
  Springer International Publishing), 1--23,
  \dodoi{10.1007/978-3-319-30648-3_100-1}

\bibitem[{{Lenorzer} {et~al.}(2004){Lenorzer}, {Mokiem}, {de Koter}, \&
  {Puls}}]{2004A&A...422..275L}
{Lenorzer}, A., {Mokiem}, M.~R., {de Koter}, A., \& {Puls}, J. 2004, \aap, 422,
  275, \dodoi{10.1051/0004-6361:20047174}

\bibitem[{Limbach {et~al.}(2020)Limbach, Schmidt, DePoy, Mason, Scobey, Brown,
  Taylor, \& Marshall}]{10.1117/12.2562371}
Limbach, M.~A., Schmidt, L.~M., DePoy, D.~L., {et~al.} 2020, in Ground-based
  and Airborne Instrumentation for Astronomy VIII, ed. C.~J. Evans, J.~J.
  Bryant, \& K.~Motohara, Vol. 11447, International Society for Optics and
  Photonics (SPIE), 1634 -- 1647, \dodoi{10.1117/12.2562371}

\bibitem[{{Liu} {et~al.}(2019){Liu}, {Jiang}, {He}, {Chen}, {Liu}, {Gao}, \&
  {Han}}]{2019arXiv190803265L}
{Liu}, L., {Jiang}, H., {He}, P., {et~al.} 2019, arXiv e-prints,
  arXiv:1908.03265.
\newblock \doarXiv{1908.03265}

\bibitem[{{Luo} {et~al.}(2021){Luo}, {Zhao}, {Li}, {Guo}, \& {Liu}}]{Luo2021}
{Luo}, F., {Zhao}, Y.-H., {Li}, J., {Guo}, Y.-J., \& {Liu}, C. 2021, arXiv
  e-prints, arXiv:2108.11120.
\newblock \doarXiv{2108.11120}

\bibitem[{{Massey} {et~al.}(2009){Massey}, {Zangari}, {Morrell}, {Puls},
  {DeGioia-Eastwood}, {Bresolin}, \& {Kudritzki}}]{2009ApJ...692..618M}
{Massey}, P., {Zangari}, A.~M., {Morrell}, N.~I., {et~al.} 2009, \apj, 692,
  618, \dodoi{10.1088/0004-637X/692/1/618}

\bibitem[{{Ness} {et~al.}(2015){Ness}, {Hogg}, {Rix}, {Ho}, \&
  {Zasowski}}]{2015ApJ...808...16N}
{Ness}, M., {Hogg}, D.~W., {Rix}, H.~W., {Ho}, A. Y.~Q., \& {Zasowski}, G.
  2015, \apj, 808, 16, \dodoi{10.1088/0004-637X/808/1/16}

\bibitem[{{Nidever} {et~al.}(2015){Nidever}, {Holtzman}, {Allende Prieto},
  {Beland}, {Bender}, {Bizyaev}, {Burton}, {Desphande}, {Fleming}, {Garc{\'\i}a
  P{\'e}rez}, {Hearty}, {Majewski}, {M{\'e}sz{\'a}ros}, {Muna}, {Nguyen},
  {Schiavon}, {Shetrone}, {Skrutskie}, {Sobeck}, \& {Wilson}}]{Nidever2015}
{Nidever}, D.~L., {Holtzman}, J.~A., {Allende Prieto}, C., {et~al.} 2015, \aj,
  150, 173, \dodoi{10.1088/0004-6256/150/6/173}

\bibitem[{{Nieva} \& {Przybilla}(2007)}]{Nieva2007}
{Nieva}, M.~F., \& {Przybilla}, N. 2007, \aap, 467, 295,
  \dodoi{10.1051/0004-6361:20065757}

\bibitem[{{O'Briain} {et~al.}(2021){O'Briain}, {Ting}, {Fabbro}, {Yi}, {Venn},
  \& {Bialek}}]{2021ApJ...906..130O}
{O'Briain}, T., {Ting}, Y.-S., {Fabbro}, S., {et~al.} 2021, \apj, 906, 130,
  \dodoi{10.3847/1538-4357/abca96}

\bibitem[{{P{\'a}pics} {et~al.}(2017){P{\'a}pics}, {Tkachenko}, {Van Reeth},
  {Aerts}, {Moravveji}, {Van de Sande}, {De Smedt}, {Bloemen}, {Southworth},
  {Debosscher}, {Niemczura}, \& {Gameiro}}]{Papics2017}
{P{\'a}pics}, P.~I., {Tkachenko}, A., {Van Reeth}, T., {et~al.} 2017, \aap,
  598, A74, \dodoi{10.1051/0004-6361/201629814}

\bibitem[{{Pedersen} {et~al.}(2018){Pedersen}, {Aerts}, {P{\'a}pics}, \&
  {Rogers}}]{Pedersen2018}
{Pedersen}, M.~G., {Aerts}, C., {P{\'a}pics}, P.~I., \& {Rogers}, T.~M. 2018,
  \aap, 614, A128, \dodoi{10.1051/0004-6361/201732317}

\bibitem[{{Pedersen} {et~al.}(2021){Pedersen}, {Aerts}, {P{\'a}pics},
  {Michielsen}, {Gebruers}, {Rogers}, {Molenberghs}, {Burssens}, {Garcia}, \&
  {Bowman}}]{Pedersen2021}
{Pedersen}, M.~G., {Aerts}, C., {P{\'a}pics}, P.~I., {et~al.} 2021, Nature
  Astronomy, 5, 715, \dodoi{10.1038/s41550-021-01351-x}

\bibitem[{{Piskunov} \& {Valenti}(2017)}]{Piskunov2017}
{Piskunov}, N., \& {Valenti}, J.~A. 2017, \aap, 597, A16,
  \dodoi{10.1051/0004-6361/201629124}

\bibitem[{{Poggio} {et~al.}(2021){Poggio}, {Drimmel}, {Cantat-Gaudin}, {Ramos},
  {Ripepi}, {Zari}, {Andrae}, {Blomme}, {Chemin}, {Clementini}, {Figueras},
  {Fouesneau}, {Fr{\'e}mat}, {Lobel}, {Marshall}, {Muraveva}, \&
  {Romero-G{\'o}mez}}]{2021A&A...651A.104P}
{Poggio}, E., {Drimmel}, R., {Cantat-Gaudin}, T., {et~al.} 2021, \aap, 651,
  A104, \dodoi{10.1051/0004-6361/202140687}

\bibitem[{{Puls} {et~al.}(2005){Puls}, {Urbaneja}, {Venero}, {Repolust},
  {Springmann}, {Jokuthy}, \& {Mokiem}}]{Puls2005}
{Puls}, J., {Urbaneja}, M.~A., {Venero}, R., {et~al.} 2005, \aap, 435, 669,
  \dodoi{10.1051/0004-6361:20042365}

\bibitem[{{Ram{\'\i}rez-Preciado} {et~al.}(2020){Ram{\'\i}rez-Preciado},
  {Roman-Lopes}, {Rom{\'a}n-Z{\'u}{\~n}iga}, {Hern{\'a}ndez},
  {Garc{\'\i}a-Hern{\'a}ndez}, {Stassun}, {Stringfellow}, \&
  {Kim}}]{Ramirez-Preciado2020}
{Ram{\'\i}rez-Preciado}, V.~G., {Roman-Lopes}, A., {Rom{\'a}n-Z{\'u}{\~n}iga},
  C.~G., {et~al.} 2020, \apj, 894, 5, \dodoi{10.3847/1538-4357/ab8127}

\bibitem[{{Raskin} {et~al.}(2011){Raskin}, {van Winckel}, {Hensberge},
  {Jorissen}, {Lehmann}, {Waelkens}, {Avila}, {de Cuyper}, {Degroote},
  {Dubosson}, {Dumortier}, {Fr{\'e}mat}, {Laux}, {Michaud}, {Morren}, {Perez
  Padilla}, {Pessemier}, {Prins}, {Smolders}, {van Eck}, \&
  {Winkler}}]{Raskin2011}
{Raskin}, G., {van Winckel}, H., {Hensberge}, H., {et~al.} 2011, \aap, 526,
  A69, \dodoi{10.1051/0004-6361/201015435}

\bibitem[{{Richards} {et~al.}(2002){Richards}, {Fan}, {Newberg}, {Strauss},
  {Vanden Berk}, {Schneider}, {Yanny}, {Boucher}, {Burles}, {Frieman}, {Gunn},
  {Hall}, {Ivezi{\'c}}, {Kent}, {Loveday}, {Lupton}, {Rockosi}, {Schlegel},
  {Stoughton}, {SubbaRao}, \& {York}}]{2002AJ....123.2945R}
{Richards}, G.~T., {Fan}, X., {Newberg}, H.~J., {et~al.} 2002, \aj, 123, 2945,
  \dodoi{10.1086/340187}

\bibitem[{{Roman-Lopes} {et~al.}(2018){Roman-Lopes},
  {Rom{\'a}n-Z{\'u}{\~n}iga}, {Tapia}, {Chojnowski}, {G{\'o}mez Maqueo Chew},
  {Garc{\'\i}a-Hern{\'a}ndez}, {Borissova}, {Minniti}, {Covey},
  {Longa-Pe{\~n}a}, {Fernandez-Trincado}, {Zamora}, \&
  {Nitschelm}}]{Roman-Lopes2018}
{Roman-Lopes}, A., {Rom{\'a}n-Z{\'u}{\~n}iga}, C., {Tapia}, M., {et~al.} 2018,
  \apj, 855, 68, \dodoi{10.3847/1538-4357/aaac27}

\bibitem[{{Sana} {et~al.}(2012){Sana}, {de Mink}, {de Koter}, {Langer},
  {Evans}, {Gieles}, {Gosset}, {Izzard}, {Le Bouquin}, \&
  {Schneider}}]{Sana2012}
{Sana}, H., {de Mink}, S.~E., {de Koter}, A., {et~al.} 2012, Science, 337, 444,
  \dodoi{10.1126/science.1223344}

\bibitem[{{Sana} {et~al.}(2013){Sana}, {de Koter}, {de Mink}, {Dunstall},
  {Evans}, {H{\'e}nault-Brunet}, {Ma{\'\i}z Apell{\'a}niz},
  {Ram{\'\i}rez-Agudelo}, {Taylor}, {Walborn}, {Clark}, {Crowther}, {Herrero},
  {Gieles}, {Langer}, {Lennon}, \& {Vink}}]{Sana2013}
{Sana}, H., {de Koter}, A., {de Mink}, S.~E., {et~al.} 2013, \aap, 550, A107,
  \dodoi{10.1051/0004-6361/201219621}

\bibitem[{{Serenelli} {et~al.}(2021){Serenelli}, {Weiss}, {Aerts}, {Angelou},
  {Baroch}, {Bastian}, {Beck}, {Bergemann}, {Bestenlehner}, {Czekala},
  {Elias-Rosa}, {Escorza}, {Van Eylen}, {Feuillet}, {Gandolfi}, {Gieles},
  {Girardi}, {Lebreton}, {Lodieu}, {Martig}, {Miller Bertolami}, {Mombarg},
  {Morales}, {Moya}, {Nsamba}, {Pavlovski}, {Pedersen}, {Ribas}, {Schneider},
  {Silva Aguirre}, {Stassun}, {Tolstoy}, {Tremblay}, \&
  {Zwintz}}]{2021A&ARv..29....4S}
{Serenelli}, A., {Weiss}, A., {Aerts}, C., {et~al.} 2021, \aapr, 29, 4,
  \dodoi{10.1007/s00159-021-00132-9}

\bibitem[{{Shulyak} {et~al.}(2004){Shulyak}, {Tsymbal}, {Ryabchikova},
  {St{\"u}tz}, \& {Weiss}}]{Shulyak2004}
{Shulyak}, D., {Tsymbal}, V., {Ryabchikova}, T., {St{\"u}tz}, C., \& {Weiss},
  W.~W. 2004, \aap, 428, 993, \dodoi{10.1051/0004-6361:20034169}

\bibitem[{{Smee} {et~al.}(2013){Smee}, {Gunn}, {Uomoto}, {Roe}, {Schlegel},
  {Rockosi}, {Carr}, {Leger}, {Dawson}, {Olmstead}, {Brinkmann}, {Owen},
  {Barkhouser}, {Honscheid}, {Harding}, {Long}, {Lupton}, {Loomis}, {Anderson},
  {Annis}, {Bernardi}, {Bhardwaj}, {Bizyaev}, {Bolton}, {Brewington}, {Briggs},
  {Burles}, {Burns}, {Castander}, {Connolly}, {Davenport}, {Ebelke}, {Epps},
  {Feldman}, {Friedman}, {Frieman}, {Heckman}, {Hull}, {Knapp}, {Lawrence},
  {Loveday}, {Mannery}, {Malanushenko}, {Malanushenko}, {Merrelli}, {Muna},
  {Newman}, {Nichol}, {Oravetz}, {Pan}, {Pope}, {Ricketts}, {Shelden},
  {Sandford}, {Siegmund}, {Simmons}, {Smith}, {Snedden}, {Schneider},
  {SubbaRao}, {Tremonti}, {Waddell}, \& {York}}]{2013AJ....146...32S}
{Smee}, S.~A., {Gunn}, J.~E., {Uomoto}, A., {et~al.} 2013, \aj, 146, 32,
  \dodoi{10.1088/0004-6256/146/2/32}

\bibitem[{{Sobol}(1967)}]{Sobol1967}
{Sobol}, I.~M. 1967, USSR Comp. Math. and Math. Phys., 7, 86.
\newblock \url{http://dx.doi.org/10.1016/0041-5553(67)90144-9}

\bibitem[{{Stoughton} {et~al.}(2002){Stoughton}, {Lupton}, {Bernardi},
  {Blanton}, {Burles}, {Castander}, {Connolly}, {Eisenstein}, {Frieman},
  {Hennessy}, {Hindsley}, {Ivezi{\'c}}, {Kent}, {Kunszt}, {Lee}, {Meiksin},
  {Munn}, {Newberg}, {Nichol}, {Nicinski}, {Pier}, {Richards}, {Richmond},
  {Schlegel}, {Smith}, {Strauss}, {SubbaRao}, {Szalay}, {Thakar}, {Tucker},
  {Vanden Berk}, {Yanny}, {Adelman}, {Anderson}, {Anderson}, {Annis},
  {Bahcall}, {Bakken}, {Bartelmann}, {Bastian}, {Bauer}, {Berman},
  {B{\"o}hringer}, {Boroski}, {Bracker}, {Briegel}, {Briggs}, {Brinkmann},
  {Brunner}, {Carey}, {Carr}, {Chen}, {Christian}, {Colestock}, {Crocker},
  {Csabai}, {Czarapata}, {Dalcanton}, {Davidsen}, {Davis}, {Dehnen},
  {Dodelson}, {Doi}, {Dombeck}, {Donahue}, {Ellman}, {Elms}, {Evans}, {Eyer},
  {Fan}, {Federwitz}, {Friedman}, {Fukugita}, {Gal}, {Gillespie}, {Glazebrook},
  {Gray}, {Grebel}, {Greenawalt}, {Greene}, {Gunn}, {de Haas}, {Haiman},
  {Haldeman}, {Hall}, {Hamabe}, {Hansen}, {Harris}, {Harris}, {Harvanek},
  {Hawley}, {Hayes}, {Heckman}, {Helmi}, {Henden}, {Hogan}, {Hogg}, {Holmgren},
  {Holtzman}, {Huang}, {Hull}, {Ichikawa}, {Ichikawa}, {Johnston}, {Kauffmann},
  {Kim}, {Kimball}, {Kinney}, {Klaene}, {Kleinman}, {Klypin}, {Knapp},
  {Korienek}, {Krolik}, {Kron}, {Krzesi{\'n}ski}, {Lamb}, {Leger},
  {Limmongkol}, {Lindenmeyer}, {Long}, {Loomis}, {Loveday}, {MacKinnon},
  {Mannery}, {Mantsch}, {Margon}, {McGehee}, {McKay}, {McLean}, {Menou},
  {Merelli}, {Mo}, {Monet}, {Nakamura}, {Narayanan}, {Nash}, {Neilsen},
  {Newman}, {Nitta}, {Odenkirchen}, {Okada}, {Okamura}, {Ostriker}, {Owen},
  {Pauls}, {Peoples}, {Peterson}, {Petravick}, {Pope}, {Pordes}, {Postman},
  {Prosapio}, {Quinn}, {Rechenmacher}, {Rivetta}, {Rix}, {Rockosi}, {Rosner},
  {Ruthmansdorfer}, {Sandford}, {Schneider}, {Scranton}, {Sekiguchi}, {Sergey},
  {Sheth}, {Shimasaku}, {Smee}, {Snedden}, {Stebbins}, {Stubbs}, {Szapudi},
  {Szkody}, {Szokoly}, {Tabachnik}, {Tsvetanov}, {Uomoto}, {Vogeley}, {Voges},
  {Waddell}, {Walterbos}, {Wang}, {Watanabe}, {Weinberg}, {White}, {White},
  {Wilhite}, {Wolfe}, {Yasuda}, {York}, {Zehavi}, \& {Zheng}}]{Stoughton2002}
{Stoughton}, C., {Lupton}, R.~H., {Bernardi}, M., {et~al.} 2002, \aj, 123, 485,
  \dodoi{10.1086/324741}

\bibitem[{{Strauss} {et~al.}(2002){Strauss}, {Weinberg}, {Lupton}, {Narayanan},
  {Annis}, {Bernardi}, {Blanton}, {Burles}, {Connolly}, {Dalcanton}, {Doi},
  {Eisenstein}, {Frieman}, {Fukugita}, {Gunn}, {Ivezi{\'c}}, {Kent}, {Kim},
  {Knapp}, {Kron}, {Munn}, {Newberg}, {Nichol}, {Okamura}, {Quinn}, {Richmond},
  {Schlegel}, {Shimasaku}, {SubbaRao}, {Szalay}, {Vanden Berk}, {Vogeley},
  {Yanny}, {Yasuda}, {York}, \& {Zehavi}}]{2002AJ....124.1810S}
{Strauss}, M.~A., {Weinberg}, D.~H., {Lupton}, R.~H., {et~al.} 2002, \aj, 124,
  1810, \dodoi{10.1086/342343}

\bibitem[{{Sundqvist} {et~al.}(2019){Sundqvist}, {Bj{\"o}rklund}, {Puls}, \&
  {Najarro}}]{Sundqvist2019}
{Sundqvist}, J.~O., {Bj{\"o}rklund}, R., {Puls}, J., \& {Najarro}, F. 2019,
  \aap, 632, A126, \dodoi{10.1051/0004-6361/201936580}

\bibitem[{{Thomas} {et~al.}(2021){Thomas}, {Richardson}, {Eldridge},
  {Schaefer}, {Monnier}, {Sana}, {Moffat}, {Williams}, {Corcoran}, {Stevens},
  {Weigelt}, {Zainol}, {Anugu}, {Le Bouquin}, {ten Brummelaar}, {Campos},
  {Couperus}, {Davies}, {Ennis}, {Eversberg}, {Garde}, {Gardner}, {Fl{\'o}},
  {Kraus}, {Labdon}, {Lanthermann}, {Leadbeater}, {Lester}, {Maki}, {McBride},
  {Ozuyar}, {Ribeiro}, {Setterholm}, {Stober}, {Wood}, \&
  {Zurm{\"u}hl}}]{2021MNRAS.504.5221T}
{Thomas}, J.~D., {Richardson}, N.~D., {Eldridge}, J.~J., {et~al.} 2021, \mnras,
  504, 5221, \dodoi{10.1093/mnras/stab1181}

\bibitem[{{Ting} {et~al.}(2019){Ting}, {Conroy}, {Rix}, \&
  {Cargile}}]{2019ApJ...879...69T}
{Ting}, Y.-S., {Conroy}, C., {Rix}, H.-W., \& {Cargile}, P. 2019, \apj, 879,
  69, \dodoi{10.3847/1538-4357/ab2331}

\bibitem[{{Tkachenko}(2015)}]{Tkachenko2015}
{Tkachenko}, A. 2015, \aap, 581, A129, \dodoi{10.1051/0004-6361/201526513}

\bibitem[{{Tkachenko} {et~al.}(2020){Tkachenko}, {Pavlovski}, {Johnston},
  {Pedersen}, {Michielsen}, {Bowman}, {Southworth}, {Tsymbal}, \&
  {Aerts}}]{Tkachenko2020}
{Tkachenko}, A., {Pavlovski}, K., {Johnston}, C., {et~al.} 2020, \aap, 637,
  A60, \dodoi{10.1051/0004-6361/202037452}

\bibitem[{{Torres} {et~al.}(2010){Torres}, {Andersen}, \&
  {Gim{\'e}nez}}]{Torres2010}
{Torres}, G., {Andersen}, J., \& {Gim{\'e}nez}, A. 2010, \aapr, 18, 67,
  \dodoi{10.1007/s00159-009-0025-1}

\bibitem[{{Tsymbal}(1996)}]{Tsymbal1996}
{Tsymbal}, V. 1996, in Astronomical Society of the Pacific Conference Series,
  Vol. 108, M.A.S.S., Model Atmospheres and Spectrum Synthesis, ed. S.~J.
  {Adelman}, F.~{Kupka}, \& W.~W. {Weiss}, 198

\bibitem[{{Valenti} \& {Piskunov}(1996)}]{Valenti1996}
{Valenti}, J.~A., \& {Piskunov}, N. 1996, \aaps, 118, 595

\bibitem[{{Vanderspek}(2019)}]{2019ESS.....433312V}
{Vanderspek}, R. 2019, in AAS/Division for Extreme Solar Systems Abstracts,
  Vol.~51, AAS/Division for Extreme Solar Systems Abstracts, 333.12

\bibitem[{{Wilson} {et~al.}(2019){Wilson}, {Hearty}, {Skrutskie}, {Majewski},
  {Holtzman}, {Eisenstein}, {Gunn}, {Blank}, {Henderson}, {Smee}, {Nelson},
  {Nidever}, {Arns}, {Barkhouser}, {Barr}, {Beland}, {Bershady}, {Blanton},
  {Brunner}, {Burton}, {Carey}, {Carr}, {Colque}, {Crane}, {Damke}, {Davidson},
  {Dean}, {Di Mille}, {Don}, {Ebelke}, {Evans}, {Fitzgerald}, {Gillespie},
  {Hall}, {Harding}, {Harding}, {Hammond}, {Hancock}, {Harrison}, {Hope},
  {Horne}, {Karakla}, {Lam}, {Leger}, {MacDonald}, {Maseman}, {Matsunari},
  {Melton}, {Mitcheltree}, {O'Brien}, {O'Connell}, {Patten}, {Richardson},
  {Rieke}, {Rieke}, {Roman-Lopes}, {Schiavon}, {Sobeck}, {Stolberg}, {Stoll},
  {Tembe}, {Trujillo}, {Uomoto}, {Vernieri}, {Walker}, {Weinberg}, {Young},
  {Anthony-Brumfield}, {Bizyaev}, {Breslauer}, {De Lee}, {Downey}, {Halverson},
  {Huehnerhoff}, {Klaene}, {Leon}, {Long}, {Mahadevan}, {Malanushenko},
  {Nguyen}, {Owen}, {S{\'a}nchez-Gallego}, {Sayres}, {Shane}, {Shectman},
  {Shetrone}, {Skinner}, {Stauffer}, \& {Zhao}}]{2019PASP..131e5001W}
{Wilson}, J.~C., {Hearty}, F.~R., {Skrutskie}, M.~F., {et~al.} 2019, \pasp,
  131, 055001, \dodoi{10.1088/1538-3873/ab0075}

\bibitem[{{Xiang} {et~al.}(2021){Xiang}, {Rix}, {Ting}, {Kudritzki}, {Conroy},
  {Zari}, {Shi}, {Przybilla}, {Ramirez-Tannus}, {Tkachenko}, {Gebruers}, \&
  {Liu}}]{Xiang2021}
{Xiang}, M., {Rix}, H.-W., {Ting}, Y.-S., {et~al.} 2021, arXiv e-prints,
  arXiv:2108.02878.
\newblock \doarXiv{2108.02878}

\bibitem[{{Zari} {et~al.}(2021){Zari}, {Rix}, {Frankel}, {Xiang}, {Poggio},
  {Drimmel}, \& {Tkachenko}}]{Zari2021}
{Zari}, E., {Rix}, H.~W., {Frankel}, N., {et~al.} 2021, \aap, 650, A112,
  \dodoi{10.1051/0004-6361/202039726}

\end{thebibliography}
\bibliographystyle{aasjournal}



\end{document}